\documentclass[a4paper,11pt]{article}
\usepackage{jheppub}
\makeatletter
\gdef\@fpheader{}%
\makeatother
\pdfoutput=1
\usepackage{xcolor}
\usepackage[utf8]{inputenc}
\usepackage{amsmath,amssymb,amsbsy,amstext,amsthm,simplewick,amsfonts}
\usepackage{graphicx}
\usepackage{mathrsfs}
\usepackage{booktabs}
\usepackage{array}
\usepackage{tabularx}
\usepackage{longtable}
\usepackage{needspace}
\usepackage[final]{microtype}

\usepackage{wrapfig}
\usepackage{upgreek}
\usepackage{bm}
\usepackage{framed}
\usepackage{bbm}
\usepackage{textcomp}
\usepackage{tikz}
\usepackage{pifont}
\usetikzlibrary{matrix,shapes,fit,tikzmark,calc}
\usepackage{adjustbox}
\usepackage{makecell}
\usepackage[most]{tcolorbox}
\usepackage{physics}
\usepackage{empheq}
\usepackage[normalem]{ulem}
\usepackage{enumitem}
\usepackage{braket} 
\usepackage{dsfont}

\usepackage[framemethod=default]{mdframed}

\definecolor{UNALVerde}{RGB}{148,180,59}
\definecolor{diycol}{RGB}{0,0,0}
\definecolor{lightblue}{RGB}{70,130,180}
\definecolor{deepgray}{RGB}{153, 153, 153}
\definecolor{headlinetextcol}{RGB}{250,250,250}
\definecolor{lightred}{RGB}{245,168,224}
\definecolor{lightgreen}{RGB}{171, 211, 189}
\definecolor{LouisBlue}{RGB}{55, 114, 202}
\definecolor{LouisOrange}{RGB}{180, 54, 22}
\definecolor{LouisColor1}{RGB}{0, 118, 63}
\definecolor{LouisColor2}{RGB}{111, 73, 189}
\usetikzlibrary{decorations.markings}
\usetikzlibrary{decorations.pathreplacing,decorations.pathmorphing}
\definecolor{cInk}{HTML}{141413}\definecolor{cSlate}{HTML}{5E5D59}\definecolor{cGrey}{HTML}{87867F}
\definecolor{cCloud}{HTML}{F0EEE6}
\definecolor{cBlue}{HTML}{3A78B8}\definecolor{cClay}{HTML}{CC5A33}\definecolor{cTeal}{HTML}{1E9A80}
\definecolor{cGold}{HTML}{B87D12}\definecolor{cViolet}{HTML}{5A4FB8}
\tikzset{
 zpl/.style={>=latex,font=\footnotesize,line cap=round,text=cInk},
 zaxis/.style={->,thin,cSlate},
 zpole/.style={very thick,cClay},
 zsing/.style={thick,cBlue},
}
\newcommand{\zcross}[2][zpole]{\draw[#1] ($(#2)+(-0.09,-0.09)$)--($(#2)+(0.09,0.09)$) ($(#2)+(-0.09,0.09)$)--($(#2)+(0.09,-0.09)$);}
\newcommand{\zdot}[1]{\fill[cBlue] (#1) circle (0.065);}
\tikzset{middlearrow/.style={
        decoration={markings,
            mark= at position #1 with {\arrow{latex}} ,
        },
        postaction={decorate}
    }
}

\usepackage{hyperref}
\hypersetup{
    colorlinks=true,
    linkcolor=LouisBlue,
    citecolor=cClay,
    urlcolor=LouisBlue,
}
\hypersetup{pdftitle={The out-state of scalar-induced gravitational waves},
 pdfauthor={Yizhou Lu}}
\newcommand{\Rres}{\mathsf R}

\makeatletter
\newlength{\apb@width}
\newcommand{\autoparbox}[2][c]{\settowidth{\apb@width}{#2}\parbox[#1]{\apb@width}{#2}}

\makeatother


\allowdisplaybreaks[1]
\definecolor{lightgray}{RGB}{240,240,240}

\newtcolorbox[auto counter,number within=section]{defi}[2][]{
colframe=UNALVerde,
enhanced,
breakable,
fonttitle=\bfseries,
title=Definition \thetcbcounter: #2,
#1}

\newtcolorbox[auto counter,number within=section]{exe}[2][]{
colframe=ocre,
enhanced,
breakable,
fonttitle=\bfseries,
title=Exercise: \thetcbcounter: #2,
#1}

\usepackage{cleveref}

\usepackage{placeins}
\def\dd{\mathrm{d}}
\def\norm#1{\left\lVert#1\right\rVert}

\newcommand{\TT}{\mathrm{TT}}
\newcommand{\out}{\mathrm{out}}
\newcommand{\refr}{\mathrm{src}}
\newcommand{\Qcal}{\mathcal{Q}}

\theoremstyle{plain}
\newtheorem{theorem}{Theorem}[section]

\newtheorem{proposition}{Proposition}[section]

\theoremstyle{definition}
\newtheorem{definition}{Definition}[section]
\theoremstyle{remark}

\begin{document}

\title{The out-state of scalar-induced gravitational waves}
\author[a,b]{Yizhou Lu}
\affiliation[a]{Shanghai Institute for Mathematics and Interdisciplinary Sciences (SIMIS), Shanghai 200433, China}
\affiliation[b]{Research Institute of Intelligent Complex Systems, Fudan University, Shanghai 200433, China}
\emailAdd{luyz@simis.cn}

\abstract{
Scalar-induced second-order tensor perturbations comprise both a forced response and freely propagating gravitational waves, the radiative signal sought in observations.
We define this free radiation as the out-state, whose amplitudes are given by the light-cone residue of the kernel for each scalar pair via a classical LSZ-type reduction formula.
We prove a classical equivalence theorem: tensor redefinitions whose shifts have power-law envelopes and no tensor-frequency component leave the out-state invariant.
For any constant equation of state $0<w<1$, the out-state is consequently universal across all intrinsic gauges, in which the slicing and spatial coordinates are fixed locally by matter and geometry without an external clock.
This class includes the comoving and uniform-density gauges, whose spectra grow without bound at late times in radiation domination.
Over this range of $w$, the late-time Newtonian-gauge kernel yields the out-state, justifying its standard use in computing the gravitational-wave energy spectrum.
In matter domination, by contrast, the out-state isolates the free wave from the static tidal response that asymptotically dominates the Newtonian-gauge tensor perturbation.
}


\maketitle
\flushbottom

\section{Introduction}\label{sec:intro}

Scalar-induced gravitational waves (SIGWs) link primordial scalar perturbations generated during inflation \cite{Starobinsky:1980te,Guth:1980zm,Linde:1981mu,Mukhanov:1981xt} to stochastic gravitational-wave backgrounds, probing small scales largely inaccessible to cosmic microwave background observations \cite{Ananda:2006af,Baumann:2007zm,Saito:2008jc,Sasaki:2018dmp,Espinosa:2018eve,Inomata:2018epa,Alabidi:2013lya,Domenech:2021ztg,Yuan:2023ofl,Yuan:2026krm,Chen:2026hnv}.
Upon horizon reentry, scalar perturbations large enough to source an observable SIGW signal can also collapse into primordial black holes (PBHs) \cite{Zeldovich:1967lct,Hawking:1971ei,Carr:1974nx,Carr:1975qj}, which may account for part or all of the dark matter and contribute to the binary black-hole mergers observed by LIGO and Virgo \cite{Bird:2016dcv,Sasaki:2016jop,Clesse:2016vqa,Carr:2020xqk,Carr:2020gox,Green:2020jor}.
SIGW searches thus place stringent bounds on the PBH abundance \cite{Saito:2008jc,Saito:2009jt,Bugaev:2009zh,Sasaki:2018dmp,Jiang:2024aju,Jiang:2026krg}.
More recently, interpreting the nanohertz background reported by pulsar timing arrays as SIGWs \cite{NANOGrav:2023gor,EPTA:2023fyk,Reardon:2023gzh,Xu:2023wog,NANOGrav:2023hvm,Franciolini:2023pbf} has enabled constraints on small-scale non-Gaussianity \cite{Liu:2023ymk}, the early-Universe equation of state and sound speed \cite{Liu:2023pau,Liu:2023hpw}, the propagation speed of tensor modes \cite{Chen:2024fir}, and adiabatic versus isocurvature sources \cite{Chen:2024twp}.

All such tests compare data against the energy spectrum of the second-order tensor perturbation.
This perturbation contains both a free homogeneous solution and a particular solution maintained by the quadratic scalar source \cite{Ananda:2006af,Baumann:2007zm,Assadullahi:2009nf,Kohri:2018awv,Gong:2019mui,Sipp:2022kmb}.
Moreover, it changes under first-order scalar gauge transformations \cite{Matarrese:1997ay,Bruni:1996im,Malik:2008im,Hwang:2017oxa,Tomikawa:2019tvi,DeLuca:2019ufz,Inomata:2019yww,Yuan:2019fwv,Lu:2020diy,Ali:2020sfw,Chang:2020tji,Chang:2020iji,Chang:2020mky,Domenech:2020xin,Ali:2023moi,Comeau:2023mxi,Yuan:2024qfz,Yuan:2025seu,Ali:2025xvj,Xue:2025yii}.
In radiation domination, its late-time spectrum grows without bound in gauges such as the comoving and uniform-density gauges, while in Newtonian gauge it settles to a constant \cite{Tomikawa:2019tvi,Lu:2020diy,Ali:2023moi}.

The resulting debate has centered on which gauge, or which tensor perturbation, should be used.
References~\cite{Inomata:2019yww,Yuan:2019fwv} find that late-time spectra agree across several gauges.
Domènech and Sasaki show that this agreement is only approximate and holds when the source becomes inactive and the slicing is well behaved on subhorizon scales \cite{Domenech:2020xin}.
Whether the excluded gauges carry the same radiation was left open.
Gauge-invariant combinations have also been proposed \cite{Chang:2020iji,Chang:2020mky,Chang:2020tji}, yet gauge invariance alone does not identify a physical observable \cite{Domenech:2020xin}, motivating dedicated constructions \cite{Comeau:2023mxi,Clarkson:2003af,Clarkson:2011td,Ota:2021fdv}.
A complementary approach defines the observable through the measurement: Ref.~\cite{DeLuca:2019ufz} computes it in the transverse-traceless (TT) frame adapted to the detector, and Domènech, Pi and Wang find that the strain measured by geodesic observers exchanging light signals is the Newtonian-gauge tensor perturbation \cite{Domenech:2025ccu}.

The measured strain, however, still contains the response maintained by the source.
The energy spectrum $\Omega_{\rm GW}$ is computed with the Isaacson prescription \cite{Isaacson:1968zza}, whose averaged stress tensor applies only to short-wavelength waves obeying the homogeneous wave equation.
This forced response does not satisfy this premise.
Standard calculations nevertheless evaluate $\Omega_{\rm GW}$ from the late-time Newtonian-gauge tensor perturbation \cite{Kohri:2018awv,Domenech:2019quo}, even though in matter domination it is asymptotically dominated by a static tidal response \cite{Kohri:2018awv,Assadullahi:2009nf,Gong:2019mui} rather than a free wave.
Two questions therefore remain: which part of a sourced tensor perturbation is free radiation, and when is that part gauge independent?

A direct approach identifies free radiation with the component propagating at the speed of light \cite{Yuan:2025seu,Ali:2025xvj}.
Because this selection acts term by term on the subhorizon asymptotic expansion \cite{Yuan:2025seu} or on explicit analytic kernels \cite{Ali:2025xvj}, it applies only when the late-time form of every kernel is known.\footnote{In Refs.~\cite{Yuan:2025seu,Ali:2025xvj} the selection is made in Fourier space, so the retained terms are those whose phase velocity equals the speed of light and whose envelope decays like that of a free wave.}
Moreover, its gauge independence rests on a structural argument: our earlier work~\cite{Yuan:2025seu} discards the entire gauge transformation on the grounds that products of first-order transfer functions do not oscillate at the tensor frequency.
This argument, however, lacks control over the frequency content of the gauge shift (the change in the tensor perturbation under a gauge transformation).
Specifically, a product of two scalar modes can itself resonate with the tensor frequency.
In addition, a gauge whose transformation from Newtonian gauge has generators decoupled from the scalar modes can inject that frequency for every scalar pair.

In this work, we isolate the free radiation in SIGWs and determine when it is independent of the gauge.
We define the free radiation through the light-cone residue, that is, the residue of the Laplace transform at the tensor frequency $\omega=k$ approached from the upper half of the complex frequency plane.
This frequency is the on-shell pole of the free tensor mode, which we call the light-cone pole.
The resulting homogeneous solution, termed the \emph{out-state}, is the wave that remains after the source is switched off adiabatically \cite{Gell-Mann:1951ooy}, and its amplitudes obey a classical reduction formula of the Lehmann--Symanzik--Zimmermann (LSZ) type \cite{Lehmann:1954rq}.
For a purely induced Newtonian-gauge solution with zero initial tensor data, this out-state equals the retarded minus the advanced response, paralleling the radiation field in classical electrodynamics \cite{Dirac:1938nz}.

Our main result is a classical equivalence theorem for the second-order tensor perturbation: any redefinition whose late-time shift has power-law envelopes and no tensor-frequency component leaves the out-state invariant, providing the classical counterpart of on-shell amplitude invariance under field redefinitions \cite{Chisholm:1961tha,Kamefuchi:1961sb}.
In \emph{intrinsic gauges}, the spatial coordinates and the observers normal to the time slices are fixed locally by matter and geometry.
Applying the theorem to these gauges shows that, for a perfect fluid with constant equation of state $0<w<1$, the out-state is identical across all intrinsic gauges on a given physical solution.
This class includes the Newtonian, flat, comoving, uniform-density, and uniform-expansion gauges, as well as the fully fixed synchronous gauge \cite{Ma:1995ey}.\footnote{The harmonic (Lorenz--de Donder) gauge is not intrinsic, since its coordinates obey a wave equation of their own.
Theorem~\ref{thm:equivalence} nevertheless applies to it directly, and its out-state is the same (Appendix~\ref{ap-sec:generator-estimates}).}
The comoving and uniform-density gauges, whose spectra grow at late times in radiation domination, thus have the same out-state as Newtonian gauge.
For $0<w<1$, the Newtonian-gauge tensor perturbation approaches the out-state asymptotically, justifying the standard practice of computing $\Omega_{\rm GW}$ from the late-time Newtonian-gauge kernel.

To see why a residue isolates the free wave, consider a driven harmonic oscillator $\ddot x+\Omega^2x=f(t)$ with initial data at $t=0$.
Its one-sided transform $\tilde x(\omega)=\int_0^\infty\mathrm{e}^{\mathrm{i}\omega t}x\,\dd t$ converges in the upper half plane and reads
\begin{equation}\label{eq:driven-oscillator}
 \tilde x(\omega)=\frac{\tilde f(\omega)+\dot x(0)-\mathrm{i}\omega x(0)}{\Omega^2-\omega^2}.
\end{equation}
Here, free oscillations arise from the oscillator poles at $\omega=\pm\Omega$, while forced motion originates from the singularities of $\tilde f(\omega)$ at the driving frequencies.
A singularity at $\omega_*$ yields an asymptotic $\mathrm{e}^{-\mathrm{i}\omega_*t}$ term, and the residue of a simple pole gives its amplitude.
Accordingly, the residues at $\omega=\pm\Omega$ combine the initial data with the on-shell values $\tilde f(\pm\Omega)$, capturing the permanent free amplitudes left behind by the drive.

Extracting the free-pole residue, however, requires separating the pole from the drive singularities.
Approaching the real axis along $\omega+\mathrm{i}\epsilon$, equivalent to switching the drive off adiabatically as $\mathrm{e}^{-\epsilon t}$, separates them only when their distance exceeds $\epsilon$.
This separation fails at exact resonance, where a drive singularity coincides with the free pole, producing a double pole and secular $t\,\mathrm{e}^{-\mathrm{i}\Omega t}$ growth.
It also becomes delicate when the drive forms a continuous spectrum crossing $\pm\Omega$, because drive singularities then lie within any distance $\epsilon$ of the pole, much like a discrete level coupled to a continuum in scattering theory.

The induced tensor perturbation is an oscillator of this kind, with a quadratic scalar drive.
Its free radiation is therefore given by the residue at the tensor frequency, and the distance of a source singularity from this pole measures the detuning between the source and tensor frequencies.
A product of two scalar modes contains oscillations at the sum of their sound frequencies, which equals the tensor frequency on the source resonance surface
\begin{equation}\label{eq:resonance_surf}
     c_s(q+p)=k,\qquad k=|\mathbf q+\mathbf p|,
\end{equation}
where $c_s$ is the sound speed and $\mathbf q$ and $\mathbf p$ are the momenta of the two scalar modes.
For a continuous scalar spectrum, source singularities reach the pole along this surface, realizing the continuous-spectrum case described above.
We therefore take the residue of the kernel for each scalar pair before performing the momentum integral, and then integrate over pairs to obtain the free amplitudes.

The same complex-frequency picture governs gauge transformations, since a gauge shift affects the free radiation solely through its own residue at the tensor frequency.
An intrinsic gauge has no external clock or residual gauge freedom, while its observers have no independent oscillatory modes and move with at most power-law growth in time.
Consequently, the generators of transformations between intrinsic gauges oscillate only at the scalar sound frequencies, apart from nonoscillating power-law terms.
The second-order tensor shifts they induce reach the tensor frequency only on the source resonance surface~\eqref{eq:resonance_surf} and the gauge resonance surfaces $c_sq=k$ and $c_sp=k$ (Sec.~\ref{sec:generator-frequencies}), all of zero volume in the space of scalar momenta $(q,p)$.
When a gauge shift grows at late times, as in the comoving and uniform-density gauges during radiation domination, near-resonant pairs carry growing power.
If the scalar spectrum has weight on the source resonance surface, the residue of the momentum-integrated tensor perturbation therefore does not exist.
Nevertheless, the residue of each individual pair remains well defined, and the out-state assembled from the pairs is the limit of the residues obtained when the near-resonant pairs are removed (Appendix~\ref{app:residue-existence}).

Gauges outside the class of intrinsic gauges, by contrast, can introduce tensor-frequency oscillations.
A gauge driven by an external clock carries the tensor frequency for every scalar pair, independently of the matter (Sec.~\ref{sec:outside-class}).
Synchronous gauge lies outside the class until its residual freedom is fixed.
Once those residual modes are specified, either by early-time regularity or by the subhorizon condition of Ref.~\cite{Domenech:2020xin}, the gauge becomes intrinsic and yields the same out-state.

In matter domination, the out-state isolates the free wave of the exact Newtonian kernel from the static tidal response \cite{Kohri:2018awv,Assadullahi:2009nf,Gong:2019mui,Ali:2020sfw}.
The strain measured in Ref.~\cite{Domenech:2025ccu} contains both contributions, and the out-state identifies the part that enters the energy spectrum.
Section~\ref{sec:disc-relation} compares the out-state with the measured strain and with earlier prescriptions.

The paper is organized as follows.
Sections~\ref{sec:setup}--\ref{sec:outstate} set up the equations and construct the residue.
Section~\ref{sec:kernel-outstate} proves the classical equivalence theorem and the invariance of the out-state across all intrinsic gauges.
Section~\ref{sec:observables} evaluates the amplitudes and spectrum in closed form, using the Legendre-function integrals of Ref.~\cite{Domenech:2019quo}.
Section~\ref{sec:conclusion} discusses measurements and limits, and Sec.~\ref{sec:conclusions} concludes.
The appendices provide the technical estimates and a separate classification of field residues.
Table~\ref{tab:notation}, placed after the appendices, collects the notation.

\section{Scalar-induced tensor perturbation and physical initial data}\label{sec:setup}

\subsection{Background and perturbation conventions}
We work in four-dimensional Einstein gravity with signature $(-+++)$ and a spatially flat FLRW background filled with a single perfect fluid,
\begin{equation}\label{eq:background}
 \dd\bar s^2=a^2\bigl(-\dd\eta^2+\dd\mathbf X^2\bigr),\qquad
 a(\eta)>0,\qquad \eta>0,\qquad \mathcal H\equiv a'/a.
\end{equation}
The equation-of-state parameter is $w\equiv\bar p/\bar\rho$.
We take $w$ to be constant with $0<w<1$ and adiabatic sound speed $c_s^2=w$, obtaining the dust case $w=0$ as a limit (Sec.~\ref{sec:observables}).
A prime denotes a derivative with respect to conformal time $\eta$.
Spatial indices are raised and lowered with $\delta_{ij}$.

For any quantity $\mathcal V$, we adopt the perturbative expansion
\begin{equation}\label{eq:order-convention}
 \mathcal V=\bar{\mathcal V}+\mathcal V^{(1)}+\tfrac12\mathcal V^{(2)}+\cdots,
\end{equation}
so the ordinary quadratic coefficient is $\mathcal V^{(2)}/2$.
The first-order perturbations consist solely of the regular growing scalar mode, with no vectors or tensors, and we include no independent second-order tensor anisotropic stress.
Up to the orders used in this work, the metric reads
\begin{equation}\label{eq:perturbed-metric}
 \dd s^2=a^2\Bigl[-(1+2\phi)\dd\eta^2+2\partial_iB\,\dd\eta\,\dd X^i
 +\bigl((1-2\psi)\delta_{ij}+2\partial_i\partial_jE+h_{ij}\bigr)\dd X^i\dd X^j\Bigr],
\end{equation}
where $\phi$, $B$, $\psi$ and $E$ are first-order scalar perturbations and $h_{ij}$ is the second-order tensor perturbation, which is transverse and traceless.
Second-order scalar and vector perturbations are omitted, since they do not enter the tensor equation of motion.
On a scalar perturbation, a subscript $*$ denotes its initial amplitude, while a superscript $*$ denotes complex conjugation.

For comoving coordinates $X^i$, the inverse Fourier transform uses $\mathrm{e}^{\mathrm{i}\mathbf k\cdot\mathbf X}$ and the measure $[\dd k]\equiv\dd^3k/(2\pi)^3$.
In a convolution, $\mathbf p\equiv\mathbf k-\mathbf q$, $p\equiv|\mathbf p|$ and $q\equiv|\mathbf q|$.
We restrict the external momentum to $k\equiv|\mathbf k|>0$ and use
\begin{gather}
\begin{aligned}
 P_i{}^j&\equiv\delta_i{}^j-\frac{k_i k^j}{k^2},\\
 \Lambda_{ij}{}^{lm}&\equiv P_i{}^{(l}P_j{}^{m)}-\tfrac12P_{ij}P^{lm},
\end{aligned}\label{eq:TT-projector}\\[2pt]
 S^{\TT}_{ij}\equiv\Lambda_{ij}{}^{lm}S_{lm},\qquad
 \norm{S}^2\equiv\sum_{i,j=1}^3|S_{ij}|^2.\label{eq:proj}
\end{gather}
The TT projector always acts on the external momentum $\mathbf k$.
We refer to the two scalar modes $\zeta_{\mathbf q}$ and $\zeta_{\mathbf p}$ (with $\mathbf p=\mathbf k-\mathbf q$) of the primordial curvature perturbation as a scalar pair, and write $u\equiv p/k$ and $v\equiv q/k$.
For each Fourier mode, we use the dimensionless time $x\equiv k\eta$, with $\partial_x=k^{-1}\partial_\eta$.
We write $f\simeq g$ for leading late-time equivalence ($f/g\to1$ as $x\to\infty$) and $f\sim g$ when only the scaling is meant.
Table~\ref{tab:notation} at the end of the paper collects the notation.

\subsection{The scalar source and the free modes}\label{sec:general-background}
The Newtonian, or Poisson, gauge \cite{Bertschinger:1993xt} sets $B=E=0$, and without anisotropic stress $\phi=\psi=\Phi$.
In terms of the expansion~\eqref{eq:order-convention}, the tensor perturbation of Eq.~\eqref{eq:perturbed-metric} is
\begin{equation}\label{eq:h-def}
 h_{ij}\equiv\frac1{2a^2}\,g_{ij}^{(2)\TT},
\end{equation}
and obeys the equation of motion \cite{Ananda:2006af,Baumann:2007zm,Kohri:2018awv,Domenech:2019quo,Domenech:2020kqm}
\begin{equation}\label{eq:sourced-history}
 \Box_kh_{ij}=S_{ij},\qquad k^ih_{ij}=\delta^{ij}h_{ij}=0,\qquad
 \Box_k\equiv\partial_\eta^2+2\mathcal H\partial_\eta+k^2,
\end{equation}
with the source
\begin{equation}\label{eq:Sg}
 S_{ij}(\eta,\mathbf k)\equiv2\Lambda_{ij}{}^{lm}(\mathbf k)
 \int[\dd q]\,q_lq_m\biggl[2\Phi(\eta,\mathbf q)\Phi(\eta,\mathbf p)
 +\frac{4\mathcal H^{-2}(\eta)}{3(1+w)}U(\eta,\mathbf q)U(\eta,\mathbf p)\biggr],
\end{equation}
where $U\equiv\Phi'+\mathcal H\Phi$.
The allowed scalar-pair domain is $u,v>0$ with $|1-v|\le u\le1+v$.

We work with the rescaled tensor perturbation $Y_{ij}\equiv ah_{ij}$, which is the tensor Mukhanov--Sasaki variable up to a constant normalization.
It obeys
\begin{equation}\label{eq:general-Y}
 Y''_{ij}+\Bigl(k^2-\frac{a''}{a}\Bigr)Y_{ij}=s_{ij}\equiv aS_{ij} .
\end{equation}
We denote this differential operator by $\mathcal L\equiv\partial_\eta^2+k^2-a''/a$.
A free tensor perturbation solves Eq.~\eqref{eq:sourced-history} with $S_{ij}=0$ and the same TT conditions.
The rescaled free solution has two amplitudes for each polarization.
We choose exact free modes $u_c,u_s$ normalized to $\cos x$ and $\sin x$ at late times.
In radiation domination, $a=A\eta$ with $A>0$, and these modes reduce to $u_c=\cos x$ and $u_s=\sin x$.

For a scalar mode, the Bardeen potential \cite{Bardeen:1980kt} takes the form
\begin{equation}\label{eq:Fw}
\begin{gathered}
 \Phi(\eta,\mathbf k)\equiv F_w(c_sk\eta)\,\Phi_*(\mathbf k),\qquad
 F_w(y)\equiv2^\nu\Gamma(\nu+1)\,y^{-\nu}J_\nu(y),\\
 \nu\equiv\frac{5+3w}{2(1+3w)},\qquad c_s^2=w,
\end{gathered}
\end{equation}
where $J_\nu$ is the Bessel function of the first kind.
The growing-mode normalization is
\begin{equation}\label{eq:Phi-star}
 \Phi_*=\frac{3(1+w)}{5+3w}\,\zeta_*,
\end{equation}
which equals $2\zeta_*/3$ in radiation domination and $3\zeta_*/5$ in matter domination, where $\zeta$ is the primordial curvature perturbation.

\subsection{Scalar statistics}\label{sec:averaging}
We take the scalar perturbation to be statistically homogeneous and isotropic, with a continuous spectrum
\begin{equation}\label{eq:gauge-scalar-spectrum}
 \langle\zeta_{\mathbf q}\zeta_{\mathbf q'}^*\rangle
 =(2\pi)^3\delta^3(\mathbf q-\mathbf q')P_\zeta(q),\qquad
 P_\zeta(q)\equiv\frac{2\pi^2}{q^3}\mathcal P_\zeta(q),
\end{equation}
where $\langle\cdot\rangle$ denotes the ensemble average over the scalar distribution.
For a zero-mean Fourier field ${\mathscr{U}}$, such as a tensor perturbation, we define its power spectrum by
\begin{equation}\label{eq:field-power-spectrum}
 \langle {\mathscr{U}}(\mathbf k)\cdot {\mathscr{U}}(\mathbf k')^*\rangle
 \equiv(2\pi)^3\delta^3(\mathbf k-\mathbf k')P_{\mathscr{U}}(\mathbf k).
\end{equation}
Here the dot denotes summation over tensor or polarization indices and any amplitude components.
We assume statistical isotropy throughout and write $P_{\mathscr{U}}(\mathbf{k})\equiv P_{\mathscr{U}}(k)$, with $k\equiv|\mathbf k|$.

For a quadratic Fourier field ${\mathscr{U}}$, we define its symmetric kernel $\mathscr K=\mathscr K[{\mathscr{U}}]$ by
\begin{equation}\label{eq:statistical-pair-kernel}
 {\mathscr{U}}(\mathbf k)=\int[\dd q]\,\mathscr K(\mathbf k,\mathbf q)
 \zeta_{\mathbf q}\zeta_{\mathbf p},\qquad
 \mathscr K(\mathbf k,\mathbf q)=\mathscr K(\mathbf k,\mathbf p).
\end{equation}
Only the symmetric part of a kernel contributes to ${\mathscr{U}}$, since the product $\zeta_{\mathbf q}\zeta_{\mathbf p}$ is symmetric.
For Gaussian scalar perturbations, the two nonzero-momentum Wick contractions give the convolution spectrum,
\begin{equation}
 P_{\mathscr{U}}(k)=2\int[\dd q]\,
 \|\mathscr K(\mathbf k,\mathbf q)\|^2P_\zeta(q)P_\zeta(p).
\end{equation}
We say that a quadratic field $\mathscr U_\epsilon$ converges to $\mathscr U$ in spectral mean square if $P_{\mathscr U_\epsilon-\mathscr U}(k)\to0$.
We consider tensor perturbations quadratic in this scalar input, allowing also for quadratic initial tensor amplitudes whose kernels are square integrable against $P_\zeta(q)P_\zeta(p)$.

Unless stated otherwise, the scalar input is the regular adiabatic mode of a single fluid, with a bounded dimensionless spectrum supported on
\begin{equation}
     \Lambda_{\rm c}^{-1}\le q/q_*\le\Lambda_{\rm c},\qquad q_*>0.
\end{equation}
Noncompact spectra are obtained by taking $\Lambda_{\rm c}\to\infty$ after taking the residue.\footnote{A monochromatic spectrum $P_\zeta\propto\delta(\log(q/q_*))$ has a single scalar pair, $q=p=q_*$.
For $k\ne2c_sq_*$, the Newtonian kernel is nonresonant and the out-state follows from the kernel residue.
A gauge whose generators have zero-frequency parts, such as synchronous gauge, also resonates at $k=c_sq_*$.
At $k=2c_sq_*$, for a narrow spectrum of width $\sigma$, the out-state spectrum grows as $\log^2(1/\sigma)$ in radiation domination and as $\sigma^{2(1-3w)/(1+3w)}$ for $w>1/3$, reproducing the integrable resonance peak of the monochromatic spectrum~\cite{Kohri:2018awv}.}
The explicit Wick-contracted spectra assume Gaussianity.

\subsection{Gauge transformations on one physical solution}\label{sec:redefinitions}
In this work, a \emph{physical solution} is an Einstein--fluid solution with its scalar and tensor initial data, modulo gauge transformations \cite{Mukhanov:1990me,Bruni:1996im,Matarrese:1997ay}.
A gauge transformation acts not only on the metric and matter variables but also on the initial data and the initial hypersurface \cite{Lu:2020diy}.

Let $\mathcal L_\xi$ denote the Lie derivative along $\xi$.
For first- and second-order generators $\xi_1,\xi_2$, the transformation of any tensor $\mathcal V$ is \cite{Bruni:1996im,Matarrese:1997ay,Malik:2008im}
\begin{equation}\label{eq:gauge-shift-general}
 \widetilde{\mathcal V}^{(2)}-\mathcal V^{(2)}
 =2\mathcal L_{\xi_1}\mathcal V^{(1)}
 +(\mathcal L_{\xi_2}+\mathcal L_{\xi_1}^2)\bar{\mathcal V} .
\end{equation}
A first-order scalar generator has components $\xi_1=(\alpha,\partial^i\beta)$.
The second-order generator contributes spatial gradients that vanish under the TT projection.
From Eq.~\eqref{eq:gauge-shift-general}, we obtain the gauge transformation of the second-order tensor perturbation \cite{Malik:2008im,Lu:2020diy}
\begin{equation}\label{eq:gauge-shift-ordinary}
 \Delta h_{ij}=\Lambda_{ij}{}^{lm}\Bigl[-\partial_l\alpha\partial_m\alpha
 +\partial_l\partial_n\beta\partial_m\partial_n\beta
 +4\partial_{(l}\Phi\partial_{m)}\beta
 -4\mathcal H\partial_{(l}\alpha\partial_{m)}\beta\Bigr].
\end{equation}
We take Newtonian gauge as our reference.
Any choice of gauge $X$ then determines $\alpha$ and $\beta$ from the scalar solution, including any chosen residual constants.
We denote the corresponding rescaled tensor perturbation by $Y^X_{ij}$, that in Newtonian gauge by $Y^{\rm N}_{ij}$, and their difference by $Q^X_{ij}\equiv Y^X_{ij}-Y^{\rm N}_{ij}$.
Every gauge shift $Q_{ij}$ is quadratic in the scalar input and takes the form \cite{Lu:2020diy}
\begin{equation}\label{eq:Q-bilinear}
\begin{aligned}
 Q_{ij}(\eta,\mathbf k)\equiv{}&\int[\dd q]\,\mathscr K_{ij}[Q](\eta;\mathbf k,\mathbf q)\zeta_{\mathbf q}\zeta_{\mathbf k-\mathbf q},
\end{aligned}
\end{equation}
where $\mathscr K_{ij}[Q]$ is the symmetric kernel of $Q_{ij}$ as defined in Eq.~\eqref{eq:statistical-pair-kernel}.
It includes the TT factor $\Lambda_{ij}{}^{lm}q_lq_m$ and depends on $\eta$ and the momenta through $x$, $u$ and $v$.
The explicit expression for $\mathscr{K}_{ij}[Q]$ is not needed until Appendix~\ref{app:histories}.
We denote by $\Qcal$ the linear space of such quadratic fields.
Writing $Q^{XX'}_{ij}=Q^X_{ij}-Q^{X'}_{ij}$ gives
\begin{equation}\label{eq:gauge-problem}
 Y^X_{ij}-Y^{X'}_{ij}=Q^{XX'}_{ij}\in\Qcal .
\end{equation}

\section{Separating free radiation from the forced response}\label{sec:outstate}
\subsection{Free-mode amplitudes of a sourced tensor perturbation}\label{sec:outstate-def}
We seek a decomposition $Y_{ij}=Y^\out_{ij}+Y^\refr_{ij}$, where $Y^\out_{ij}$ is the homogeneous solution representing free radiation and $Y^\refr_{ij}$ is a particular solution carrying zero free amplitude.
At fixed $k$, the equation of motion and the free-mode normalization are
\begin{equation}
\begin{gathered}
 \partial_x^2Y_{ij}+\bigl(1-(\partial_x^2a)/a\bigr)Y_{ij}=s_{ij}/k^2,\\
 W[f,g]\equiv f\partial_xg-(\partial_xf)g,\quad W[u_c,u_s]=1,
\end{gathered}
\end{equation}
where $W[f,g]$ is the Wronskian.
For a homogeneous solution, the coefficients of \(u_c\) and \(u_s\) are constant.
To follow their evolution in the presence of a source, we use variation of constants and write
\begin{equation}
\begin{aligned}
 Y_{ij}&=\mathsf h_c[Y_{ij}]u_c+\mathsf h_s[Y_{ij}]u_s,\\
 \partial_xY_{ij}&=\mathsf h_c[Y_{ij}]\partial_xu_c+\mathsf h_s[Y_{ij}]\partial_xu_s.
\end{aligned}
\end{equation}
At each time, these two relations uniquely determine the coefficients by matching both the field and its first derivative.
Using \(W[u_c,u_s]=1\), we obtain
\begin{equation}\label{eq:wronskian-pair}
\mathsf h[Y_{ij}]\equiv
\begin{pmatrix}
\mathsf h_c[Y_{ij}]\\
\mathsf h_s[Y_{ij}]
\end{pmatrix}
=
\begin{pmatrix}
W[Y_{ij},u_s]\\
W[u_c,Y_{ij}]
\end{pmatrix}.
\end{equation}
The sourced equation then gives
\begin{equation}\label{eq:h-evolution}
\partial_x\mathsf h[Y_{ij}]
=\frac{s_{ij}}{k^2}
\begin{pmatrix}
-u_s\\
u_c
\end{pmatrix}.
\end{equation}
Thus the source drives the evolution of these two coefficients, which become constant whenever the source vanishes.
The components \(\mathsf h[Y_{ij}](x_*)\) are therefore the amplitudes of the free wave that would remain if the source were switched off at \(x_*\).

We define the exact complex free mode by
\begin{equation}\label{eq:psi-def}
 \varphi\equiv u_c+\mathrm{i}u_s,\qquad W[\varphi,\varphi^*]=-2\mathrm{i},\qquad
 \varphi(x)\simeq\mathrm{e}^{\mathrm{i}x},
\end{equation}
where the last relation is our late-time normalization.
In radiation domination, $\varphi(x)=\mathrm{e}^{\mathrm{i}x}$ exactly.
We then define the complex amplitude by
\begin{equation}\label{eq:complex-amplitude}
\begin{aligned}
 \mathsf A[Y_{ij}]&\equiv W[\varphi,Y_{ij}]=\mathsf h_s[Y_{ij}]-\mathrm{i}\,\mathsf h_c[Y_{ij}],\\
 \frac{\dd\mathsf A[Y_{ij}]}{\dd x}&=\frac{\varphi\,s_{ij}}{k^2}.
\end{aligned}
\end{equation}
For a free solution, this amplitude is constant,
\begin{equation}
 Y_{ij}=\mathsf a_{c,ij}u_c+\mathsf a_{s,ij}u_s\quad\Longrightarrow\quad
 \mathsf A[Y_{ij}]=\mathsf c_{ij}\equiv\mathsf a_{s,ij}-\mathrm{i}\,\mathsf a_{c,ij}.
\end{equation}
We assume that $\mathsf A[Y_{ij}](x)$ grows at most polynomially at late times and take its Laplace transform in $x$,\footnote{Laplace representations of mode functions have recently been used to reduce cosmological correlators to flat-space integrals, with energy singularities in the Laplace variable \cite{Belrhali:2026ygh,Belrhali:2026jqe,Belrhali:2026uxn}.
Here the transform acts on the tensor amplitude, and its pole at the tensor frequency defines the out-state.}
\begin{equation}\label{eq:rotating-transform}
 \widetilde{\mathsf A}[Y_{ij}](z)\equiv\int_0^\infty {\mathrm{e}}^{\mathrm{i}zx}\,\mathsf A[Y_{ij}](x)\,\dd x ,
\end{equation}
where the initial data are imposed at $x=0$.
Because $\mathsf A$ already removes the free phase, the complex frequency $z$ conjugate to $x$ is measured from the tensor frequency in units of $k$.
The tensor frequency thus sits at $z=0$, the light-cone pole.
The transform is analytic for $\operatorname{Im}z>0$.
Integrating by parts with Eq.~\eqref{eq:complex-amplitude} then gives
\begin{equation}\label{eq:transform-identity}
\boxed{\begin{gathered}
 \widetilde{\mathsf A}[Y_{ij}](z)=\frac{\mathrm{i}}{z}\Bigl[\mathsf A[Y_{ij}](0)+\widehat s_{ij}(z)\Bigr],\\
 \widehat s_{ij}(z)\equiv\frac1{k^2}\int_0^\infty\dd x\,\mathrm{e}^{\mathrm{i}zx}\,\varphi(x)\,s_{ij}(x).
\end{gathered}}
\end{equation}
A free wave with amplitude $\mathsf c_{ij}$ therefore contributes $\mathrm i\mathsf c_{ij}/z$, placing its pole at $z=0$.

In radiation domination, the amplitude of a mode $Y_{\omega,ij}=C\mathrm{e}^{\mathrm{i}\omega x}$ with constant $C$ follows from
\begin{equation}\label{eq:frequency-transform}
\begin{aligned}
 \mathsf A[Y_{ij}]&=\mathrm{e}^{\mathrm{i}x}(\partial_x-\mathrm{i})Y_{ij},\\
 \mathsf A[Y_{\omega,ij}]&=\mathrm{i}(\omega-1)C\mathrm{e}^{\mathrm{i}(1+\omega)x},\\
 \widetilde{\mathsf A}[Y_{\omega,ij}](z)&=-\frac{(\omega-1)C}{z+1+\omega}.
\end{aligned}
\end{equation}
For homogeneous modes, only $\omega=-1$ contributes at $z=0$, while the $\omega=+1$ branch is annihilated by the Wronskian.

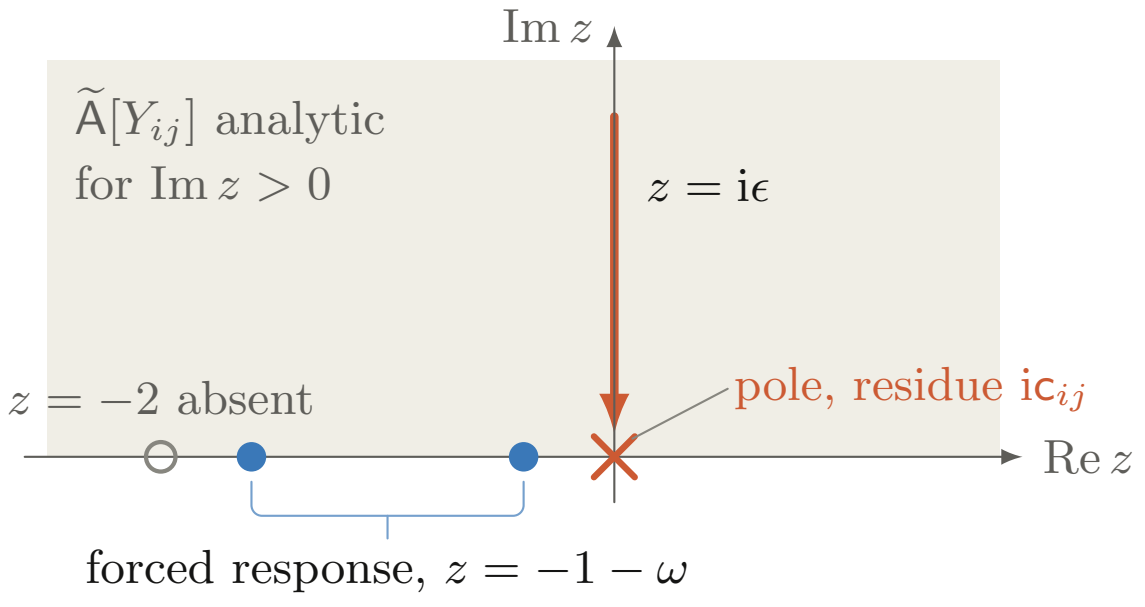
\begin{figure}[t]
\centering
\begin{tikzpicture}[zpl,x=1.5cm,y=1.5cm]
\begin{scope}
\fill[cCloud] (-1.5,0) rectangle (2.7,1.75);
\draw[cClay,line width=1.6pt,-latex] (1,1.5)--(1,0.1);
\node[cInk,anchor=west] at (1.05,1.2) {$z=\mathrm{i}\epsilon$};
\draw[zaxis] (-1.6,0)--(2.8,0) node[right] {$\operatorname{Re}z$};
\draw[zaxis] (1,-0.2)--(1,1.9) node[anchor=east] {$\operatorname{Im}z$};
\node[cSlate,align=left,anchor=north west] at (-1.47,1.72) {$\widetilde{\mathsf A}[Y_{ij}]$ analytic\\for $\operatorname{Im}z>0$};
\zcross{1,0}
\draw[cGrey] (1.08,0.08)--(1.5,0.3) node[anchor=west,cClay,inner sep=1pt] {pole, residue $\mathrm{i}\mathsf c_{ij}$};
\zdot{0.6,0} \zdot{-0.6,0}
\draw[cBlue!70,rounded corners=2pt] (-0.6,-0.11)--(-0.6,-0.26)--(0.6,-0.26)--(0.6,-0.11);
\draw[cBlue!70] (0,-0.26)--(0,-0.36);
\node[cInk,anchor=north,inner sep=2pt] at (0,-0.36) {forced response, $z=-1-\omega$};
\draw[cGrey,thick] (-1,0) circle (0.065);
\node[cSlate,anchor=south] at (-1.0,0.08) {$z=-2$ absent};
\end{scope}
\end{tikzpicture}
\caption{Frequency plane of $\widetilde{\mathsf A}[Y_{ij}]$.
The free wave yields the pole $\mathrm{i}\mathsf c_{ij}/z$, and a source frequency $\omega k$ produces a singularity at $z=-1-\omega$.
The orange ray is the adiabatic approach $z=\mathrm{i}\epsilon\to0$ from the analytic upper half plane.
The open circle marks $z=-2$, the second free frequency, which is absent because $W[\varphi,\varphi]=0$.}
\label{fig:zplane}
\end{figure}

\subsection{The out-state as a residue}\label{sec:residue-def}
Equation~\eqref{eq:transform-identity} identifies the free wave with the pole $\mathrm{i}\mathsf c_{ij}/z$, but $\widetilde{\mathsf A}[Y_{ij}]$ is in general not meromorphic at $z=0$ because of $\widehat s_{ij}(z)$.
For one scalar pair, the source is the product of two scalar modes, which by Eq.~\eqref{eq:Fw} oscillate at late times as
\begin{equation}
 F_w(y)\propto y^{-\nu-1/2}\cos\bigl(y-\tfrac\pi2(\nu+\tfrac12)\bigr).
\end{equation}
In radiation domination, the source is therefore a finite sum of oscillating terms with power-law envelopes,
\begin{equation}
 s_{ij}\supset k^2s_{\omega,ij}\,\mathrm{e}^{\mathrm{i}\omega x},\qquad
 \omega k=c_s(\sigma_1q+\sigma_2p),\qquad\sigma_{1,2}=\pm1.
\end{equation}
For a constant envelope, Eq.~\eqref{eq:complex-amplitude} with $\varphi=\mathrm{e}^{\mathrm{i}x}$ gives
\begin{equation}\label{eq:forced-term}
\begin{aligned}
 \frac{\dd\mathsf A[Y_{ij}]}{\dd x}&\supset s_{\omega,ij}\,\mathrm{e}^{\mathrm{i}\delta x},\\
 \mathsf A[Y_{ij}]&\supset-\frac{\mathrm{i}s_{\omega,ij}}{\delta}\,\mathrm{e}^{\mathrm{i}\delta x},\\
 \widetilde{\mathsf A}[Y_{ij}](z)&\supset\frac{s_{\omega,ij}}{\delta(z+\delta)},\qquad\delta\equiv1+\omega.
\end{aligned}
\end{equation}
This is the mode $Y_{\omega,ij}$ of Eq.~\eqref{eq:frequency-transform} with $C=s_{\omega,ij}/(1-\omega^2)$, now driven by the source.
It has a pole at $z=-\delta$, and a continuous source spectrum places such poles arbitrarily close to $z=0$ (Fig.~\ref{fig:zplane}).
We therefore replace the ordinary residue by the limit of $z\widetilde{\mathsf A}[Y_{ij}](z)$ as $z\to0$ along the positive imaginary axis $z=\mathrm{i}\epsilon$, staying within the upper half plane where the transform is analytic, and define the out-amplitude by
\begin{equation}\label{eq:residue-definition}
\boxed{
 \mathsf c[Y_{ij}]\equiv-\mathrm{i}\operatorname*{Res}_{z=0}\widetilde{\mathsf A}[Y_{ij}](z)
 \equiv\lim_{\epsilon\to0^+}\epsilon\,\widetilde{\mathsf A}[Y_{ij}](\mathrm{i}\epsilon).}
\end{equation}
This reduces to the usual residue when the transform is meromorphic with at most a simple pole.
The same amplitude follows from the tensor perturbation itself.
Since $\mathsf A[Y_{ij}]=\varphi\,\partial_xY_{ij}-(\partial_x\varphi)Y_{ij}$ and $\partial_x\varphi\simeq\mathrm{i}\varphi$ at late times, an integration by parts in Eq.~\eqref{eq:residue-definition} gives
\begin{equation}\label{eq:tensor-transform-form}
 \mathsf c[Y_{ij}]=-2\mathrm{i}\lim_{\epsilon\to0^+}\epsilon\int_0^\infty\mathrm{e}^{-\epsilon x}\,\varphi(x)\,Y_{ij}(x)\,\dd x .
\end{equation}
This is the transform of $Y_{ij}$ at the tensor frequency, multiplied by the inverse free propagator (Sec.~\ref{sec:lsz}).
We work with $\mathsf A[Y_{ij}]$ because it is exactly constant for a free solution, so a free wave contributes only the pole $\mathrm{i}\mathsf c_{ij}/z$.
Equations~\eqref{eq:rotating-transform} and \eqref{eq:transform-identity} give
\begin{equation}\label{eq:abel-form}
\begin{aligned}
 \mathsf c[Y_{ij}]&=\lim_{\epsilon\to0^+}\epsilon\int_0^\infty {\mathrm{e}}^{-\epsilon x}\,\mathsf A[Y_{ij}](x)\,\dd x\\
 &=\mathsf A[Y_{ij}](0)+\lim_{\epsilon\to0^+}\frac1{k^2}\int_0^\infty\dd x\,\mathrm{e}^{-\epsilon x}\varphi(x)s_{ij}(x).
\end{aligned}
\end{equation}
The right-hand side is the amplitude left by adiabatic switching, $s_{ij}\mapsto\mathrm{e}^{-\epsilon x}s_{ij}$ \cite{Gell-Mann:1951ooy}.
If $\mathsf A[Y_{ij}](x)$ settles to a constant at late times, the out-amplitude is that constant.
If it continues to oscillate (possibly with a growing envelope, as for the gauge shifts of Sec.~\ref{sec:kernel-outstate}), the residue extracts its constant part.
In terms of the source transform of Eq.~\eqref{eq:transform-identity}, this reads
\begin{equation}\label{eq:residue-reduction}
 \mathsf c[Y_{ij}]=\mathsf A[Y_{ij}](0)+\widehat s_{ij}(0),\qquad
 \widehat s_{ij}(0)\equiv\lim_{\epsilon\to0^+}\widehat s_{ij}(\mathrm{i}\epsilon).
\end{equation}
The forced term of Eq.~\eqref{eq:forced-term} contributes $s_{\omega,ij}\,\epsilon/[\delta(\delta+\mathrm{i}\epsilon)]$ to $\epsilon\widetilde{\mathsf A}[Y_{ij}](\mathrm{i}\epsilon)$, which vanishes as $\epsilon\to0^+$ at fixed $\delta\neq0$.
Appendix~\ref{app:residue-existence} shows that it still vanishes for envelopes with power-law growth, $\mathsf A\supset \mathsf b_{\nu,ij}(x)\mathrm{e}^{\mathrm{i}\delta_{\nu}x}$ with $|\mathsf b_{\nu,ij}|\lesssim x^M(\log x)^L$.
The free wave, at $\delta=0$, survives with its full amplitude.

\begin{figure}[t]
\centering
\includegraphics[width=\textwidth]{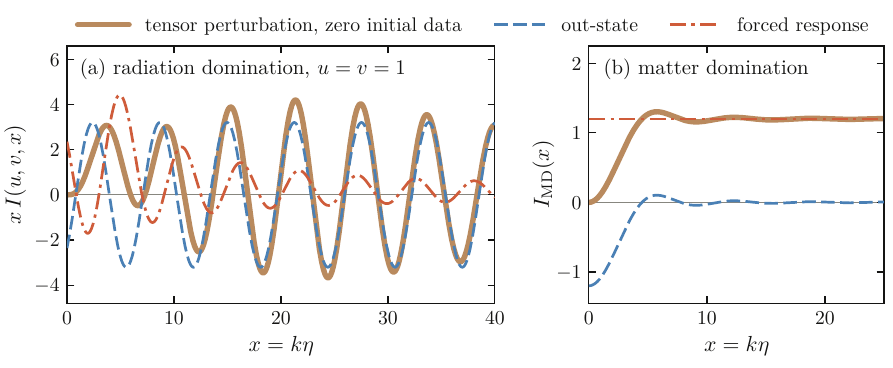}
\caption{Out-state (dashed) and forced response (dash-dotted) of the purely induced Newtonian tensor perturbation with zero initial data (solid), for the Newtonian-gauge kernel $I$ of Eq.~\eqref{eq:rd-full-kernel}.
(a) Radiation domination, scalar pair $u=v=1$, where $Y_{ij}\propto x\,I(u,v,x)$.
The forced response decays as $x^{-1}$ and $Y_{ij}$ approaches $Y^\out_{ij}$.
(b) Matter domination, where $h_{ij}\propto I_{\rm MD}(x)$ of Eq.~\eqref{eq:md-explicit-split}.
The forced response is static and the out-state decays as $x^{-2}$.
In both panels the forced response cancels the out-state at $x=0$.}
\label{fig:schematic}
\end{figure}

Applying the same limit to both entries of $\mathsf h$ defines the out-amplitudes of the real free modes $u_c$ and $u_s$,
\begin{equation}\label{eq:real-out-amplitudes}
\begin{gathered}
\mathsf a[Y_{ij}]\equiv\lim_{\epsilon\to0^+}\epsilon\int_0^\infty\mathrm{e}^{-\epsilon x}\mathsf h[Y_{ij}](x)\,\dd x,\\
 \mathsf c[Y_{ij}]=\mathsf a_s[Y_{ij}]-\mathrm{i}\mathsf a_c[Y_{ij}],
\end{gathered}
\end{equation}
and the exact homogeneous solution
\begin{equation}\label{eq:outstate}
 Y^\out_{ij}(x)\equiv \mathsf a_c[Y_{ij}]\,u_c(x)+\mathsf a_s[Y_{ij}]\,u_s(x).
\end{equation}
The amplitude depends on the entire time evolution of the source, but the homogeneous solution is defined for every $x>0$.
Subtracting the out-state from the full solution gives the forced response (Fig.~\ref{fig:schematic}),
\begin{equation}\label{eq:forced-remainder}
\begin{gathered}
 Y^\refr_{ij}\equiv Y_{ij}-Y^\out_{ij},\qquad
 \operatorname*{Res}_{z=0}\widetilde{\mathsf A}[Y^\refr_{ij}]=0,\\
 \bigl(\partial_x^2+1-(\partial_x^2a)/a\bigr)Y^\refr_{ij}=s_{ij}/k^2,\\
 \widetilde{\mathsf A}[Y^\refr_{ij}](z)=\widetilde{\mathsf A}[Y_{ij}](z)-\frac{\mathrm{i}\mathsf c[Y_{ij}]}{z}.
\end{gathered}
\end{equation}
This decomposition is unique: the difference between two zero-residue responses to the same source is a homogeneous solution with vanishing out-amplitudes $\mathsf a_{ij}=(\mathsf a_{c,ij},\mathsf a_{s,ij})$, and therefore vanishes.
Accordingly, $Y^\refr_{ij}$ depends only on the source, while changes in the tensor initial data affect only $Y^\out_{ij}$.
A vanishing residue does not require the source to decay or become inactive.
In matter domination, the tensor source $S_{ij}$ in Eq.~\eqref{eq:Sg} is constant, so $s_{ij}=aS_{ij}\propto x^2$ and the forced response is the static strain $h^\refr_{ij}=S_{ij}/k^2$ (that is, $Y^\refr_{ij}=s_{ij}/k^2$).
Its complex amplitude oscillates as $\mathrm{e}^{\mathrm{i}x}$ with a polynomial envelope, so its transform is singular only at $z=-1$ and its residue vanishes even though $Y^\refr_{ij}$ grows (Fig.~\ref{fig:schematic}(b) and Sec.~\ref{sec:rd-md}).
When $\widehat s_{ij}(0)$ is finite, the forced response is the advanced integral of the Green function $G$,
\begin{equation}\label{eq:advanced-integral}
\begin{gathered}
 G(x,x')\equiv u_c(x')u_s(x)-u_s(x')u_c(x),\\
 (G_{\rm adv}*s_{ij})(x)\equiv
 -\lim_{\epsilon\to0^+}\frac1{k^2}
 \int_x^\infty
 \mathrm{e}^{-\epsilon x'}G(x,x')s_{ij}(x')\,\dd x'.
\end{gathered}
\end{equation}
By Eq.~\eqref{eq:h-evolution}, its free-mode amplitudes are the tail of the integral that defines $\widehat s_{ij}(0)$,
\begin{equation}\label{eq:advanced-tail}
 \mathsf h[G_{\rm adv}*s_{ij}](x)=-\lim_{\epsilon\to0^+}\int_x^\infty\mathrm{e}^{-\epsilon x'}\,\partial_{x'}\mathsf h[Y_{ij}](x')\,\dd x'.
\end{equation}
The limit~\eqref{eq:real-out-amplitudes} of this tail vanishes, so the advanced response has zero out-amplitude and zero residue.
Let $Y^{\rm ini}_{ij}$ be the free solution with the initial data of $Y_{ij}$ at an initial time $x_{\rm ini}\ge0$,
\begin{equation}
\begin{gathered}
 Y^{\rm ini}_{ij}\equiv\mathsf h_c[Y_{ij}](x_{\rm ini})\,u_c+\mathsf h_s[Y_{ij}](x_{\rm ini})\,u_s,\\
 Y_{ij}=Y^{\rm ini}_{ij}+G_{\rm ret}*s_{ij},\\
 (G_{\rm ret}*s_{ij})(x)\equiv k^{-2}\int_{x_{\rm ini}}^xG(x,x')s_{ij}(x')\,\dd x'.
\end{gathered}
\end{equation}
The out-state is then
\begin{equation}\label{eq:ret-adv-free}
\boxed{\begin{aligned}
 Y^\out_{ij}&=Y_{ij}-G_{\rm adv}*s_{ij}\\
 &=Y^{\rm ini}_{ij}+(G_{\rm ret}-G_{\rm adv})*s_{ij}\\
 &=Y^{\rm ini}_{ij}+\lim_{\epsilon\to0^+}\frac1{k^2}
 \int_{x_{\rm ini}}^\infty\mathrm{e}^{-\epsilon x'}G(x,x')s_{ij}(x')\,\dd x'.
\end{aligned}}
\end{equation}
Equation~\eqref{eq:ret-adv-free} holds in any gauge, with $s_{ij}$ and $Y^{\rm ini}_{ij}$ taken in the same gauge, whenever $\widehat s_{ij}(0)$ is finite.
In flat space, the same split separates the free on-shell radiation field of scattered sources from the advanced Coulomb field of the outgoing sources \cite{Caron-Huot:2023vxl}.

\subsection{Gauge transformations and initial amplitudes}\label{sec:history-labels}
The out-state of Sec.~\ref{sec:residue-def} is defined for the tensor perturbation in any given gauge, with Newtonian gauge taken as the reference.
Another gauge $X$ describes the same physical solution via $Y^X_{ij}=Y^{\rm N}_{ij}+Q^X_{ij}$ (Sec.~\ref{sec:redefinitions}).
We now examine how $\mathsf c[Y^X_{ij}]$ depends on $X$.
In gauge $X$, the source is $s^X_{ij}=s^{\rm N}_{ij}+\mathcal LQ^X_{ij}$, and the initial amplitudes are transported from Newtonian gauge,
\begin{equation}\label{eq:initial-amplitude-transport}
\begin{aligned}
 \mathsf a^{X,{\rm ini}}_{ij}&\equiv\mathsf h[Y^X_{ij}](x_{\rm ini})
 =\mathsf a^{{\rm N},{\rm ini}}_{ij}+\mathsf h[Q^X_{ij}](x_{\rm ini}),\\
 \mathsf a^{{\rm N},{\rm ini}}_{ij}&\equiv\mathsf h[Y^{\rm N}_{ij}](x_{\rm ini}).
\end{aligned}
\end{equation}
Since $\mathsf h$ is linear in the tensor perturbation, the amplitudes in the two gauges differ by those of the shift at all times,
\begin{equation}\label{eq:hX-additive}
 \mathsf h[Y^X_{ij}](x)=\mathsf h[Y^{\rm N}_{ij}](x)+\mathsf h[Q^X_{ij}](x).
\end{equation}
The transforms of the complex amplitudes are therefore additive in the shift,
\begin{equation}\label{eq:boundary-term}
 \widetilde{\mathsf A}[Y^X_{ij}](z)-\widetilde{\mathsf A}[Y^{\rm N}_{ij}](z)
 =\widetilde{\mathsf A}[Q^X_{ij}](z)
 =\int_{x_{\rm ini}}^\infty\mathrm{e}^{\mathrm{i}zx}\,W[\varphi,Q^X_{ij}]\,\dd x.
\end{equation}
Starting the transform at $x_{\rm ini}$ instead of $0$ adds an entire function of $z$ and leaves the residue unchanged.
Taking the residue gives
\begin{equation}\label{eq:gauge-criterion}
\boxed{\;
 \mathsf c[Y^X_{ij}]-\mathsf c[Y^{\rm N}_{ij}]=\mathsf c[Q^X_{ij}].
\;}
\end{equation}
The out-state is therefore gauge independent if and only if the gauge shift has zero residue, $\mathsf c[Q^X_{ij}]=0$.
This criterion applies directly to each scalar pair (Sec.~\ref{sec:kernel-def}).
Resetting the initial amplitudes to zero in gauge $X$, instead of transporting zero Newtonian data, adds the free amplitude $-\mathsf h[Q^X_{ij}](x_{\rm ini})$ and changes the physical solution.
In radiation domination, the reset amplitude vanishes as $x_{\rm ini}\to0$ precisely when $Q^X_{ij}(0^+)=\partial_xQ^X_{ij}(0^+)=0$.
Appendix~\ref{app:histories} checks this condition for the standard gauges and treats synchronous gauge, whose residual modes make $Q^X_{ij}$ depend on integration constants.

\subsection{Continuous spectra: residues pair by pair}\label{sec:kernel-def}
\begin{figure}[t]
\centering
\resizebox{\textwidth}{!}{%
\begin{tikzpicture}[zpl,x=1.4cm,y=1.4cm]
\def\cs{0.577}%
\begin{scope}[xshift=0cm]
\fill[cCloud] (-0.3,0) rectangle (2.3,0.9);
\draw[zaxis] (-0.4,0)--(2.45,0) node[right] {$\operatorname{Re}z$};
\draw[zaxis] (1,-0.2)--(1,1.0) node[anchor=east] {$\operatorname{Im}z$};
\node[anchor=south,font=\small] at (1.0,1.1) {one pair, $c_s=1/\sqrt3$};
\draw[cClay,line width=1.2pt,-latex] (1,0.8)--(1,0.14);
\node[cInk,anchor=west] at (1.03,0.55) {$z=\mathrm{i}\epsilon$};
\zcross{1,0}
\node[cInk,anchor=south west] at (1.05,0.06) {free wave};
\node[cInk,below] at (1,-0.06) {$0$};
\zdot{1.6,0} \zdot{0.35,0}
\node[cInk,below] at (1.6,-0.06) {$-1-\omega_1$};
\node[cInk,below] at (0.35,-0.06) {$-1-\omega_2$};
\node[font=\small] at (-0.3,-0.7) {(a)};
\end{scope}
\begin{scope}[xshift=5.25cm]
\fill[cCloud] (-0.3,0) rectangle (2.3,0.9);
\draw[zaxis] (-0.4,0)--(2.45,0) node[right] {$\operatorname{Re}z$};
\draw[zaxis] (1,-0.2)--(1,1.0) node[anchor=east] {$\operatorname{Im}z$};
\node[anchor=south,font=\small] at (1.0,1.1) {all pairs, $c_s=1/\sqrt3$};
\draw[line width=1.6pt,cBlue!60] (\cs,0)--(2.3,0);
\draw[line width=3.6pt,cBlue,line cap=butt] (0.88,0)--(1.12,0);
\fill[cBlue] (\cs,0) circle (0.05);
\draw[cClay,line width=1.2pt,-latex] (1,0.8)--(1,0.14);
\zcross{1,0}
\node[cInk,anchor=south west] at (1.05,0.06) {free wave};
\node[cInk,anchor=south] at (\cs,0.06) {$c_s-1$};
\node[cInk,anchor=north west] at (1.02,-0.04) {$|z|\lesssim\epsilon$};
\node[cInk,below] at (2.12,-0.06) {source};
\node[font=\small] at (-0.3,-0.7) {(b)};
\end{scope}
\begin{scope}[xshift=10.5cm]
\fill[cCloud] (-0.3,0) rectangle (2.3,0.9);
\draw[zaxis] (-0.4,0)--(2.45,0) node[right] {$\operatorname{Re}z$};
\draw[zaxis] (1,-0.2)--(1,1.0) node[anchor=east] {$\operatorname{Im}z$};
\node[anchor=south,font=\small] at (1.0,1.1) {matter era, $c_s=0$};
\draw[cClay,line width=1.2pt,-latex] (1,0.8)--(1,0.14);
\fill[cBlue] (0,0) circle (0.075);
\node[cInk,anchor=south] at (0,0.1) {static source};
\node[cInk,below] at (0,-0.06) {$-1$};
\zcross{1,0}
\node[cInk,anchor=south west] at (1.05,0.06) {free wave};
\node[cInk,below] at (1,-0.06) {$0$};
\draw[<->,cSlate] (0.12,-0.4)--(0.88,-0.4);
\node[font=\small] at (-0.3,-0.7) {(c)};
\end{scope}
\end{tikzpicture}}
\caption{Singularities of the transform $\widetilde{\mathsf A}$ in the complex detuning plane.
The free-wave pole at $z=0$ is approached along $z=\mathrm{i}\epsilon$.
(a) Nonresonant radiation-era pair: the source singularities at $z=-1-\omega_\nu$ stay away from $z=0$.
(b) Scalar spectrum on all $q>0$: the source singularity that can reach $z=0$ sweeps $[c_s-1,\infty)$ and crosses $z=0$ on the source resonance surface $c_s(q+p)=k$.
Pairs with $|z|\lesssim\epsilon$ are not separated from the free wave at distance $\epsilon$, so the residue is taken pair by pair.
(c) Matter domination: the static source keeps its singularity at $z=-1$, away from $z=0$.}
\label{fig:zplane-continuum}
\end{figure}
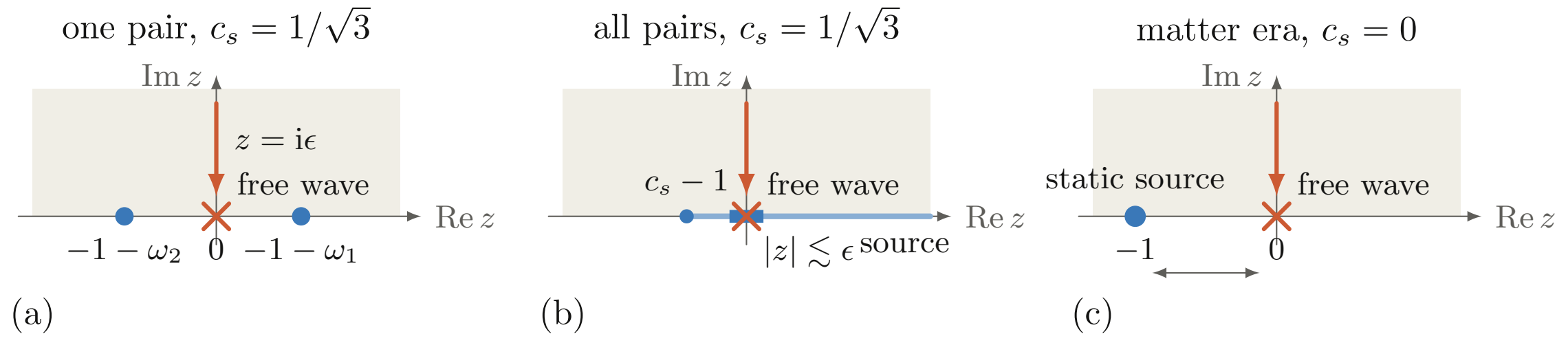

The source associated with a single scalar pair $\zeta_{\mathbf q}\zeta_{\mathbf p}$ in Eq.~\eqref{eq:forced-term} has finitely many frequencies, while the tensor perturbation with momentum $\mathbf k$ is a convolution over all scalar pairs.
As $(q,p)$ varies in this integral, the frequencies $\omega_\nu$ and detunings $\delta_\nu=1+\omega_\nu$ of the pairs form a continuum.
In particular, the term with frequency $\omega k=-c_s(q+p)$ has its singularity at $z=c_s(q+p)/k-1$.
For the Newtonian source, where $\sigma_{1,2}=\pm1$, it is the only term that can reach $z=0$, since the other detunings satisfy $|\delta_\nu|\ge1-c_s$.
Because $q+p\ge k$, its singularity ranges over $[c_s-1,\infty)$ when the scalar spectrum is supported on all $q>0$.
If the spectral support intersects $c_s(q+p)=k$, these singularities approach the light-cone pole arbitrarily closely (Fig.~\ref{fig:zplane-continuum}).
An isolated-pole argument therefore does not justify interchanging the residue with the momentum integral, so we take the residue pair by pair before integrating over momenta.
We express the second-order tensor perturbation in terms of the kernel~\eqref{eq:statistical-pair-kernel},
\begin{equation}\label{eq:history-pair-kernel}
 Y_{ij}(x,\mathbf k)\equiv\int[\dd q]\,\mathscr K_{ij}[Y](x;\mathbf k,\mathbf q)\,
 \zeta_{\mathbf q}\zeta_{\mathbf p},\qquad\mathbf p\equiv\mathbf k-\mathbf q.
\end{equation}
The symmetric kernel includes the TT projector and $a$.
We call $\mathsf A[\mathscr K_{ij}[Y]]$ the kernel amplitude and its transform the pair transform.
For each nonresonant pair with $\min_\nu|\delta_\nu|>0$, the residue at $z=0$ defines the out-amplitude kernel $\mathscr K_{0,ij}=(\mathscr K_{0,c,ij},\mathscr K_{0,s,ij})$, with the subscript $0$ marking the pole,
\begin{equation}\label{eq:kernel-outstate-limit}
 \mathscr K_{0,s,ij}(\mathbf k,\mathbf q)-\mathrm{i}\,\mathscr K_{0,c,ij}(\mathbf k,\mathbf q)
 \equiv-\mathrm{i}\operatorname*{Res}_{z=0}
 \widetilde{\mathsf A}[\mathscr K_{ij}[Y]](z;\mathbf k,\mathbf q).
\end{equation}
We call $\mathscr K_{0,ij}$ the kernel residue.

Integrating these kernels over all pairs defines the out-amplitude of the tensor perturbation, and for Gaussian input Wick contraction gives its power spectrum,
\begin{equation}\label{eq:kernel-outstate-assembly}
 \begin{aligned}
 \mathsf a_{ij}(\mathbf k)
 &\equiv\int[\dd q]\,\mathscr K_{0,ij}(\mathbf k,\mathbf q)
 \zeta_{\mathbf q}\zeta_{\mathbf p},\\
 P_{\mathsf a}(k)
 &=2\int[\dd q]\,
 \|\mathscr K_0(\mathbf k,\mathbf q)\|^2P_\zeta(q)P_\zeta(p).
 \end{aligned}
\end{equation}
We therefore define the out-state of SIGWs as follows.

\begin{definition}[Out-state]\label{def:outstate}
The out-state of a tensor perturbation $Y_{ij}$ is the free solution
\begin{equation*}
 Y^\out_{ij}(x,\mathbf k)=\mathsf a_{c,ij}(\mathbf k)\,u_c(x)+\mathsf a_{s,ij}(\mathbf k)\,u_s(x),
\end{equation*}
whose amplitudes $\mathsf a_{ij}=(\mathsf a_{c,ij},\mathsf a_{s,ij})$ are given by the light-cone residue taken for every nonresonant scalar pair before the momentum integral,
\begin{equation*}
 \mathsf a_{ij}(\mathbf k)
 =\int[\dd q]\,\Bigl[\lim_{\epsilon\to0^+}\epsilon\int_0^\infty\mathrm{e}^{-\epsilon x}\,\mathsf h[\mathscr K_{ij}[Y]](x;\mathbf k,\mathbf q)\,\dd x\Bigr]\zeta_{\mathbf q}\zeta_{\mathbf p}.
\end{equation*}
The bracket is the kernel residue $\mathscr K_{0,ij}$ of Eq.~\eqref{eq:kernel-outstate-limit}.
For a single scalar pair, the momentum integral is absent and the amplitudes reduce to Eq.~\eqref{eq:real-out-amplitudes}.
\end{definition}
For a continuous scalar spectrum, we write $\mathsf a[Y_{ij}]$ for these assembled amplitudes.

The amplitude~\eqref{eq:kernel-outstate-assembly} is a limit of residues of the field itself, taken on tensor perturbations whose pair singularities stay away from the pole.
Appendix~\ref{app:residue-existence} constructs these tensor perturbations by discarding the pairs within a detuning gap $\delta_0$ of the pole at fixed scalar cutoff $\Lambda_{\rm c}$, and shows that their residues converge to $\mathsf a_{ij}$ in spectral mean square as $\delta_0\to0$.
The limits are taken in the order $\epsilon\to0^+$, $\delta_0\to0$, and finally $\Lambda_{\rm c}\to\infty$ (Appendix~\ref{sec:rd-cutoff-removal}).
However, zero spectral weight on the resonance surfaces does not guarantee that the discarded layer $\min_\nu|\delta_\nu|<\delta_0$ contributes no residue to the field.
For this reason, the limit is taken on the kernel amplitudes.

Taking the residue after the momentum integral instead defines the field residue $\mathsf a_{\rm f}[Y_{ij}]$.
This residue can fail to exist for growing shifts in a shrinking resonance layer, as in the comoving gauge for $w\ge1/9$ and in the uniform-density gauge (Fig.~\ref{fig:limits} in Appendix~\ref{ap-app:fixed-k-proof}).
Whenever the field residue exists, it coincides with the amplitude~\eqref{eq:kernel-outstate-assembly} (Appendix~\ref{ap-sec:fixed-k-space}), so the pair-by-pair definition of the out-state does not rely on its existence.

\subsection{Relation to the LSZ reduction formula}\label{sec:lsz}
The reduction~\eqref{eq:residue-reduction} has the structure of the LSZ reduction formula \cite{Lehmann:1954rq,Peskin:1995ev,Schwartz:2014sze}, which becomes explicit when we transform the tensor perturbation itself rather than its amplitude.
For $\operatorname{Im}z>0$, let $\varphi_z$ be the free mode of frequency $1+z$,
\begin{equation}
\begin{gathered}
 [\partial_x^2+(1+z)^2-(\partial_x^2a)/a]\varphi_z=0,\\
 \varphi_z\simeq\mathrm{e}^{\mathrm{i}(1+z)x},\qquad\varphi_0=\varphi.
\end{gathered}
\end{equation}
The Wronskian then obeys
\begin{equation}
    \partial_xW[\varphi_z,Y_{ij}]=\varphi_zs_{ij}/k^2+z(2+z)\varphi_zY_{ij}.
\end{equation}
Integrating over $x$ from $0$ to $\infty$ gives
\begin{equation}\label{eq:jost-identity}
 \widetilde Y_{ij}(z)\equiv\int_0^\infty\varphi_zY_{ij}\,\dd x
 \quad\Rightarrow\quad -z(2+z)\,\widetilde Y_{ij}(z)=\widetilde s_{ij}(z)+W[\varphi_z,Y_{ij}](0),
\end{equation}
where
\begin{equation}
 \widetilde s_{ij}(z)\equiv\frac1{k^2}\int_0^\infty\varphi_z\,s_{ij}\,\dd x .
\end{equation}
In radiation domination,
\begin{equation}
 \varphi_z=\mathrm{e}^{\mathrm{i}(1+z)x},\qquad
 W[\varphi_z,Y_{ij}](0)=\partial_xY_{ij}(0)-\mathrm{i}(1+z)Y_{ij}(0),
\end{equation}
so Eq.~\eqref{eq:jost-identity} reproduces the driven-oscillator transform~\eqref{eq:driven-oscillator} for $\Omega=1$ and $\omega=1+z$, with the Wronskian at $x=0$ carrying the initial data.
The factor $-z(2+z)=1-(1+z)^2$ is the inverse free propagator, whose zeros at $z=0$ and $z=-2$ correspond to the two free frequencies $\pm k$.
Only the first survives in $\widetilde{\mathsf A}$, since $W[\varphi,\varphi]=0$ removes the $\varphi$ component of $Y_{ij}$.
At late times,
\begin{equation}
 \varphi_z=\mathrm{e}^{\mathrm{i}zx}\varphi\,\bigl[1+O(z/x)\bigr],
\end{equation}
so $\widetilde s_{ij}(z)$ and $\widehat s_{ij}(z)$ have the same limit at $z=0$ for nonresonant pairs.
Multiplying $\widetilde Y_{ij}$ by the inverse propagator and going on shell therefore gives
\begin{equation}\label{eq:jost-reduction}
 \lim_{\epsilon\to0^+}\bigl[1-(1+\mathrm{i}\epsilon)^2\bigr]
 \widetilde Y_{ij}(\mathrm{i}\epsilon)
 =\mathsf A[Y_{ij}](0)+\widehat s_{ij}(0)=\mathsf c[Y_{ij}],
\end{equation}
which recovers Eq.~\eqref{eq:residue-reduction}.
As in the LSZ formula, the inverse propagator cancels the free pole and leaves the on-shell source amplitude.
For a continuous scalar spectrum, the scalar-pair kernels play the role of matrix elements with two scalar quanta, and the reduction is applied to each pair before the momentum integral, as in Sec.~\ref{sec:kernel-def}.
In flat space, LSZ-type reductions also extract classical observables and gravitational waveforms from scattering amplitudes \cite{Kosower:2018adc,Cristofoli:2021vyo,Caron-Huot:2023vxl}.
The comparison with weak asymptotic limits \cite{Haag:1958vt,Haag:1996hvx} is discussed in Appendix~\ref{ap-sec:fixed-k-space}.

The LSZ reduction assumes that the interaction switches off sufficiently fast at late times.
In Coulomb scattering, the $1/r$ tail of the potential violates this condition, and the asymptotic states require the Dollard modification \cite{Dollard:1964cok,Reed:1979ne,Morchio:2014zga}.
The radiation-era source has a similar resonant $x^{-1}$ tail, whose singular contribution is
\begin{equation}\label{eq:lsz-coulomb}
\begin{aligned}
 \int_1^\infty\frac{\mathrm{e}^{\mathrm{i}\delta x}}{x}\,\dd x
 ={}&-\gamma_{\rm E}-\log|\delta|+\frac{\mathrm{i}\pi}{2}\operatorname{sgn}\delta+O(\delta),
\end{aligned}
\end{equation}
where $\gamma_{\rm E}$ is Euler's constant and $\delta=1-c_s(q+p)/k$.
Here $\log|\delta|$ and its branch give the logarithmic and step terms in the radiation-era amplitude \cite{Kohri:2018awv}.
At exact resonance, the amplitude grows as $\log x$.
Off resonance, this growth saturates at $x\sim1/|\delta|$ and leaves the kernel singularity $-\log|\delta|$, whose square is integrable across $\delta=0$, so the spectrum $P_{\mathsf a}$ remains finite.
The Dollard modification is a phase acting at all momenta.
Here, by contrast, the resonant tail is an additive forced contribution, and its near-resonant part vanishes in spectral mean square under the bounds of Appendix~\ref{ap-app:gaussian-reference}.
For general $0<w<1$, the tail scales as $x^{-\gamma}$ with $\gamma=2/(1+3w)$.
The kernel singularity then takes the form $\Gamma(1-\gamma)(-\mathrm{i}\delta)^{\gamma-1}$ for $\gamma<1$ ($w>1/3$), while for $\gamma>1$ the source integral remains finite at $\delta=0$.
In matter domination, the source is nonoscillating and its singularity stays at $z=-1$.

\section{The equivalence theorem and gauge invariance}\label{sec:kernel-outstate}
By Eq.~\eqref{eq:gauge-criterion}, the out-state is gauge independent if the gauge shift has zero residue for almost every scalar pair.
Here and below, ``almost every pair'' means every pair outside a set of zero volume in the scalar-pair domain $(u,v)$, which then also carries zero spectral weight.
Section~\ref{sec:completeness} gives sufficient conditions on the late-time behavior of a shift for its residue to vanish.
Section~\ref{sec:gauge-independence} defines the intrinsic gauges, and Sec.~\ref{sec:generator-frequencies} shows that they satisfy these conditions.
Section~\ref{sec:outside-class} discusses gauges outside the class, and Sec.~\ref{sec:main-result} states the invariance result and the resulting free--forced split.

\subsection{A classical equivalence theorem}\label{sec:completeness}
The theorem concerns shifts whose kernel amplitude, as for the source in Sec.~\ref{sec:residue-def}, is a finite sum of oscillating terms with power-law envelopes.
\begin{theorem}[Classical equivalence theorem for tensor perturbations]\label{thm:equivalence}
For a fixed mode $\mathbf k$, let $Y_{ij}$ and $Y_{ij}+Q_{ij}$, with $Q_{ij}\in\Qcal$, be two tensor perturbations on the same physical solution.
Suppose that for almost every scalar pair, the kernel amplitude of the shift has the late-time form
\begin{equation}\label{eq:shift-late-form}
 \mathsf A[\mathscr K_{ij}[Q]](x;\mathbf k,\mathbf q)
 =\sum_{\nu=1}^J\mathrm{e}^{\mathrm{i}\delta_{\nu}x}\,\mathsf b_{\nu,ij}(x)+\mathfrak e_{ij}(x),\qquad
 \lim_{x\to\infty}\mathfrak e_{ij}=0,
\end{equation}
with $\delta_\nu=1+\omega_\nu$ for shift frequencies $\omega_{\nu}k$, satisfying
\begin{itemize}
\item[(E1)] \emph{power-law growth}: $|\partial_x^l\mathsf b_{\nu,ij}|\le C_lx^{M-l}(1+\log x)^L$ for every $l\ge0$ at late times, with finite $M,L$,
\item[(E2)] \emph{no tensor-frequency component}: $\delta_\nu\neq0$ for every $\nu$.
\end{itemize}
Then the pair transform of the shift has no light-cone pole,
\begin{equation}
 \lim_{\epsilon\to0^+}\epsilon\,\widetilde{\mathsf A}[\mathscr K_{ij}[Q]](\mathrm{i}\epsilon)=0 .
\end{equation}
For almost every pair, the residue of $Y_{ij}+Q_{ij}$ exists if and only if that of $Y_{ij}$ does, in which case
\begin{equation}
 \mathscr K_{0,ij}[Y+Q]=\mathscr K_{0,ij}[Y],\qquad
 \mathsf a[Y_{ij}+Q_{ij}]=\mathsf a[Y_{ij}].
\end{equation}
In particular, their spectra agree.
\end{theorem}

\begin{proof}
Fix a nonresonant pair and a sufficiently late time $x_0$.
The interval before $x_0$ contributes $O(\epsilon)$ to $\epsilon\,\widetilde{\mathsf A}[\mathscr K_{ij}[Q]](\mathrm{i}\epsilon)$, since the amplitude is integrable there.
For an integer $n>M+1$, integration by parts gives
\begin{equation}\label{eq:shift-ibp-bound}
 \left|\epsilon\int_{x_0}^\infty
 \mathrm{e}^{(-\epsilon+\mathrm{i}\delta_\nu)x}\mathsf b_{\nu,ij}(x)\,\dd x\right|
 \le\epsilon\left[
 \sum_{l=0}^{n-1}\frac{|\partial_x^l\mathsf b_{\nu,ij}(x_0)|}{|\delta_\nu|^{l+1}}
 +\frac{\int_{x_0}^\infty|\partial_x^n\mathsf b_{\nu,ij}|\,\dd x}{|\delta_\nu|^n}\right]
 \longrightarrow0.
\end{equation}
The expression in the bracket is independent of $\epsilon$ and finite, since (E1) with $M-n<-1$ makes $\partial_x^n\mathsf b_{\nu,ij}$ integrable at late times.
For $x_1>x_0$, we have
\begin{equation}
 \left|\epsilon\int_{x_0}^\infty \mathrm{e}^{-\epsilon x}\mathfrak e_{ij}(x)\,\dd x\right|
 \le\epsilon\int_{x_0}^{x_1}|\mathfrak e_{ij}(x)|\,\dd x
 +\sup_{x\ge x_1}|\mathfrak e_{ij}(x)|.
\end{equation}
At fixed $x_1$, the first term vanishes as $\epsilon\to0^+$, so the limit superior of the left-hand side is at most $\sup_{x\ge x_1}|\mathfrak e_{ij}|$, which tends to zero as $x_1\to\infty$ since $\mathfrak e_{ij}\to0$.
With transported initial data, Eq.~\eqref{eq:gauge-criterion} gives
\begin{equation}
 \mathsf c[\mathscr K_{ij}[Y+Q]]=\mathsf c[\mathscr K_{ij}[Y]]+\mathsf c[\mathscr K_{ij}[Q]]=\mathsf c[\mathscr K_{ij}[Y]],
\end{equation}
so $\mathscr K_{0,ij}[Y+Q]=\mathscr K_{0,ij}[Y]$.
Since this holds for almost every pair, the assembled amplitudes~\eqref{eq:kernel-outstate-assembly} of $Y_{ij}$ and $Y_{ij}+Q_{ij}$ are given by the same momentum integral.
\end{proof}

The kernel statement holds for each pair separately: it uses no statistics of $\zeta$, and the constants in (E1) may depend on $(u,v)$.
For Gaussian input, the corresponding identity for the assembled amplitudes is
\begin{equation}\label{eq:two-level}
 P_{\mathsf a[Y+Q]-\mathsf a[Y]}(k)
 =2\int[\dd q]\,\|\mathscr K_0[Y+Q]-\mathscr K_0[Y]\|^2 P_\zeta(q)P_\zeta(p)=0.
\end{equation}

\begin{figure}[t]
\centering
\begin{tikzpicture}[zpl,x=1.5cm,y=1.5cm]
\begin{scope}
\fill[cCloud] (-1.0,0) rectangle (2.3,1.75);
\draw[zaxis] (-1.1,0)--(2.45,0) node[right] {$\operatorname{Re}z$};
\draw[zaxis] (1,-0.2)--(1,1.9) node[anchor=east] {$\operatorname{Im}z$};
\node[cSlate,align=left,anchor=north west] at (-0.95,1.72) {$\widetilde{\mathsf A}[\mathscr K_{ij}[Q]]$ of a\\gauge shift\\for one pair};
\draw[cClay,thick,dashed] (1,0) circle (0.12);
\node[cInk,anchor=south east] at (0.92,0.12) {no pole};
\zdot{-0.6,0} \zdot{0.42,0} \zdot{1.8,0}
\node[cInk,below] at (-0.6,-0.07) {$-\delta_1$};
\node[cInk,below] at (0.42,-0.07) {$-\delta_2$};
\node[cInk,below] at (1.8,-0.07) {$-\delta_3$};
\draw[->,thick,cBlue] (1.78,0.12) to[bend right=45] (1.17,0.12);
\node[cInk,align=left,anchor=west] at (1.08,0.85) {$\delta_3\to0$ near a\\resonance surface};
\draw[<->,cSlate] (1,-0.62)--(1.8,-0.62) node[midway,below] {$|\delta_3|$};
\end{scope}
\end{tikzpicture}
\caption{Frequency plane of the transform of a gauge shift for one pair.
Its singularities sit at $z=-\delta_\nu$ with $\delta_\nu=1+\omega_\nu$, and there is no pole at $z=0$ (dashed circle), so its residue vanishes for $\delta_\nu\ne0$.
Near a resonance surface, one detuning $\delta_3$ approaches zero and its singularity moves toward the light-cone pole (arrow).}
\label{fig:zplane-shift}
\end{figure}
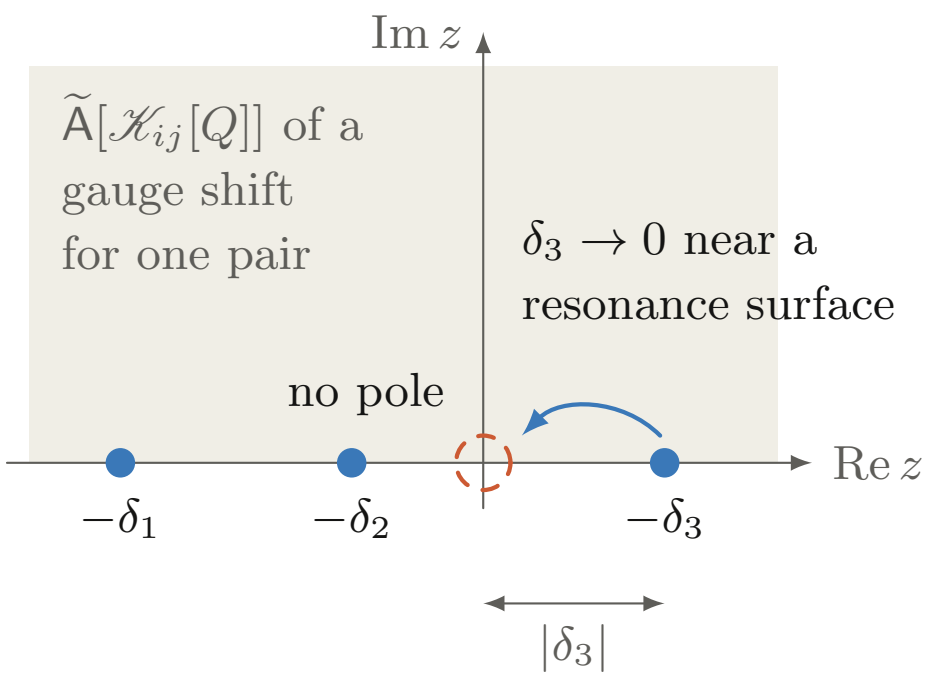

A single term with a power-law envelope illustrates both cases of condition (E2).
For $m>-1$,
\begin{equation}\label{eq:shift-term-transform}
 \int_0^\infty\mathrm{e}^{\mathrm{i}zx}\,x^m\mathrm{e}^{\mathrm{i}\delta_{\nu}x}\,\dd x
 =\frac{\Gamma(m+1)}{[-\mathrm{i}(z+\delta_\nu)]^{m+1}},\qquad
 \operatorname{Im}z>0,
\end{equation}
so that (Fig.~\ref{fig:zplane-shift})
\begin{equation}
 \epsilon\widetilde{\mathsf A}(\mathrm{i}\epsilon)\to0\quad(\delta_\nu\ne0),\qquad
 \epsilon\widetilde{\mathsf A}(\mathrm{i}\epsilon)=\Gamma(m+1)\,\epsilon^{-m}\quad(\delta_\nu=0).
\end{equation}
A term with $\delta_\nu=0$ whose envelope tends to a nonzero constant, $\mathsf b_{\nu,ij}(x)\to \mathsf b^\infty_{\nu,ij}\ne0$, gives $\epsilon\widetilde{\mathsf A}(\mathrm{i}\epsilon)\to \mathsf b^\infty_{\nu,ij}$, so on a set of positive spectral weight it changes the out-amplitude.
Resonant growth of such an envelope, $\mathsf b_{\nu,ij}\sim x^m(\log x)^l$ with $m>0$, or $m=0$ and $l>0$, prevents a finite residue.
Shifts confined to a finite time interval contribute $O(\epsilon)$ and do not affect the theorem.

Theorem~\ref{thm:equivalence} is the classical counterpart of the equivalence theorem in quantum field theory, which states that on-shell amplitudes are invariant under a field redefinition whose added term has no one-particle pole \cite{Chisholm:1961tha,Kamefuchi:1961sb}.
A gauge shift is quadratic in the scalar perturbations and contains no term linear in the free tensor perturbation.
It therefore also preserves the normalization of the pole, which corresponds to a unit field-strength renormalization, $Z=1$.
The source resonance surface expresses a three-wave resonance condition: the frequencies of the two scalar modes add up to the tensor frequency $k$ while their momenta add up to $\mathbf k$, just as energy and momentum are conserved at a vertex with on-shell lines.

\subsection{Intrinsic gauges}\label{sec:gauge-independence}
By Eq.~\eqref{eq:gauge-shift-ordinary}, $Q^X_{ij}$ is quadratic in the generators $\alpha,\beta$ and the scalar perturbations, so applying conditions (E1) and (E2) to a gauge transformation constrains the frequencies and envelopes of $\alpha$ and $\beta$.
A scalar gauge fixes a slicing and spatial labels.
We define the observers associated with a gauge to follow the unit normals to its time slices.
Their worldlines coincide with the curves of constant spatial coordinates only when the shift vector vanishes.
In Newtonian gauge the shift vector vanishes, so the Newtonian observers sit at fixed spatial coordinates and their normal congruence has zero shear (Table~\ref{tab:normal-rules}).
This zero-shear gauge \cite{Hwang:2001qk} leaves no residual freedom, and its metric potentials are Bardeen's gauge-invariant potentials \cite{Bardeen:1980kt}.
Relative to Newtonian gauge, the first-order generator $\xi_1=(\alpha,\partial^i\beta)$ of Sec.~\ref{sec:redefinitions} has
\begin{equation}\label{eq:generator-profiles}
\begin{aligned}
 \alpha(\eta,\mathbf k)&=T_\alpha(\eta,k)\,\zeta_{\mathbf k},\\
 \beta(\eta,\mathbf k)&=T_\beta(\eta,k)\,\zeta_{\mathbf k}.
\end{aligned}
\end{equation}
In Newtonian coordinates, the slices of gauge $X$ are determined by $\eta-\alpha(\eta,\mathbf X)=\mathrm{const}$.
Their unit normal $n_X^\mu$ is the four-velocity of the gauge-$X$ observers, and to first order their spatial three-velocity relative to the Newtonian observers is \cite{Ellis:1998ct,Tsagas:2007yx}
\begin{equation}\label{eq:tilt-velocity}
 v_X^i\equiv\frac{n_X^i}{n_X^0}=\partial^i\alpha .
\end{equation}
The first term of Eq.~\eqref{eq:gauge-shift-ordinary} represents the Lorentz contraction seen by the tilted observers,
\begin{equation}\label{eq:lorentz-contraction}
 \Delta h_{ij}\supset-\Lambda_{ij}{}^{lm}\,v_{Xl}\,v_{Xm} .
\end{equation}
The remaining terms of Eq.~\eqref{eq:gauge-shift-ordinary} arise from the spatial reparametrization and its combination with the tilt \cite{Lu:2020diy}.

Equation~\eqref{eq:generator-profiles} leaves the time dependence of $T_\alpha$ and $T_\beta$ to the gauge rule, so (E1) and (E2) are conditions on the rule itself.
Standard cosmological gauges fix their observers by a physical condition on the solution: the observers comove with the matter, see uniform density or uniform expansion, have flat slices or zero shear, or fall freely.
Such a rule reads the slicing from the matter and geometry, so it introduces no time scale of its own.
The generator is then driven by the scalar perturbations, oscillates at their sound frequencies, and inherits the power-law time dependence of the background.
We use these properties to define the class of intrinsic gauges, and show below that they ensure the late-time form required by (E1) and (E2).

\begin{definition}[Intrinsic gauge]\label{def:intrinsic}
On a constant-$w$ background, an \emph{intrinsic gauge} fixes its normal observers and spatial coordinates by rules satisfying the following requirements, in the precise sense of Definition~\ref{def:intrinsic-precise}.
\begin{itemize}
\item[(P1)] \emph{Locality}.
The rule uses local fields and background quantities at one point, with no prescribed function of time and hence no external clock.
\item[(P2)] \emph{No modes of their own}.
The observers and spatial labels have no independent oscillatory dynamics.
They are fixed either kinematically by the local state, as for the comoving, flat and uniform-density slicings, or through a local first-order force law, as for freely falling synchronous observers.
\item[(P3)] \emph{Temperedness}.\footnote{The name is borrowed from tempered distributions, the continuous linear functionals on Schwartz test functions, which the Wightman axioms require of quantum fields \cite{Haag:1996hvx}.
A function growing at most as a power of time defines such a distribution.}
The observer motion and spatial-coordinate shifts grow at most as a power of time, and any negative damping in a force law is at most of the order of the Hubble rate.
\item[(P4)] \emph{Complete fixing}.
The initial observers and spatial labels are specified once, leaving no residual gauge freedom, and the same coordinate choice is retained when the solution is expressed in another reference gauge.
\end{itemize}
\end{definition}
These conditions constrain both the observers and the spatial-coordinate rule.
They are sufficient for (E1) and (E2), but not necessary.
A gauge outside the class may still share the out-state, and for such a gauge (E1) and (E2) must be checked directly, as Appendix~\ref{ap-sec:generator-estimates} does for the harmonic gauge.

For the standard gauges, the determining equations are explicit (Table~\ref{tab:normal-rules}).
These six slicings encompass every slicing compared in Refs.~\cite{Hwang:2017oxa,Tomikawa:2019tvi,Lu:2020diy,Ali:2023moi,Domenech:2020xin,Inomata:2019yww}.
The total-matter and comoving-orthogonal gauges of Refs.~\cite{Lu:2020diy,Ali:2023moi} both use comoving slices, and the TT gauge of Ref.~\cite{Inomata:2019yww}, defined by $\tilde\phi=\tilde B=0$, is synchronous gauge.
The comoving and uniform-density gauges, whose late-time spectral growth prompted the debate over gauge dependence \cite{Tomikawa:2019tvi,Lu:2020diy,Ali:2023moi}, therefore belong to this class.
Synchronous gauge satisfies (P4) once its residual modes are fixed, either by early-time regularity \cite{Lu:2020diy,Ali:2020sfw} or by requiring good behavior on subhorizon scales, which selects nonzero residual constants \cite{Domenech:2020xin} (Appendix~\ref{app:histories}).
A tilde marks the first-order scalar perturbations of Eq.~\eqref{eq:perturbed-metric} in gauge $X$, together with the fluid velocity $\delta\tilde u^i=a^{-1}\partial^i\tilde v$, the density perturbation $\delta\tilde\rho$ and the perturbation $\delta\tilde\theta$ of the normal expansion.
They follow from the Newtonian values by Eq.~\eqref{eq:firstorder-convention-compact}, where Newtonian gauge has $\phi=\psi=\Phi$, $B=E=0$, velocity potential $v_{\rm N}$ and density perturbation $\delta\rho_{\rm N}$.
\begin{table}[t]
\centering\footnotesize
\setlength{\tabcolsep}{5pt}
\begin{tabular}{lllll}
\toprule
Gauge & Rule for $n_X$ & \begin{tabular}[b]{@{}l@{}}Slicing\\condition\end{tabular} & \begin{tabular}[b]{@{}l@{}}Spatial\\condition\end{tabular} & Generator from Newtonian gauge\\
\midrule
Newtonian & $\sigma_{\mu\nu}[n]=0$ & $\tilde E'-\tilde B=0$ & $\tilde E=0$ & $\alpha=0$, $\beta=0$\\
Flat & ${}^{(3)}\!R$ unperturbed & $\tilde\psi=0$ & $\tilde E=0$ & $\mathcal H\alpha=\Phi$, $\beta=0$\\
Comoving & $n=u$ & $\tilde v+\tilde B=0$ & $\tilde E=0$ & $\alpha=v_{\rm N}$, $\beta=0$\\
Uniform density & $n\parallel\nabla\rho$ & $\delta\tilde\rho=0$ & $\tilde E=0$ & $\bar\rho'\alpha=-\delta\rho_{\rm N}$, $\beta=0$\\
Uniform expansion & $\theta[n]=3\mathcal H/a$ & $\delta\tilde\theta=0$ & $\tilde E=0$ & $[q^2+3(\mathcal H^2-\mathcal H')]\alpha=-3U$, $\beta=0$\\
Synchronous & $a^\mu[n]=0$ (free fall) & $\tilde\phi=0$ & $\tilde B=0$ & $\alpha'+\mathcal H\alpha=-\Phi$, $\beta'=\alpha$\\
\bottomrule
\end{tabular}
\caption{Standard intrinsic gauges and the equations determining their generators.
Here $a^\mu[n]$, $\theta[n]$ and $\sigma_{\mu\nu}[n]$ are the acceleration, expansion and shear of the normal congruence, and $u^\mu$ is the fluid four-velocity.
The first five time generators are algebraic, while synchronous gauge requires two successive first-order steps with rates $\mathcal H$ and zero, whose integration constants are the fixed residual data.
Appendix~\ref{ap-sm:intrinsic-full} verifies the conditions for arbitrary reference perturbations.}
\label{tab:normal-rules}
\end{table}

\subsection{Frequency content of the generators}\label{sec:generator-frequencies}
We now show that the generators of intrinsic gauges oscillate only at the scalar frequencies, with power-law envelopes.
By Definition~\ref{def:intrinsic} (and explicitly for Table~\ref{tab:normal-rules}), the gauge conditions are solved by two kinds of steps.
Algebraic division by the background coefficients changes only the envelope.
A first-order equation $v_X'+rv_X=F$ gives
\begin{equation}\label{eq:first-order-solution}
\begin{aligned}
 v_X(\eta)&=\mathrm{e}^{-\varrho(\eta)}
 \Bigl[C+\int^\eta\mathrm{e}^{\varrho(\eta')}F(\eta')\,\dd\eta'\Bigr],\\
 \varrho(\eta)&\equiv\int^\eta r\,\dd\eta',
\end{aligned}
\end{equation}
with a nonoscillating homogeneous term obeying
\begin{equation}\label{eq:hubble-slow}
 -r\le\frac{c_0(q)}\eta
 \quad\Longrightarrow\quad
 |v_{X,\rm hom}(\eta)|\le C\eta^{c_0(q)}.
\end{equation}
For a drive $F=\mathrm{e}^{\mathrm{i}\omega k\eta}f$, the controlled expansion begins with
\begin{equation}\label{eq:forced-expansion}
 v_{X,\rm forced}=\mathrm{e}^{\mathrm{i}\omega k\eta}
 \biggl[\frac{f}{r+\mathrm{i}\omega k}
 -\frac1{r+\mathrm{i}\omega k}\partial_\eta
 \Bigl(\frac{f}{r+\mathrm{i}\omega k}\Bigr)+\cdots\biggr],
\end{equation}
where $\omega\ne0$.
At late times, the scalar transfer function has a Bessel asymptotic expansion with oscillatory factors \(\mathrm e^{\pm\mathrm ic_sq\eta}\).
This expansion remains valid under differentiation, so the nonzero driving frequencies in Eq.~\eqref{eq:forced-expansion} are \(\omega k=\pm c_sq\).
Applying the algebraic and first-order steps successively gives
\begin{equation}\label{eq:late-time-form}
\begin{gathered}
 T(\eta,q)=\sum_{\omega k\in\{0,\pm c_sq\}}\mathrm{e}^{\mathrm{i}\omega k\eta}\,g_\omega(\eta,q),\\
 |\partial_\eta^lg_\omega(\eta,q)|\le C_l(q)\,\eta^{M(q)-l}
 \bigl(1+\log\eta\bigr)^{L(q)},\qquad l\ge0.
\end{gathered}
\end{equation}
This form holds for both $T_\alpha$ and $T_\beta$.
The envelope exponents $M,L$ and the constants $C_l$ may depend on $q$.
Appendix~\ref{ap-sm:intrinsic-full} derives these bounds from the expansion~\eqref{eq:forced-expansion}, including strong positive damping and zero-frequency drives.
At first order, the generator taking one intrinsic gauge to another is the difference between their generators relative to Newtonian gauge.
Since subtraction preserves the allowed frequencies and envelope bounds, the same late-time form holds between any two intrinsic gauges.

Substituting Eq.~\eqref{eq:late-time-form} into the quadratic shift~\eqref{eq:gauge-shift-ordinary} gives the form~\eqref{eq:shift-late-form}, because $\mathsf A=W[\varphi,\cdot]$ multiplies each term by the late-time phase $\mathrm{e}^{\mathrm{i}x}$, with
\begin{equation}\label{eq:intrinsic-detunings}
 \delta_\nu=1+\frac{c_s}{k}(\sigma_1q+\sigma_2p),\qquad
 \sigma_{1,2}\in\{0,\pm1\}.
\end{equation}
Products and derivatives preserve (E1).
For $0<c_s<1$, each equation $\delta_\nu=0$ with a nonempty solution set in the scalar-pair domain defines a resonance surface of zero volume, so (E2) holds almost everywhere.
We call $c_s(q+p)=k$ the source resonance surface, on which the Newtonian source itself resonates with the tensor mode, and $c_sq=k$ and $c_sp=k$ the gauge resonance surfaces, which only gauge shifts reach.
Off the resonance surfaces, Theorem~\ref{thm:equivalence} therefore gives
\begin{equation}\label{eq:kernel-equality-intrinsic}
 \mathscr K^X_{0,ij}(\mathbf k,\mathbf q)=\mathscr K^{\rm N}_{0,ij}(\mathbf k,\mathbf q).
\end{equation}
The out-amplitudes $\mathsf a^X_{ij}$ and $\mathsf a^{\rm N}_{ij}$ of Eq.~\eqref{eq:kernel-outstate-assembly} are then given by the same momentum integral, so all intrinsic gauges share the same out-state, and by Eq.~\eqref{eq:two-level} the same spectrum.

\subsection{Gauges outside the class}\label{sec:outside-class}
The invariance among intrinsic gauges cannot extend to all gauge choices.
Since the gauge rule alone fixes the time dependence of $T_\alpha$ and $T_\beta$, nothing in the definition of a gauge prevents a coordinate rule from introducing a tensor-frequency component into the shift $Q^X_{ij}$.
Such a component is itself a free tensor perturbation, so it adds to the out-state, and the example below constructs one explicitly.
An invariance statement covering all gauges would therefore be false.
Restricting invariance to a specified class of transformations is standard for asymptotic observables.
The equivalence theorem of quantum field theory holds only for field redefinitions that leave the one-particle poles, and hence the asymptotic states, invariant \cite{Chisholm:1961tha,Kamefuchi:1961sb}.
In asymptotically flat spacetimes, the radiation field and the fluxes at null infinity are defined for metrics obeying asymptotic falloff conditions, and the transformations preserving those conditions form the Bondi--Metzner--Sachs group \cite{Bondi:1962px,Sachs:1962wk,Wald:1999wa}.
In both cases, the admissible class is fixed by the asymptotic structure to which the observable refers.
Here the out-state is defined at the light-cone pole, (E1) and (E2) are the conditions under which a shift has no residue there, and (P1)--(P4) provide sufficient conditions stated directly in terms of the observers and spatial coordinates.

Gauges outside this class can violate (E1) or (E2), because their rules introduce a time dependence that the physical solution does not contain.
A rule that reads time from an external clock, rather than from the scalar perturbations, violates (E2).
For example, in radiation domination with $a=A\eta$, fix a target tensor momentum $k$ and choose
\begin{equation}\label{eq:clock-example}
 \alpha(\eta,\mathbf q)=\eta^{-1/2}\cos\frac{k\eta}{2}\,\zeta_{\mathbf q},\qquad \beta=0.
\end{equation}
These observers keep time at half the tensor frequency, and the Lorentz contraction~\eqref{eq:lorentz-contraction} gives the kernel of the shift
\begin{equation}
 \mathscr K_{ij}[Q]=-A\cos^2(k\eta/2)\,\Lambda_{ij}{}^{lm}q_lq_m
 =-\frac{A}{2}(1+\cos k\eta)\,\Lambda_{ij}{}^{lm}q_lq_m.
\end{equation}
Its oscillating part is an exact free tensor perturbation of constant amplitude.
The gauge transformation alone therefore adds a free wave of momentum $k$ for every pair, not only on a set of zero volume.
In position space, this corresponds to the local rule $\alpha=\eta^{-1/2}\cos(k\eta/2)\,\zeta(\mathbf X)$ with a fixed clock frequency.
When the clock frequency is held fixed, the added free wave appears only in tensor modes whose momentum equals twice that frequency, making their out-state depend on the choice of clock.
Observers whose three-velocity obeys a second-order equation with a restoring force, such as $v_X''+\varpi^2v_X=F$, admit homogeneous solutions $\mathrm{e}^{\pm\mathrm{i}\varpi\eta}$ and oscillate at their own frequency.
A first-order law $v_X'+rv_X=F$ with negative damping stronger than the expansion, i.e. $\eta r\to-\infty$, makes the generators grow faster than any power of time.
Each of these rules attaches to the observers a clock, an oscillator or an instability that the physical solution does not contain.
The clock frequency $k/2$ in Eq.~\eqref{eq:clock-example} and the restoring frequency $\varpi$ are introduced by the choice of observers, and the instability rate is not set by the expansion.
What such a shift adds to the out-state therefore reflects the observers rather than the physical solution, and the definition of an intrinsic gauge excludes it by requiring the coordinates to be fixed by the matter and geometry.

\subsection{Gauge invariance and the free--forced split}\label{sec:main-result}
The gauge criterion~\eqref{eq:gauge-criterion} and the kernel equality~\eqref{eq:kernel-equality-intrinsic} combine into the invariance of the out-state among intrinsic gauges.
\begin{proposition}[Invariance of the out-state among intrinsic gauges]\label{prop:main}
Fix $0<w<1$ and compare intrinsic gauges on the same physical solution, with transported tensor data at an initial time $x_{\rm ini}$.
Under the scalar assumptions of Sec.~\ref{sec:averaging}, the kernel residues agree off the resonance surfaces of Eq.~\eqref{eq:intrinsic-detunings}, which have zero volume, and all intrinsic gauges have the same out-state,
\begin{equation}\label{eq:intrinsic-invariance}
 \boxed{\mathsf a[Y^X_{ij}]=\mathsf a[Y^{\rm N}_{ij}]=\mathsf a^{{\rm N},{\rm ini}}_{ij}+\mathsf a^{\rm ind}_{ij}}.
\end{equation}
Here $\mathsf a^{{\rm N},{\rm ini}}_{ij}=\mathsf h[Y^{\rm N}_{ij}](x_{\rm ini})$ is the free amplitude of the transported initial data, and $\mathsf a^{\rm ind}_{ij}\equiv\mathsf a[G_{\rm ret}*s^{\rm N}_{ij}]$ is the out-amplitude of the purely induced Newtonian tensor perturbation.
\end{proposition}
In Eq.~\eqref{eq:abel-form}, with the lower limit moved to $x_{\rm ini}$, $\mathsf a^{{\rm N},{\rm ini}}_{ij}$ arises from the initial amplitude $\mathsf A[Y^{\rm N}_{ij}](x_{\rm ini})$ and $\mathsf a^{\rm ind}_{ij}$ from the source integral.
For non-Gaussian scalar input, the same equality holds when the four-point function defines a finite quadratic form on the kernels and its connected part assigns no weight to the resonance surfaces.

In Newtonian gauge, the tensor perturbation splits into its out-state and the forced response, both pair by pair and in spectral mean square at fixed scalar cutoff,
\begin{equation}\label{eq:physical-free-forced}
\boxed{\begin{gathered}Y^{\rm N}_{ij}=Y^\out_{ij}+Y^{{\rm N},\refr}_{ij},\\
 Y^\out_{ij}=\mathsf a_{c,ij}u_c+\mathsf a_{s,ij}u_s,\qquad
 Y^{{\rm N},\refr}_{ij}=G_{\rm adv}*s^{\rm N}_{ij}\end{gathered}},
\end{equation}
where the forced response has zero light-cone residue.
In any other intrinsic gauge, both the tensor perturbation and its kernel obey the same split, with the same out-state:
\begin{equation}\label{eq:all-gauge-split}
\begin{gathered}
Y^X_{ij}=Y^\out_{ij}+(Y^{{\rm N},\refr}_{ij}+Q^X_{ij}),\\
 \mathscr K_{0,ij}[Y^{{\rm N},\refr}+Q^X]=0\quad\text{almost everywhere}.
\end{gathered}
\end{equation}
The vanishing kernel residue of the forced part fixes the out-state uniquely.
A radiative observable, such as the energy spectrum $\Omega_{\rm GW}$ of a stochastic background, pertains solely to the free radiation, since the Isaacson energy density used to compute $\Omega_{\rm GW}$ applies only to waves obeying the homogeneous equation.
Such an observable is therefore determined by the out-state $Y^\out_{ij}$ alone.
In contrast, a strain or redshift measurement at finite time can also record the forced response (Sec.~\ref{sec:disc-relation}).

\FloatBarrier

\section{The out-state for a constant equation of state}\label{sec:observables}
By Proposition~\ref{prop:main}, the out-state is common to all intrinsic gauges, so we evaluate it in Newtonian gauge, where the scalar source takes its standard form.

\subsection{Amplitude of a scalar pair}\label{sec:pair-amplitude}
Write
\begin{equation}
 b\equiv\frac{1-3w}{1+3w},\qquad\text{so that}\qquad
 a\propto\eta^{1+b},\qquad\mathcal H=\frac{1+b}\eta,\qquad\nu=b+\tfrac32,\qquad1+b=\gamma
\end{equation}
in Eq.~\eqref{eq:Fw}.
Our $h_{ij}$ is one half of the tensor perturbation used in Refs.~\cite{Kohri:2018awv,Espinosa:2018eve,Domenech:2019quo}.
Using $u=p/k$, $v=q/k$, and $x=k\eta$ as before, we define the kernel $I$ of the purely induced Newtonian-gauge tensor perturbation following these references,
\begin{equation}\label{eq:rd-full-kernel}
\begin{gathered}
 h^{\rm N}_{ij}(x,\mathbf k)=\frac2{k^2}\Lambda_{ij}{}^{lm}\int[\dd q]\,q_lq_m\zeta_{\mathbf q}\zeta_{\mathbf p}I(u,v,x),\\
 \partial_x^2I+\frac{2(1+b)}x\partial_xI+I=f,
\end{gathered}
\end{equation}
with $I=\partial_xI=0$ at $x=0$.
The source function $f$ is given by
\begin{equation}\label{eq:rd-source}
\begin{aligned}
 f(u,v,x)&=\left(\frac{3+3w}{5+3w}\right)^2\biggl[2F_w(c_sux)F_w(c_svx)
 +\frac4{3(1+w)}G_w(c_sux)G_w(c_svx)\biggr],\\
 G_w(y)&\equiv F_w(y)+\frac{y\,\partial_yF_w(y)}{1+b}.
\end{aligned}
\end{equation}
The residue for a scalar pair $(u,v)$ is
\begin{equation}\label{eq:rd-residue-source}
\begin{aligned}
 \mathsf c(u,v)&=\lim_{\epsilon\to0^+}\int_0^\infty\mathrm{e}^{-\epsilon x}
 \varphi(x)\,x^{1+b}f(u,v,x)\,\dd x,\\
 \varphi(x)&=\mathrm{e}^{\mathrm{i}(1+b)\pi/2}\sqrt{\frac{\pi x}2}
 H^{(1)}_{b+1/2}(x),
\end{aligned}
\end{equation}
where $\varphi=u_c+\mathrm{i}u_s$ is the free mode of Sec.~\ref{sec:outstate-def}, and $H^{(1)}$ is the Hankel function of the first kind.
Here $u,v>0$ obey the triangle inequality $|u-v|\le1\le u+v$, and we exclude the resonance $u+v=1/c_s$.

With
\begin{equation}
\begin{gathered}
 y\equiv\frac{u^2+v^2-1/w}{2uv},\qquad Z\equiv2uv\,|1-y^2|^{1/2},\\
 k_b\equiv\frac{2+b}{1+b}=\frac{3(1+w)}2,
\end{gathered}
\end{equation}
the integral~\eqref{eq:rd-residue-source} evaluates to
\begin{equation}\label{eq:rd-residue-result}
\begin{aligned}
 \mathsf c(u,v)&=\mathrm{e}^{\mathrm{i}b\pi/2}\,
 \frac{\mathcal N_b}{(uv)^{1+b}}
 \Bigl[I_Y(u,v)+\frac{\mathrm{i}\pi}2I_J(u,v)\Bigr],\\[2pt]
 \mathcal N_b&=\frac{3\cdot2^{b+1}(1+b)(2+b)\,\Gamma(b+\tfrac32)^2}
 {\pi(1-b)(b+\tfrac32)},\\[2pt]
 I_J&=Z^b\bigl[\mathsf P^{-b}_b(y)+k_b\mathsf P^{-b}_{b+2}(y)\bigr]
 \Theta(u+v-1/c_s),\\
 I_Y&=Z^b\bigl[\mathsf Q^{-b}_b(y)+k_b\mathsf Q^{-b}_{b+2}(y)\bigr]
 \Theta(u+v-1/c_s)\\
 &\;-Z^b\bigl[\mathcal Q^{-b}_b(-y)+2k_b\mathcal Q^{-b}_{b+2}(-y)\bigr]
 \Theta(1/c_s-u-v).
\end{aligned}
\end{equation}
Here $\Theta$ is the Heaviside step function, $\mathsf P^\mu_\nu$ and $\mathsf Q^\mu_\nu$ are Ferrers functions (the Legendre functions on the cut $|y|<1$) \cite[Chap.~14]{Olver:2010ouy}, and $\mathcal Q^\mu_\nu(y)\equiv\mathrm{e}^{-\mathrm{i}\mu\pi}Q^\mu_\nu(y)/\Gamma(\mu+\nu+1)$ is the real Legendre function of the second kind for $y>1$.
Because $\Gamma(3)/\Gamma(1)=2$, rewriting the last line in terms of $\mathrm{e}^{-\mathrm{i}\mu\pi}Q^\mu_\nu$ restores the coefficients $1$ and $k_b$, matching the preceding two lines.
Reference~\cite{Domenech:2019quo} evaluates the late-time Newtonian kernel for constant $w$ and obtains the same functions $I_J$ and $I_Y$, with its index $\beta$ equal to $b+\tfrac12$.
Since the Newtonian forced response decays at late times for $0<w<1$, these asymptotic amplitudes agree with those of the out-state, including the phase $\mathrm{e}^{\mathrm{i}b\pi/2}$.
The out-state kernel is therefore
\begin{equation}\label{eq:free-kernel-reconstruction}
 I_\out(u,v,x)=x^{-1-b}\bigl[\mathsf a_c(u,v)u_c(x)
 +\mathsf a_s(u,v)u_s(x)\bigr].
\end{equation}
For $w\ne1/3$, the term $a''/a=b(1+b)/\eta^2$ in Eq.~\eqref{eq:general-Y} does not vanish, so $u_c$ and $u_s$ differ from $\cos x$ and $\sin x$.
Adopting the latter is a subhorizon approximation with $O(x^{-1})$ corrections.
Near the source resonance surface, the amplitude behaves as
\begin{equation}
\begin{gathered}
 |\mathsf c|=\begin{cases}
 O(1),&w<1/3,\\
 O(1+|\log|\delta||),&w=1/3,\\
 O(|\delta|^b),&1/3<w<1,
 \end{cases}\\
 b=\gamma-1,\qquad\delta=1-c_s(u+v),
\end{gathered}
\end{equation}
since the phase $\mathrm{e}^{\mathrm{i}\delta x}$ cuts off the source integral $\int^xx'^{-\gamma}\dd x'$ at $x\sim1/|\delta|$, in agreement with the kernel singularity of Sec.~\ref{sec:lsz}.
The $I_J$ term turns on at $u+v=1/c_s$.

\subsection{Radiation and matter domination}\label{sec:rd-md}
In radiation domination, $b=0$ and $\mathcal N_0=2$, and the integer-degree Legendre functions reproduce the coefficients of Ref.~\cite{Kohri:2018awv}, including the logarithm and step at $u+v=\sqrt3$.
Here $u_c=\cos x$ and $u_s=\sin x$ exactly.

Matter domination, $w=0$, is reached as the limit $w\to0^+$.
In this limit, $b\to1$, $y\to-\infty$, and $I_Y=O(w)$, which cancels the factor $(1-b)^{-1}$ in $\mathcal N_b$.
The resulting amplitude and exact matter-era kernel are \cite{Kohri:2018awv,Assadullahi:2009nf,Gong:2019mui}
\begin{equation}\label{eq:md-explicit-split}
\begin{gathered}
 \mathsf c_{\rm MD}=-\frac{18}{5}\mathrm{i},\qquad
 I_{\rm MD}(0)=\partial_xI_{\rm MD}(0)=0,\\
 u_c=\cos x-\frac{\sin x}{x},\qquad
 u_s=\sin x+\frac{\cos x}{x},\\[2pt]
 I_{\rm MD}(x)=\underbrace{\frac65}_{I_\refr}
 +\underbrace{\frac{18}{5x^2}\left(\cos x-\frac{\sin x}{x}\right)}_{I_\out}.
\end{gathered}
\end{equation}
Thus $h^\refr_{ij}=S_{ij}/k^2$ is static, since the matter-era source $S_{ij}$ in Eq.~\eqref{eq:Sg} is constant, while $h^\out_{ij}$ is the exact free wave and redshifts as $a^{-1}$ inside the horizon.
With $c_s=0$, the source singularity stays at $z=-1$ and the light-cone pole remains isolated even after the momentum integral.
Every intrinsic generator is then nonoscillating, so its shift has $\delta_\nu=1$ for every pair, and Theorem~\ref{thm:equivalence} gives the same kernel residue in all intrinsic gauges without needing to exclude a resonance surface.

\subsection{Energy spectrum}\label{sec:energy-spectrum}
On subhorizon scales, the fractional energy density of the out-state is
\begin{equation}\label{eq:Omega-out}
 \Omega^\out_{\rm GW}(k,\eta)=\frac1{24}\left(\frac{k}{\mathcal H}\right)^2\overline{\mathcal P^\out},
\end{equation}
where $\mathcal P^\out$ is the power spectrum of a single polarization of $\chi^\out_{ij}\equiv2h^\out_{ij}$ in the convention of Ref.~\cite{Kohri:2018awv}.
The polarization tensors are normalized as $e^\lambda_{ij}e^{\lambda'}_{ij}=\delta^{\lambda\lambda'}$, and the prefactor $1/24$ accounts for the sum over both polarizations.
The bar denotes the average over oscillations.
Equivalently,
\begin{equation}
 \Omega^\out_{\rm GW}=\frac1{12}\left(\frac{k}{\mathcal H}\right)^2\overline{\mathcal P_{h^\out}},\qquad
 \mathcal P_{h^\out}\equiv\frac{k^3}{2\pi^2}P_{h^\out},
\end{equation}
where $P_{h^\out}$ is the spectrum~\eqref{eq:field-power-spectrum}, which sums both tensor indices.
Explicitly,
\begin{equation}\label{eq:rd-omega-out}
 \Omega^\out_{\rm GW}(k,\eta)=\frac{x^{-2b}}{12(1+b)^2}
 \int_0^\infty\dd v\int_{|1-v|}^{1+v}\dd u\,|\mathsf c(u,v)|^2
 \biggl[\frac{4v^2-(1+v^2-u^2)^2}{4uv}\biggr]^2
 \mathcal P_\zeta(uk)\mathcal P_\zeta(vk).
\end{equation}
The factor $x^{-2b}\propto a^{3w-1}$ accounts for the dilution of radiation relative to the background energy density.
For $w=1/3$, this reduces to the standard late-time radiation-era spectrum \cite{Kohri:2018awv,Domenech:2021ztg}.
For a scale-invariant spectrum $\mathcal P_\zeta=\mathcal A_\zeta$ in radiation domination, Eq.~\eqref{eq:rd-omega-out} gives $\Omega^\out_{\rm GW}\approx0.82\mathcal A_\zeta^2$ \cite{Kohri:2018awv}.
Applied to a generic second-order tensor perturbation, the same expression would also include contributions from the forced response and the gauge shift, neither of which represents free radiation.

\section{Discussion}\label{sec:conclusion}
\subsection{Relation to the measured strain and to earlier prescriptions}\label{sec:disc-relation}
The geodesic light-signal measurement of Ref.~\cite{Domenech:2025ccu} yields the Newtonian-gauge tensor perturbation,
\begin{equation}
 h^{\rm N}_{ij}=h^\out_{ij}+h^{{\rm N},\refr}_{ij},\qquad
 h^\out_{ij}=Y^\out_{ij}/a,\qquad h^{{\rm N},\refr}_{ij}=Y^{{\rm N},\refr}_{ij}/a.
\end{equation}
These geodesic observers coincide with synchronous-gauge observers, since their first-order velocity relative to Newtonian observers obeys the synchronous rule $(a\alpha)'=-a\Phi$ in Table~\ref{tab:normal-rules}, with the residual constant fixed by their initial velocity.
The measured strain is nevertheless the Newtonian-gauge tensor perturbation rather than the synchronous-gauge one, since signal propagation depends only on the gauge-invariant combination that reduces to $h_{ij}$ in Newtonian gauge (Eq.~(10) of Ref.~\cite{Domenech:2025ccu}).
The motion of the observers enters through separate boundary terms.
With this residual constant fixed, synchronous gauge is intrinsic, so the two tensor perturbations have the same out-state.

At late times for $0<w<1$, the forced response decays relative to the free wave, and the energy spectrum obtained from $h^{\rm N}_{ij}$ via Eq.~\eqref{eq:Omega-out} approaches the out-state spectrum~\eqref{eq:rd-omega-out}.
In matter domination, $h^\refr_{ij}=S_{ij}/k^2$ is static and dominates the late-time strain.
The redshift and the time delay are linear in $h^{\rm N}_{ij}$, so each splits into contributions from $h^\out_{ij}$ and $h^\refr_{ij}$.
The propagation term in the redshift involves $\partial_\eta h^{\rm N}_{ij}$ \cite{Domenech:2025ccu}, so the static $h^\refr_{ij}$ of matter domination drops out.
In the fractional time delay, it leaves only a constant offset once the static strain has formed, which does not contribute to the time variation.
The time-varying tensor contribution in matter domination therefore comes from $h^\out_{ij}$, while the scalar and observer terms of Ref.~\cite{Domenech:2025ccu} enter separately.

The static contribution also drops out of an energy density constructed from time derivatives ($\rho_{\rm GW}\propto\langle h'_{ij}h'_{ij}\rangle$ \cite{Domenech:2025ccu}), so at short wavelengths this energy density coincides with that of the out-state.
Since the out-state is a free wave, its energy density follows from the conventional Isaacson definition, and its spectrum $P_{\mathsf a}$ is the same in all intrinsic gauges.

Earlier calculations find that the late-time spectra agree across several gauges, while in others the tensor perturbation develops growing gauge shifts \cite{Hwang:2017oxa,Tomikawa:2019tvi,Inomata:2019yww,Yuan:2019fwv,Lu:2020diy,Ali:2020sfw,Ali:2023moi,Kugarajh:2025pjl}.
Examples include the comoving and uniform-density gauges.
The field residue, taken after the momentum integral, fails to yield the out-state in the comoving gauge for $w\ge1/9$ and in the uniform-density gauge.
In these gauges, the near-resonant layer carries nonvanishing power whenever the scalar spectrum has weight on the source resonance surface (Table~\ref{ap-tab:shell-orders} and Appendix~\ref{ap-app:fixed-k-proof}).
Their out-state is nevertheless the same as in Newtonian gauge.
Reference~\cite{Domenech:2020xin} handles such growth by restricting the gauge.
Let $m$ denote the resonance order of a gauge shift, defined as the power of $x$ with which its kernel amplitude grows on a resonance surface (Appendix~\ref{ap-app:fixed-k-proof}).
Under the controlled frequency expansion used here, the reasonable-gauge bounds of Ref.~\cite{Domenech:2020xin} (Eq.~(16) there) give $m\le-2/(1+3w)$ for $Q_{ij}=a\Delta h_{ij}$, and its condition for an invariant spectrum (Eq.~(24) there) corresponds to $m\le0$.
By contrast, the out-state is invariant in every intrinsic gauge, regardless of the resonance order $m$ of its shift.

Gauge-invariant second-order tensor perturbations \cite{Chang:2020iji,Chang:2020mky,Chang:2020tji} add terms bilinear in the first-order perturbations that cancel the gauge shift, and each such variable equals the tensor perturbation in the gauge where the compensating first-order fields vanish.
Distinct constructions therefore differ by gauge shifts of the form $Q^X_{ij}$.
When the compensating fields define an intrinsic gauge, Proposition~\ref{prop:main} implies that all of them share the same out-state.
Gauge invariance does not remove the forced response, however, since the construction that reduces to $h^{\rm N}_{ij}$ in Newtonian gauge equals $h^\out_{ij}+h^{{\rm N},\refr}_{ij}$.
The free radiation must therefore still be extracted from a gauge-invariant tensor perturbation via its out-state, just as from the tensor perturbation in any intrinsic gauge.

Finally, for nonresonant radiation-era pairs, the light-speed-filtered kernel of Ref.~\cite{Yuan:2025seu} (Eq.~(B.2) there) approaches at late times the out-state kernel~\eqref{eq:free-kernel-reconstruction} with the amplitudes~\eqref{eq:rd-residue-result} at $b=0$, so the filter selects the same out-state.
The luminal projection of Ref.~\cite{Ali:2025xvj} performs an analogous selection for isocurvature kernels.
Comparing its coefficients with those of the out-state requires a multi-fluid calculation.
The residue prescription needs no closed-form kernel to prove invariance and reconstructs an exact free solution using $u_c$ and $u_s$.
In addition, at a finite time, a filtered sourced kernel can retain forced contributions, and for $w\ne1/3$ a projection onto $\cos x$ and $\sin x$ agrees with the exact free basis only for $x\gg1$.

\subsection{Order of limits, assumptions and open questions}\label{sec:disc-limits}
The construction in this work acts on the second-order tensor perturbation in the order
\begin{equation}
\begin{aligned}
 \text{perturbative expansion}&\;\longrightarrow\;\text{kernel residue}\\
 &\;\longrightarrow\;\text{momentum integral}\\
 &\;\longrightarrow\;\Lambda_{\rm c}\to\infty.
\end{aligned}
\end{equation}
Since the reduction formula~\eqref{eq:residue-reduction} uses the source at all times after $x_{\rm ini}$, the first arrow means that the residue acts on the second-order tensor perturbation rather than on the full solution.
The second arrow means that the residue is taken for the kernels before the momentum integral (Definition~\ref{def:outstate}).
With this ordering, the out-state is well defined in every intrinsic gauge, including those whose tensor spectra grow at late times.

The matter-era results of Sec.~\ref{sec:observables} arise as the $w\to0^+$ limit of the constant-$w$ expressions, in which $c_s=\sqrt w\to0$.
Pairs on the source resonance surface, $c_s(q+p)=k$, are absent from the cutoff support $q,p\le q_*\Lambda_{\rm c}$ once $2c_sq_*\Lambda_{\rm c}<k$.
The light-cone pole then remains isolated after the momentum integral for all sufficiently small $w$, as it is in matter domination.
For a noncompact spectrum, by contrast, resonant pairs at $q+p=k/c_s$ remain for every $w>0$, and the limit $w\to0^+$ need not commute with the late-time limit or with $\Lambda_{\rm c}\to\infty$.

Physical applicability of the perturbative result requires the perturbations to remain small over the relevant times.
For constant $0<w<1$, the Newtonian potential decays.
The subhorizon density contrast $\delta_{\rm N}\equiv\delta\rho_{\rm N}/\bar\rho$ of the sound waves instead scales as $a^{(3w-1)/2}$, since their energy density $\delta_{\rm N}^2\bar\rho$ redshifts as $a^{-4}$.
It therefore decays for $w<1/3$, stays bounded in radiation domination, and grows for $w>1/3$, while in matter domination $\delta_{\rm N}\propto a$.
For $w>1/3$ and in matter domination, the result applies before the density evolution becomes nonlinear.
In gauges whose tensor perturbation grows at late times, such as the uniform-density gauge, the perturbative expansion can fail while the physical perturbations remain small.
This limits the validity of the perturbative expansion in that gauge, not the out-state itself, which by Proposition~\ref{prop:main} is a property of the physical solution and agrees across all intrinsic gauges.

The energy density of Sec.~\ref{sec:disc-relation} rests on a further assumption.
The Isaacson energy density of the out-state requires subhorizon wavelengths and an averaging scale between the wavelength and the Hubble radius \cite{Isaacson:1968zza,Misner:1973prb,Giovannini:2020qta}.
Comparing it with quasilocal energies \cite{Cai:2021jbi,Cai:2021ndu} also involves the cross terms between the out-state and the forced response in $h^{\rm N}_{ij}$.

Several extensions of the present construction remain.
First, extending the proof to non-Gaussian scalar perturbations requires analyzing the connected four-point contribution, in particular its weight on the resonance surfaces.
Second, the method may extend to quantities beyond the tensor perturbation.
Since the background Weyl tensor vanishes, the second-order Weyl tensor transforms only by terms bilinear in the first-order generator and the first-order Weyl tensor \cite{Bruni:1996im}.
This suggests an analogous residue prescription.
Its electric part also depends on the observer's four-velocity, whose first-order transformation introduces additional terms requiring a separate calculation.
Similarly, non-tensor second-order contributions to the redshift and time delay require their own analysis.
Third, beyond a single fluid, multi-fluid and reheating scenarios require accounting for constrained sources and matching data \cite{Yuan:2024qfz,Ali:2025xvj,Laine:2025gin,Arroja:2009sh,Han:2025wlo}.
Fourth, these late-time criteria may also apply to other sourced tensor perturbations, such as gravitational waves from sound waves after a first-order phase transition \cite{Hindmarsh:2013xza}.

In practice, the source integral in Eq.~\eqref{eq:residue-reduction} is cut off either by a change of cosmological era or by the finite time of observation.
If an early matter era ends at reheating, the tensor perturbation and its derivative at the transition enter Eq.~\eqref{eq:residue-reduction} as initial data for the next stage, whose own scalar source must then be followed \cite{Domenech:2020xin}.
A transition can thus convert the static strain of matter domination into free radiation.
In a sudden transition, the new source enhances the induced waves \cite{Inomata:2019ivs,Inomata:2025wiv}, while in a gradual one the potential decays and the enhancement need not occur \cite{Inomata:2019zqy}.
At a finite observation time $x_{\rm obs}$, the amplitude $\mathsf A[Y_{ij}](x_{\rm obs})$ differs from the out-amplitude by the tail of the source integral beyond $x_{\rm obs}$.
The size of this remainder depends on the scalar spectrum, and obtaining a controlled error bound on it remains an open problem.

\section{Conclusions}\label{sec:conclusions}
In this paper, we have shown that the light-cone residue defines an exact free solution, the out-state, that is common to all intrinsic gauges for a given physical solution.
The proof rests on the frequency structure of the generators connecting intrinsic gauges (Sec.~\ref{sec:generator-frequencies}).
At first order, the tensor perturbation is gauge invariant \cite{Stewart:1974uz} and propagates freely when there is no anisotropic stress.
At second order, gauge shifts oscillate at scalar frequencies, which meet the tensor frequency only on resonance surfaces of zero volume.
Consequently, the out-state remains invariant even when the full tensor perturbation grows at late times.
All gauge dependence then resides in the forced response $Y^{{\rm N},\refr}_{ij}+Q^X_{ij}$ of Eq.~\eqref{eq:all-gauge-split}.
The out-state is determined by the classical LSZ-type reduction formula~\eqref{eq:residue-reduction}, and for a purely induced Newtonian-gauge tensor perturbation it takes the form $(G_{\rm ret}-G_{\rm adv})*s^{\rm N}_{ij}$ in Eq.~\eqref{eq:ret-adv-free}.
For $0<w<1$, the out-amplitudes coincide with those of the late-time Newtonian-gauge kernel, justifying the standard practice of computing the SIGW energy spectrum from that kernel.

\acknowledgments
This work was supported by a startup grant from the Shanghai Institute for Mathematics and Interdisciplinary Sciences (SIMIS).
OpenAI's GPT-6 Astra and Anthropic's Claude Opus 5.5 were used under the author's direction to assist with computational checks, drafting and revising the text, internal review of the manuscript, and checking references.
The author reviewed all AI-assisted material and scientific conclusions and takes full responsibility for the content.

\appendix
\section{Existence of the residue}\label{app:residue-existence}
This appendix shows that the residue~\eqref{eq:residue-definition} exists when the amplitude has a late-time form with power-law envelopes, first for a single scalar pair and then uniformly over the pairs kept outside a detuning gap, as used in Sec.~\ref{sec:kernel-def}.
At finite resolution $\epsilon$, the out-amplitude~\eqref{eq:real-out-amplitudes} is replaced by
\begin{equation}\label{eq:residue-at-resolution}
\begin{gathered}
\Rres_\epsilon[Y_{ij}]\equiv\epsilon\int_0^\infty\mathrm{e}^{-\epsilon x}\mathsf h[Y_{ij}](x)\,\dd x,\\
 \Rres_{\epsilon,s}-\mathrm{i}\Rres_{\epsilon,c}
 =\epsilon\widetilde{\mathsf A}[Y_{ij}](\mathrm{i}\epsilon).
\end{gathered}
\end{equation}

Let $\mathsf A[Y_{ij}]$ be locally integrable on $[x_{\rm ini},\infty)$, and let the integrals below start at $x_{\rm ini}$, with $x_{\rm ini}=0$ for data regular at the origin.
Suppose that, for some constant $\mathsf c_{ij}$, the difference $\mathsf A[Y_{ij}]-\mathsf c_{ij}$ has the late-time form~\eqref{eq:shift-late-form} satisfying (E1) and (E2).
Choose $x_0\ge\max(1,x_{\rm ini})$ beyond which these bounds hold.
Then
\begin{equation}\label{eq:transform-split}
\begin{aligned}
 \epsilon\widetilde{\mathsf A}[Y_{ij}](\mathrm{i}\epsilon)
 ={}&\epsilon\int_{x_{\rm ini}}^{x_0}\mathrm{e}^{-\epsilon x}\mathsf A[Y_{ij}]\,\dd x
 +\mathsf c_{ij}\,\mathrm{e}^{-\epsilon x_0}\\
 &+\epsilon\sum_{\nu=1}^J\int_{x_0}^\infty
 \mathrm{e}^{(-\epsilon+\mathrm{i}\delta_\nu)x}\mathsf b_{\nu,ij}\,\dd x
 +\epsilon\int_{x_0}^\infty\mathrm{e}^{-\epsilon x}\mathfrak e_{ij}\,\dd x.
\end{aligned}
\end{equation}
The initial interval contributes $O(\epsilon)$, and the constant term tends to $\mathsf c_{ij}$.
The oscillating and decaying terms vanish as $\epsilon\to0^+$ by the estimates in the proof of Theorem~\ref{thm:equivalence}.
Hence
\begin{equation}
 \epsilon\widetilde{\mathsf A}[Y_{ij}](\mathrm{i}\epsilon)\longrightarrow\mathsf c_{ij}\qquad(\epsilon\to0^+).
\end{equation}

We now construct the gapped tensor perturbations of Sec.~\ref{sec:kernel-def}, whose residues converge to the amplitude~\eqref{eq:kernel-outstate-assembly}.
We fix the scalar cutoff $\Lambda_{\rm c}$ for the spectrum (Sec.~\ref{sec:averaging}), so that the bounds above can hold uniformly over pairs, and discard the pairs within a detuning gap $\delta_0$ of the pole:
\begin{equation}\label{eq:gapped-history}
\begin{gathered}
Y_{\delta_0,ij}\equiv\int[\dd q]\,\chi_{\delta_0}\,\mathscr K_{ij}[Y]\,\zeta_{\mathbf q}\zeta_{\mathbf p},\\
 \chi_{\delta_0}\equiv\begin{cases}1, & \min_\nu|\delta_\nu|\ge\delta_0,\\ 0, & \text{otherwise}.\end{cases}
\end{gathered}
\end{equation}
The singularities of the kept pairs then lie at distance at least $\delta_0$ from $z=0$.
On these pairs, $|\mathrm{i}\epsilon+\delta_\nu|\ge\delta_0$, so the integrated-by-parts terms in the proof of Theorem~\ref{thm:equivalence} are bounded by $\epsilon$ times a constant uniform over all kept pairs.
This condition holds for the explicit kernels treated here, whose late-time Hankel expansions are uniform on the compact kept domain, and for synchronous gauge, whose residual constants are fixed and locally bounded by (P4).
The residue of the gapped tensor perturbation therefore passes through the momentum integral in spectral mean square:
\begin{equation}\label{eq:gapped-residue}
\begin{gathered}
-\mathrm{i}\operatorname*{Res}_{z=0}\widetilde{\mathsf A}[Y_{\delta_0,ij}]=\mathsf a_{\delta_0,s,ij}-\mathrm{i}\,\mathsf a_{\delta_0,c,ij},\\
 \mathsf a_{\delta_0,ij}\equiv\int[\dd q]\,\chi_{\delta_0}\,\mathscr K_{0,ij}\,\zeta_{\mathbf q}\zeta_{\mathbf p}.
\end{gathered}
\end{equation}
The discarded pairs drop out as the gap closes.
For Gaussian scalar perturbations, their contribution to the power spectrum satisfies
\begin{equation}\label{eq:gapped-error}
 P_{\mathsf a-\mathsf a_{\delta_0}}(k)=2\int[\dd q]\,\bigl(1-\chi_{\delta_0}\bigr)
 \|\mathscr K_0(\mathbf k,\mathbf q)\|^2P_\zeta(q)P_\zeta(p)
 \longrightarrow 0\qquad(\delta_0\to0).
\end{equation}
As $\delta_0\to0$, the integrand tends to zero at every pair off the resonance surfaces $\min_\nu|\delta_\nu|=0$, which carry zero spectral weight, and it is bounded by the integrand of $P_{\mathsf a}$.
Dominated convergence then yields the limit~\eqref{eq:gapped-error} and implies $\mathsf a_{\delta_0,ij}\to\mathsf a_{ij}$ in spectral mean square.
The scalar cutoff is removed in Appendix~\ref{sec:rd-cutoff-removal}.

\section{Early-time initial data and synchronous residual modes}\label{app:histories}
Section~\ref{sec:history-labels} showed that resetting zero tensor data in gauge $X$, instead of transporting them, adds the free amplitude $-\mathsf h[Q^X_{ij}](x_{\rm ini})$.
Here we show that this amplitude vanishes as $x_{\rm ini}\to0$ for the standard gauges with regular generators, and treat the synchronous residual modes, for which it does not.
In radiation domination,
\begin{equation}
\begin{gathered}
 \mathsf h[Q_{ij}]=\mathcal O(-x)\bigl(Q_{ij},\partial_xQ_{ij}\bigr)^{\mathsf T},\\
 \mathcal O(-x)=\begin{pmatrix}
 \cos x&-\sin x\\\sin x&\cos x
 \end{pmatrix}.
\end{gathered}
\end{equation}
Imposing zero initial data in both gauges selects the same solution precisely when
\begin{equation}\label{ap-eq:reset-condition}
 Q^X_{ij}(0^+)=\partial_xQ^X_{ij}(0^+)=0 .
\end{equation}
Otherwise, resetting the initial data to zero adds the free amplitude
\begin{equation}\label{ap-eq:reset-amplitude}
 \mathsf a^{X,{\rm reset}}_{ij}=-\mathsf h[Q^X_{ij}](x_{\rm ini}) .
\end{equation}
As $\eta\to0$, the Newtonian perturbations and the background obey
\begin{equation}
\begin{gathered}
 \Phi\to\Phi_*,\qquad U\to\frac{\Phi_*}{\eta},\qquad
 \frac{\delta\rho_{\rm N}}{\bar\rho}\to-2\Phi_*,\\
 \bar\rho'=-4\mathcal H\bar\rho,\qquad\mathcal H=\frac1\eta,\qquad
 \mathcal H'=-\frac1{\eta^2}.
\end{gathered}
\end{equation}
Substituting these relations into the rules of Table~\ref{tab:normal-rules} gives \cite{Malik:2008im,Lu:2020diy}
\begin{equation}\label{ap-eq:superhorizon-generators}
\begin{aligned}
 \alpha_{\rm flat}&\to\eta\Phi_*,
 &\alpha_{\rm com}&\to-\tfrac12\eta\Phi_*,\\
 \alpha_{\rm ud}&\to-\tfrac12\eta\Phi_*,
 &\alpha_{\rm ue}&\to-\tfrac12\eta\Phi_*.
\end{aligned}
\end{equation}
The regular synchronous generator, which has no residual modes, solves $\alpha'+\mathcal H\alpha=-\Phi$ with $\alpha$ finite at $\eta=0$ and takes the form
\begin{equation}\label{eq:rd-sync-generator}
\begin{gathered}
 \alpha(\eta,\mathbf q)=-\eta\Phi_*(\mathbf q)g(y),\qquad y=c_sq\eta,\\
 g(y)=\frac3{y^2}\left(1-\frac{\sin y}{y}\right).
\end{gathered}
\end{equation}
Since $g(0)=1/2$, its leading terms are
\begin{equation}
 \alpha\to-\tfrac12\eta\Phi_*,\qquad
 \beta=\int_0^\eta\alpha\,\dd\eta'=O(\eta^2),
\end{equation}
so $\alpha$ has the same leading behavior as $\alpha_{\rm com}$.
We write the transfer functions of Eq.~\eqref{eq:generator-profiles} as
\begin{equation}
 \alpha_q\equiv T_\alpha(\eta,q),\qquad
 \beta_q\equiv T_\beta(\eta,q),\qquad
 \Phi_q\equiv\frac{\Phi(\eta,\mathbf q)}{\zeta_{\mathbf q}} .
\end{equation}
Substituting them into Eq.~\eqref{eq:gauge-shift-ordinary} and symmetrizing over the pair gives the kernel of $Q^X_{ij}$,
\begin{equation}\label{ap-eq:Delta-C}
\begin{gathered}
 \mathscr K_{ij}[Q^X]=\tfrac12a\,\Delta_{qp}\,\Lambda_{ij}{}^{lm}q_lq_m,\\
 \Delta_{qp}=-2\alpha_q\alpha_p-2(\mathbf q\cdot\mathbf p)\beta_q\beta_p
 +4(\Phi_q\beta_p+\Phi_p\beta_q)
 -4\mathcal H(\alpha_q\beta_p+\alpha_p\beta_q).
\end{gathered}
\end{equation}
Every term in Eq.~\eqref{ap-eq:Delta-C} is then $O(\eta^2)$.
With $a\propto\eta$, the flat, comoving, uniform-density, uniform-expansion, and regular synchronous gauges therefore satisfy
\begin{equation}
 Q^X_{ij}=O(x^3),\qquad\partial_xQ^X_{ij}=O(x^2),
\end{equation}
and hence, for these regular generators,
\begin{equation}
 \mathsf h[Q^X_{ij}](0^+)=0,\qquad
 \mathsf a^{X,{\rm reset}}_{ij}=O(x_{\rm ini}^2).
\end{equation}
For general synchronous residual data specified at $\eta_0>0$,
\begin{equation}\label{eq:sync-general-generators}
\begin{aligned}
 \alpha&=\frac1a\left[C-\int_{\eta_0}^{\eta}a\Phi\,\dd\eta'\right],\\
 \beta&=d+\int_{\eta_0}^{\eta}\alpha\,\dd\eta'.
\end{aligned}
\end{equation}
Here $C=C_q\zeta_{\mathbf q}$ and $d=d_q\zeta_{\mathbf q}$, and we write $\Phi_{*q}\equiv\Phi_*(\mathbf q)/\zeta_{\mathbf q}$.
The time-shift residual changes both generators,
\begin{equation}
 \alpha\to\alpha+\frac Ca,\qquad
 \beta\to\beta+\int_{\eta_0}^{\eta}\frac Ca\,\dd\eta',
\end{equation}
and together these shifts contribute to the kernel~\eqref{ap-eq:Delta-C}
\begin{equation}\label{ap-eq:sync-residual-C}
 \mathscr K_{ij}[Q^X]\supset-\frac{C_qC_p}{a}\Bigl[1+4\log\frac{\eta}{\eta_0}\Bigr]\Lambda_{ij}{}^{lm}q_lq_m,
\end{equation}
which diverges as $x^{-1}\log x$ as $x\to0^+$.
The reset amplitude $-\mathsf h[Q^X_{ij}](x_{\rm ini})$ then diverges as $x_{\rm ini}\to0^+$, so imposing zero synchronous data at ever earlier times does not converge to any physical solution.
For such a tensor perturbation, initial data must be imposed at a finite $x_{\rm ini}>0$, where Eqs.~\eqref{eq:initial-amplitude-transport}--\eqref{eq:hX-additive} still apply despite the singularity of the transformation at the origin.
As in Sec.~\ref{sec:history-labels}, starting the transform at $x_{\rm ini}$ leaves the residue unchanged.
A time-independent spatial residual $\beta\to\beta+d$ adds, in the sign convention of Eq.~\eqref{ap-eq:Delta-C},
\begin{equation}\label{ap-eq:sync-residual-d}
\begin{gathered}
 \mathscr K_{ij}[Q^X]\supset c_d\,\eta\,\Lambda_{ij}{}^{lm}q_lq_m+O(\eta^2),\\
 c_d\equiv A\Bigl[3\bigl(\Phi_{*q}d_p+\Phi_{*p}d_q\bigr)-(\mathbf q\cdot\mathbf p)\,d_qd_p\Bigr],\qquad a=A\eta .
\end{gathered}
\end{equation}
The linear term combines $4\Phi\beta$ and $-4\mathcal H\alpha\beta$ with $\mathcal H\alpha\to-\Phi_*/2$.
For generic $d$, this gives, for each pair and up to the factor $\Lambda_{ij}{}^{lm}q_lq_m$,
\begin{equation}
 Q^X_{ij}(0^+)=0,\qquad
 \partial_xQ^X_{ij}(0^+)=\frac{c_d}{k}\neq0,\qquad
 \mathsf a^{X,{\rm reset}}_{ij}\to\Bigl(0,-\frac{c_d}{k}\Bigr),
\end{equation}
so the reset amplitude is finite and nonzero.
Changing $d$ relates two synchronous gauges, both intrinsic, so by Theorem~\ref{thm:equivalence} the $d$-dependent part of $Q^X_{ij}$ has zero kernel residue off the resonance surfaces.
Transporting the initial data therefore preserves the physical solution and the out-state, while resetting them adds the free amplitude $(0,-c_d/k)$ to the Newtonian solution.

\section{Intrinsic gauges and the frequency content of their generators}\label{ap-sec:generator-estimates}\label{ap-sm:intrinsic-full}
We specify the gauge rules underlying Definition~\ref{def:intrinsic} and prove Eq.~\eqref{eq:late-time-form} for $0<w<1$.
The envelope constants and late-time starting point may depend on $q$.

Let $\Pi=(\phi,\psi,B,E,\delta\rho,v)$ collect the first-order scalar perturbations.
Here $v$ is defined by
\begin{equation}
 \delta u^i=a^{-1}\partial^iv,\qquad\text{so that}\qquad\delta u_i=a\,\partial_i(v+B).
\end{equation}
A generator $(\alpha,\partial^i\beta)$ acts via $\widetilde\Pi=\Pi+\mathcal T[\alpha,\beta]$, with
\begin{equation}\label{eq:firstorder-convention-compact}
\begin{aligned}
 \phi&\to\phi+\mathcal H\alpha+\alpha',
 &\psi&\to\psi-\mathcal H\alpha,\\
 \delta\rho&\to\delta\rho+\bar\rho'\alpha,
 &B&\to B-\alpha+\beta',\\
 E&\to E+\beta,
 &v&\to v-\beta'.
\end{aligned}
\end{equation}
In general, two scalar gauge conditions determine the time and spatial shifts,
\begin{equation}\label{eq:intrinsic-conditions}
 \sum_A\sum_{n=0,1}c_{\ell A,n}(\eta,\nabla^2)\,\partial_\eta^n\tilde\Pi_A=0,\qquad \ell=1,2,
\end{equation}
where $\ell$ labels the two gauge conditions and $A$ runs over the perturbation variables.
Write the two conditions of a target gauge $X$ as $F_X[\widetilde\Pi]=0$, and let $\Pi_X$ denote the perturbations in that gauge.
For a reference scalar gauge $R$ on the same solution, linearity gives
\begin{equation}
\begin{gathered}
 \Pi_X=\Pi_R+\mathcal T[\alpha_{R\to X},\beta_{R\to X}],\\
 F_X\!\left[\mathcal T[\alpha_{R\to X},\beta_{R\to X}]\right]=-F_X[\Pi_R].
\end{gathered}
\end{equation}
The determining operator depends only on the target coordinate rule and the FLRW background, while changing the reference gauge alters its drive.

\Needspace{10\baselineskip}
\begin{definition}[Precise form of Definition~\ref{def:intrinsic}]\label{def:intrinsic-precise}
A scalar gauge~\eqref{eq:intrinsic-conditions} is intrinsic if:
\begin{itemize}
\item[(P1)] \emph{Locality}.
The conditions are local in $\tilde\Pi$ and $\partial_\eta\tilde\Pi$.
Each $c_{\ell A,n}$ is polynomial in $\nabla^2$, with coefficients that are finite sums of products of $a,\mathcal H,\mathcal H',\bar\rho,\bar\rho'$ and their powers with real numerical prefactors.
At fixed $q$, they therefore have the form $\sum_m c_m(q)\eta^{\tau_m}$ with real $\tau_m$ and real $c_m(q)$ polynomial in $q^2$.
\item[(P2)] \emph{No modes of their own}.
The conditions admit a finite ordered solution for the generators and intermediate quantities $\mathfrak g_m$, valid for arbitrary reference drives, using steps
\begin{equation}
\begin{gathered}
 A_{\mathrm{alg}}\mathfrak g_m=\mathfrak F_m\quad\text{or}\\
 A_1\partial_\eta\mathfrak g_m+A_0\mathfrak g_m=\mathfrak F_m.
\end{gathered}
\end{equation}
Here $A_{\mathrm{alg}},A_0,A_1$ have the form in (P1), with $A_{\mathrm{alg}},A_1\ne0$ for every $\eta>0$.
The drive is a finite linear combination of reference perturbations, previously determined quantities, and their derivatives, with coefficients formed by sums, products, and quotients of the allowed coefficients and nonvanishing denominators.
No step depends on a later unknown.
\item[(P3)] \emph{Temperedness}.
Every first-order rate $r=A_0/A_1$ satisfies
\begin{equation}\label{eq:intrinsic-fuchs}
 \eta r(\eta,q)\ge-c_0(q)\qquad(\eta\ge\eta_1(q)).
\end{equation}
This allows Hubble-scale negative damping, zero damping, and stronger positive damping.
\item[(P4)] \emph{Complete fixing}.
Residual generator data at the initial time $\eta_0(q)$ and the spatial labels are fixed, measurable, and locally bounded in $q>0$.
They are transported when changing the reference gauge to preserve the same coordinate choice.
The late time $\eta_1(q)$ in (P3) is independent of this initial-data prescription.
\end{itemize}
\end{definition}

Harmonic (Lorenz--de Donder) gauge provides an example outside (P2): on FLRW backgrounds, its coordinate generators satisfy a coupled second-order hyperbolic system \cite{Bicak:2007zza}.
The wave principal part gives homogeneous subhorizon phases $\mathrm{e}^{\pm\mathrm{i}q\eta}$ independent of the matter drive, so a real first-order formulation contains feedback between a coordinate displacement and its velocity rather than the ordered steps of (P2).
Fixing the coordinate Cauchy data removes the residual ambiguity but does not change this determining operator.
Harmonic gauge therefore lies outside the class of Definition~\ref{def:intrinsic}.
Its out-state is nevertheless the same, since Theorem~\ref{thm:equivalence} applies to it directly.
For the condition $\bar\nabla^\mu(\delta g_{\mu\nu}-\tfrac12\bar g_{\mu\nu}\delta g)=0$, with $\bar\nabla$ the background covariant derivative, the generators from Newtonian gauge obey
\begin{equation}\label{eq:harmonic-generators}
\begin{aligned}
 \alpha''+4\mathcal H\alpha'+(q^2-2\mathcal H')\alpha+2q^2\mathcal H\beta&=-4\Phi'-12\mathcal H\Phi,\\
 \beta''+4\mathcal H\beta'+q^2\beta-2\mathcal H\alpha&=0 .
\end{aligned}
\end{equation}
The homogeneous solutions oscillate as $\mathrm{e}^{\pm\mathrm{i}q\eta}$ with envelopes $a^{-1}$ and $a^{-3}$, and the driven solution oscillates at $\pm c_sq$ with a power-law envelope, since $c_s<1$.
With the Cauchy data fixed, the generators have the form~\eqref{eq:late-time-form} with the additional frequencies $\pm q$, and the detunings of the shift are $\delta_\nu=1+(\sigma_1c_1q+\sigma_2c_2p)/k$ with $c_{1,2}\in\{c_s,1\}$ and $\sigma_{1,2}\in\{0,\pm1\}$.
They vanish only on sets of zero volume in the scalar-pair domain, so (E1) and (E2) hold for almost every pair.
For this gauge condition, the rescaled shift $Q_{ij}$ decays at least as fast as $a^{-1}$ for every pair.

Table~\ref{tab:normal-rules} gives the standard gauge conditions and their generators from Newtonian gauge.
For the first five gauges $\beta=0$, while for synchronous gauge $\beta'=\alpha$ with fixed residual data.
The uniform-expansion row follows from the expression \cite{Malik:2008im}
\begin{equation}
 a\,\delta\theta=-3(\psi'+\mathcal H\phi)+\nabla^2(E'-B),
\end{equation}
in which the $\alpha'$ and $\beta'$ terms cancel upon transformation.
For a general reference gauge $R$, the right-hand sides in the six rows become, respectively,
\begin{equation}
 B_R-E_R',\qquad\psi_R,\qquad v_R+B_R,\qquad-\delta\rho_R,\qquad a\,\delta\theta_R,\qquad-\phi_R,
\end{equation}
while the left-hand coefficients remain unchanged.
The spatial condition gives $\beta=-E_R$ in the first five rows and $\beta'=\alpha-B_R$ in the synchronous row.
All these coordinate conditions have the background coefficients and derivative order required by (P1).
Their algebraic coefficients are nonzero:
\begin{equation}
\begin{gathered}
 \mathcal H=\frac{\gamma}{\eta}>0,\qquad\gamma=\frac{2}{1+3w},\\
 \bar\rho'=-3(1+w)\mathcal H\bar\rho\ne0,\\
 q^2+3(\mathcal H^2-\mathcal H')
 =q^2+\frac{3\gamma(\gamma+1)}{\eta^2}>0.
\end{gathered}
\end{equation}
Thus Newtonian, flat, comoving, uniform-density, and uniform-expansion gauges satisfy (P2) algebraically.
Synchronous gauge requires two successive first-order steps, with $r=\mathcal H$ and $r=0$, respectively.
Both rates satisfy (P3).
Its two residual constants are fixed at $\eta_0(q)$ as measurable, locally bounded functions of $q>0$, as required by (P4).
The remaining five gauges, including Newtonian gauge, have no integration constants.

The differentiable Hankel expansion of $F_w$ gives the Newtonian scalar phases $\mathrm{e}^{\pm\mathrm{i}c_sq\eta}$, which are preserved by the density and velocity constraints.
We call $f$ an envelope of order $(M,L)$ if, at late times,
\begin{equation}
 |\partial_\eta^lf|\le C_l\,\eta^{M-l}(1+\log\eta)^L\qquad\text{for every }l\ge0 .
\end{equation}

Let the drive $F$ be a finite sum of terms $\mathrm{e}^{\mathrm{i}\omega k\eta}f(\eta)$ with $\omega k\in\{0,\pm c_sq\}$ and envelopes of order $(M,L)$.
Each step in (P2) maps such a drive to a sum over the same frequencies.
An algebraic step divides by a real coefficient $A_{\mathrm{alg}}$ of the form allowed by (P1), which (P2) keeps nonzero for $\eta>0$.
At late times, it and its inverse behave as
\begin{equation}
 A_{\mathrm{alg}}=c_*\eta^{\tau_*}\bigl[1+O(\eta^{-\varsigma})\bigr],\qquad
 \frac1{A_{\mathrm{alg}}}=c_*^{-1}\eta^{-\tau_*}\bigl[1+O(\eta^{-\varsigma})\bigr],
\end{equation}
with $c_*\neq0$, $\varsigma>0$, and the corresponding derivative bounds.
The quotient $g=F/A_{\mathrm{alg}}$ therefore oscillates at the same frequencies, with envelopes of order $(M-\tau_*,L)$.
A first-order step solves $\partial_\eta g+rg=F$ with $r=A_0/A_1$, where $A_0$ and $A_1$ have the form allowed by (P1) and $A_1\neq0$ for $\eta>0$.
Its homogeneous solution $\exp(-\int r\,\dd\eta)$ does not oscillate, and by Eq.~\eqref{eq:intrinsic-fuchs} it grows at most as a power of $\eta$.
For damping of Hubble order or weaker, the rate, integrating factor $\mu$, and solution are
\begin{equation}
 r=\frac c\eta+O(\eta^{-1-\varsigma}),\qquad\mu\simeq\eta^c,\qquad
 g=\frac1\mu\Bigl[C+\int^\eta\mu F\,\dd\eta'\Bigr].
\end{equation}
Repeated integration by parts turns each nonzero-frequency drive $\mu F\supset\mathrm{e}^{\mathrm{i}\omega k\eta}f_\omega$ into
\begin{equation}
 \mathrm{e}^{\mathrm{i}\omega k\eta}\sum_l\frac{(-1)^lf_\omega^{(l)}}{(\mathrm{i}\omega k)^{l+1}}
\end{equation}
plus a constant that joins the homogeneous term, while a zero-frequency drive integrates to a nonoscillating envelope with at most an additional power of $\eta$ or a logarithm.
For stronger positive damping, $r\propto\eta^\tau$ with $\tau>-1$, the homogeneous mode decays faster than any power of $\eta$ and the response follows the drive adiabatically,
\begin{equation}
 g\simeq\frac{\mathrm{e}^{\mathrm{i}\omega k\eta}f}{r+\mathrm{i}\omega k},
\end{equation}
at the driving frequency.
Applying these steps successively in the order prescribed by (P2), starting from the Newtonian scalar phases $\pm c_sq$, gives Eq.~\eqref{eq:late-time-form}.
Algebraic steps retain the frequencies, while first-order steps add only frequency zero.
The zero-frequency contribution is absent for the five algebraic standard gauges and contains the fixed homogeneous terms in synchronous gauge.

This argument fixes the late-time frequency content for each scalar momentum, with constants that may depend on $q$ and may be singular as $q\to0$.
Their uniformity over pairs is discussed in Appendix~\ref{app:residue-existence}, initial-data matching in Appendix~\ref{app:histories}, and noncompact spectra in Appendix~\ref{sec:rd-cutoff-removal}.

\section{Source integrals and mean-square convergence}\label{ap-app:gaussian-reference}
We use the Gaussian quadratic fields of Sec.~\ref{sec:averaging} and a common scalar realization for all cutoffs.
For two such fields ${\mathscr U}$ and ${\mathscr U}'$ with symmetric kernels $\mathscr K$ and $\mathscr K'$, Wick contraction gives
\begin{equation}\label{ap-eq:fixed-k-isometry}
 P_{{\mathscr U}-{\mathscr U}'}(k)
 =2\int[\dd q]\,P_\zeta(q)P_\zeta(p)\,
 \bigl\|\mathscr K(\mathbf k,\mathbf q)
 -\mathscr K'(\mathbf k,\mathbf q)\bigr\|^2.
\end{equation}
Here $\mathbf p=\mathbf k-\mathbf q$, and the norm sums tensor and amplitude components.
A quadratic field therefore converges in spectral mean square if and only if its kernel converges with the weight $P_\zeta(q)P_\zeta(p)$.

\subsection{Near-resonant source pairs}\label{ap-app:general-w-reference}\label{ap-app:gaussian-resonance}
For constant $0<w<1$, the velocity term dominates the late-time source,
\begin{equation}
 a\Bigl(\frac{U}{\mathcal H}\Bigr)^2=O(x^{-\gamma}),\qquad\gamma=\frac2{1+3w},
\end{equation}
while the potential term is $O(x^{-\gamma-2})$.
Its projection on $\varphi$ is a sum of terms
\begin{equation}
 x^{-\gamma}\mathrm{e}^{\mathrm{i}\delta_{\sigma_1\sigma_2}x},\qquad
 \delta_{\sigma_1\sigma_2}=1+\frac{c_s(\sigma_1q+\sigma_2p)}k,\qquad\sigma_{1,2}=\pm1,
\end{equation}
up to corrections of order $x^{-\gamma-1}$.
Only
\begin{equation}
 \delta\equiv\delta_{--}=1-\frac{c_s(q+p)}k
\end{equation}
can vanish, since the triangle inequality gives $|\delta_{\sigma_1\sigma_2}|\ge1-c_s$ for the other branches.
The phase $\mathrm{e}^{\mathrm{i}\delta x}$ cuts off the tail of the source integral at $x'\sim1/|\delta|$, so for $x\ge1$
\begin{equation}
 \Bigl|\int_x^\infty x'^{-\gamma}\mathrm{e}^{\mathrm{i}\delta x'}\,\dd x'\Bigr|\lesssim
 \begin{cases}
  |\delta|^{\gamma-1}, & w>1/3,\\
  1+\bigl|\log|\delta|\bigr|, & w=1/3,\\
  1, & w<1/3.
 \end{cases}
\end{equation}
Its square is integrable across the source resonance surface for $0<w<1$, since the strongest singularity is $|\delta|^{2\gamma-2}$ with $\gamma>1/2$, so pairs in a shrinking layer around the surface contribute vanishing power.
For a noncompact spectrum in radiation domination, the resonance surface $q+p=\sqrt3k$ stays away from $q=0$ and $p=0$.
At soft momenta, the TT factor $\Lambda_{ij}{}^{lm}q_lq_m$ suppresses the kernel as $q^2$, since the exact transfer function of the soft mode remains bounded.
We take a spectrum $P_\zeta(q)$ that decays in the ultraviolet at least as $q^{-3}$, such as a scale-invariant spectrum or a lognormal peak.
The source is finite as $\eta\to0$, so the tensor initial data can be imposed at $x_{\rm ini}=0$ in radiation domination.
Hence the tail of the source integral, together with its supremum over $x\ge x_{\rm ini}$, is square integrable against $P_\zeta(q)P_\zeta(p)[\dd q]$.

\subsection{The forced response and its residue}\label{ap-sec:forced-field-limit}
This subsection shows that the forced response of the Newtonian tensor perturbation, Eq.~\eqref{eq:forced-remainder}, decays at late times and has zero residue in spectral mean square, so that the field residue of $Y^{\rm N}_{ij}$ exists and equals the out-state.
Initial data are imposed at $x_{\rm ini}>0$ for general $0<w<1$, and at $x_{\rm ini}=0$ in radiation domination by Appendix~\ref{ap-app:gaussian-resonance}.
For the Newtonian source $s_{ij}=s^{\rm N}_{ij}$, define
\begin{equation}\label{eq:common-wronskian-tail}
 \mathcal J_{ij}(x)=\frac1{k^2}\int_x^\infty(-u_s(x'),u_c(x'))^{\mathsf T}s_{ij}(x')\,\dd x'.
\end{equation}
Since
\begin{equation}
 \partial_x\mathsf h[Y_{ij}]=\frac{s_{ij}}{k^2}\,(-u_s,u_c)^{\mathsf T},
\end{equation}
the sum $\mathsf a_{ij}=\mathsf h[Y_{ij}]+\mathcal J_{ij}$ is constant.
The forced response has zero out-amplitude, so
\begin{equation}
 \mathsf h[Y^\refr_{ij}]=-\mathcal J_{ij}.
\end{equation}
Write $\mathcal J_{\rm pair}$ for the kernel of $\mathcal J_{ij}$, including the TT factor.
For each nonresonant pair, $\mathcal J_{\rm pair}(x)\to0$ as $x\to\infty$.
By Appendix~\ref{ap-app:general-w-reference}, its squared supremum over $x\ge x_{\rm ini}$ is bounded by a function integrable against $P_\zeta(q)P_\zeta(p)[\dd q]$.
Wick contraction and dominated convergence therefore give $\mathcal J_{ij}(x)\to0$ in spectral mean square as $x\to\infty$, showing that the forced response decays at late times.

The residue at resolution $\epsilon$ uses integrals from $x_{\rm ini}$, as in Sec.~\ref{sec:outstate-def}, so the out-state contributes $\mathsf a_{ij}\,\mathrm{e}^{-\epsilon x_{\rm ini}}\to\mathsf a_{ij}$.
Because the weight $\epsilon\mathrm{e}^{-\epsilon x}$ on $[x_{\rm ini},\infty)$ is positive and has total mass at most one, we obtain
\begin{equation}\label{ap-eq:reference-projection-bound}
\begin{aligned}
 \|\Rres_\epsilon[Y_{\refr,\rm pair}]\|^2
 &\le\epsilon\int_{x_{\rm ini}}^\infty\mathrm{e}^{-\epsilon x}
 \|\mathcal J_{\rm pair}(x)\|^2\,\dd x\\
 &\le\sup_{x\ge x_{\rm ini}}\|\mathcal J_{\rm pair}(x)\|^2.
\end{aligned}
\end{equation}
By Appendix~\ref{ap-app:general-w-reference}, this supremum is integrable with the weight $P_\zeta(q)P_\zeta(p)[\dd q]$ for compact spectra and general $w$ with $x_{\rm ini}>0$, as well as for noncompact spectra in radiation domination with $x_{\rm ini}\ge0$.
For any nonresonant pair, the middle expression tends to zero as $\epsilon\to0^+$, since $\mathcal J_{\rm pair}(x)\to0$ as $x\to\infty$.
Wick contraction and dominated convergence give
\begin{equation}\label{eq:rd-forced-field-residue}
 P_{\Rres_\epsilon[Y^\refr]}(k)
 \le 2\int[\dd q]\,P_\zeta(q)P_\zeta(p)\,
 \epsilon\int_{x_{\rm ini}}^\infty\mathrm{e}^{-\epsilon x}
 \|\mathcal J_{\rm pair}(x)\|^2\,\dd x\longrightarrow0.
\end{equation}
Since $Y^{\rm N}_{ij}=Y^\out_{ij}+Y^\refr_{ij}$ and $\Rres_\epsilon[Y^\out_{ij}]=\mathsf a_{ij}\,\mathrm{e}^{-\epsilon x_{\rm ini}}$, it follows that
\begin{equation}\label{eq:general-w-field-residue}
 \Rres_\epsilon[Y^{\rm N}_{ij}]\longrightarrow\mathsf a_{ij}
 \quad\text{in spectral mean square},
\end{equation}
so the field residue of $Y^{\rm N}_{ij}$ exists and equals the out-state.

\subsection{Removing scalar cutoffs}\label{sec:rd-cutoff-removal}
The scalar cutoff $\Lambda_{\rm c}$ of Sec.~\ref{sec:averaging} is removed last, on the same realization of the scalar field.
Inside the cutoff every kernel residue is unchanged, so the cutoff out-amplitudes converge in spectral mean square to the amplitude~\eqref{eq:kernel-outstate-assembly} as $\Lambda_{\rm c}\to\infty$.
Since the kernel residues of all intrinsic gauges agree for almost every scalar pair, so do these limits.
In radiation domination, the bounds of Appendix~\ref{ap-app:gaussian-resonance} hold uniformly in time, so the decay of the Newtonian forced response and its vanishing residue also survive the removal of the cutoff.
Because the tensor perturbation depends only on gradients of the potential and the TT factor suppresses soft scalar modes as $q^2$, the full out-state is the $\Lambda_{\rm c}\to\infty$ limit of the cutoff out-states.

\section{Field residues of the tensor perturbation in different gauges}\label{ap-app:fixed-k-proof}
\begin{figure}[t]
\centering
\resizebox{\textwidth}{!}{%
\begin{tikzpicture}[>=latex,font=\small,text=cInk,
 box/.style={draw=cSlate,text=cInk,rounded corners=2pt,align=center,inner sep=4pt,minimum height=1.05cm},
 lab/.style={text=cInk,font=\footnotesize,align=center}]
\node[box] (K) at (0,0) {pair transform\\$\widetilde{\mathsf A}[\mathscr K_{ij}[Y^X]](z;\mathbf k,\mathbf q)$};
\node[box] (Kinf) at (5.0,1.35) {limiting kernel $\mathscr K_{0,ij}$\\[1pt]{\footnotesize same for every intrinsic $X$}};
\node[box,very thick,draw=cInk] (aK) at (10.6,1.35) {out-state\\$\mathsf a_{ij}(\mathbf k)$};
\node[box] (PT) at (5.0,-1.35) {field transform\\$\widetilde{\mathsf A}[Y^X_{ij}](z;\mathbf k)$};
\node[box,dashed] (aY) at (10.6,-1.35) {field residue of $Y^X_{ij}$\\$\mathsf a_{\rm f}[Y^X_{ij}]$, if it exists};
\draw[->,cSlate] (K.north) |- node[lab,left,pos=0.3] {residue at $z=0$\\pair by pair} (Kinf.west);
\draw[->,cSlate] (Kinf) -- node[lab,above] {$\int[\dd q]\cdots\zeta_{\mathbf q}\zeta_{\mathbf p}$} (aK);
\draw[->,cSlate] (K.south) |- node[lab,below,pos=0.75] {$\int[\dd q]\cdots\zeta_{\mathbf q}\zeta_{\mathbf p}$} (PT.west);
\draw[->,cSlate] (PT) -- node[lab,below] {field\\residue} (aY);
\draw[->,very thick,cClay,shorten <=2pt,shorten >=2pt] (aY) -- node[lab,right] {equal\\when it exists} (aK);
\end{tikzpicture}}
\caption{Two orders of taking the residue.
Top: the residue is taken pair by pair and then integrated over $\mathbf q$, defining the gauge-independent out-state $\mathsf a_{ij}$, Eq.~\eqref{eq:kernel-outstate-assembly}.
Bottom: the momentum integral is performed first and the field residue is defined in spectral mean square, Eq.~\eqref{eq:fixed-k-spectral-convergence}.
Near-resonant layers can prevent the field residue from existing.
Whenever it exists, it equals $\mathsf a_{ij}$.}
\label{fig:limits}
\end{figure}
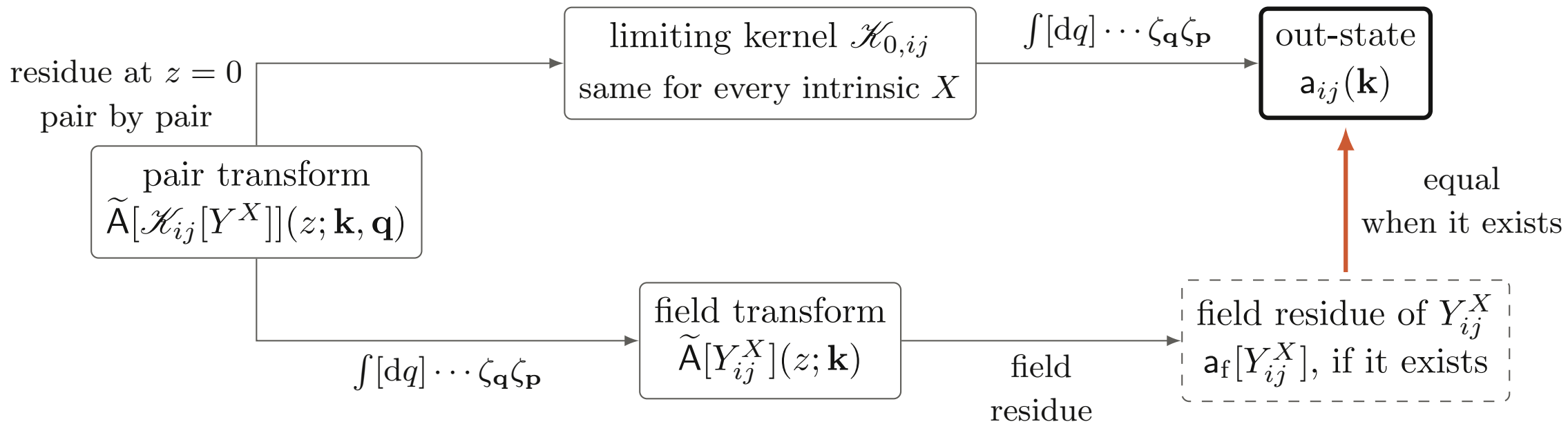

We determine when the tensor perturbation in a given gauge has a residue in spectral mean square after the momentum integral (Fig.~\ref{fig:limits}).
Let $m$ be the growth exponent of the gauge transformation on a resonance surface.
Under the resonance expansion of Appendix~\ref{ap-sec:uniform-shell-form}, this field residue exists and equals the out-state for $m<1/2$, and fails to exist for $m\ge1/2$ when the leading coefficient is nonzero.

\subsection{The field residue and the resonance threshold}\label{ap-sec:field-residue}
\begin{figure}[t]
\centering
\includegraphics[width=\textwidth]{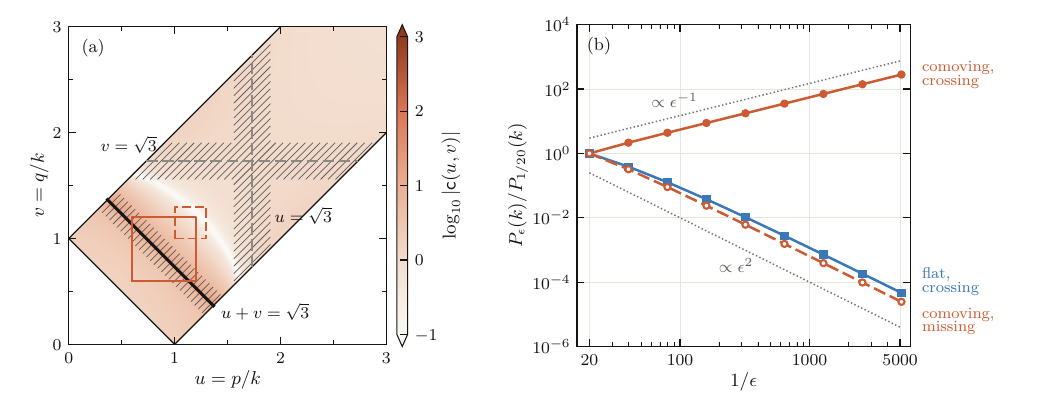}
\caption{(a) Resonance surfaces in the scalar-pair plane $(u,v)=(p/k,q/k)$ for radiation domination, $c_s=1/\sqrt3$, over the shading $\log_{10}|\mathsf c(u,v)|$ of Eq.~\eqref{eq:rd-residue-result} at $w=1/3$.
Scalar pairs obey $|u-v|\le1\le u+v$.
Black line: source resonance surface $u+v=\sqrt3$.
Grey dashed lines: gauge resonance surfaces $u=\sqrt3$ and $v=\sqrt3$.
Hatched regions: resonance layers $|c_s(u+v)-1|,\,|c_su-1|,\,|c_sv-1|<\epsilon$ around these surfaces for $\epsilon=0.1$.
Orange squares: supports $u,v\in[0.6,1.2]$ (solid) and $u,v\in[1.0,1.3]$ (dashed) of the top-hat scalar spectra in (b).
(b) Power spectrum $P_\epsilon(k)\equiv P_{\Rres_\epsilon[Q^X]}(k)$ of the resolution-$\epsilon$ residue, Eq.~\eqref{eq:abel-form}, of the radiation-era gauge shift.
For the solid square, which crosses the source resonance surface, the power of the comoving shift ($m=1$, orange solid) grows as $\epsilon^{-1}$, and that of the flat shift ($m=-1$, blue) decays as $\epsilon^2$, the floor of Eq.~\eqref{ap-eq:fixed-k-upper-bound}.
For the dashed square, which misses every resonance surface, the power of the comoving shift (orange dashed) also decays as $\epsilon^2$.}
\label{ap-fig:shell-variance}
\end{figure}

For Gaussian input, let
\begin{equation}
 {\mathscr U}_\epsilon=\epsilon\widetilde{\mathsf A}[Y_{ij}](\mathrm{i}\epsilon),
\end{equation}
and let ${\mathscr U}_0$ have kernel $\mathscr K_{0,ij}$ on the same scalar realization and initial data.
The field-residue condition is
\begin{equation}\label{eq:fixed-k-spectral-convergence}
 P_{{\mathscr{U}}_\epsilon-{\mathscr{U}}_0}(k)\longrightarrow0\qquad(\epsilon\to0^+),
\end{equation}
and ${\mathscr{U}}_0$ is then the field residue of Sec.~\ref{sec:kernel-def}, with real components $\mathsf a_{\rm f}[Y_{ij}]=(\mathsf a_{{\rm f},c},\mathsf a_{{\rm f},s})$.
By Eq.~\eqref{ap-eq:fixed-k-isometry}, this condition requires the kernel of ${\mathscr U}_\epsilon$ to converge to $\mathscr K_{0,ij}$ with the weight $P_\zeta(q)P_\zeta(p)$.
Pointwise convergence alone does not control a shrinking layer around any of the resonance surfaces
\begin{equation}\label{eq:shell}
 c_s(q+p)=k,\qquad c_sq=k,\qquad c_sp=k .
\end{equation}
The first is the source resonance surface, and the other two are the gauge resonance surfaces of Sec.~\ref{sec:generator-frequencies}, on which a single scalar mode of the shift resonates with the tensor mode.
For a shift amplitude growing as $x^m$, the layer $|\delta|\lesssim\epsilon$ (with $\delta$ the detuning~\eqref{ap-eq:fixed-k-detunings} from the surface) has width $O(\epsilon)$ and resolved amplitude $O(\epsilon^{-m})$.
Its spectrum therefore scales as
\begin{equation}\label{eq:layer-variance}
 P_{\Delta\mathsf a}(k)\sim \epsilon^{-2m}\times\epsilon=\epsilon^{1-2m},\qquad
 \Delta\mathsf a_{ij}=\epsilon\widetilde{\mathsf A}[Q_{ij}](\mathrm{i}\epsilon).
\end{equation}
Under the resonance expansion of Appendix~\ref{ap-sec:uniform-shell-form}, $m<1/2$ gives convergence to the out-state.
For $m\ge1/2$, a spectrum continuous across the resonance surfaces reached by the shift, together with $C_{\rm res}(k)>0$, implies nonexistence (Fig.~\ref{ap-fig:shell-variance}).
Table~\ref{ap-tab:shell-orders} lists $m$ for each standard gauge.

\begin{table}[ht]
\centering\small
\begin{tabularx}{\textwidth}{@{}l l >{\raggedright\arraybackslash}X@{}}
\toprule
Gauge & resonance order $m$ & field residue\\
\midrule
Flat & $-2/(1+3w)$ & equals the out-state\\
Comoving (total matter) & $6w/(1+3w)$ & equals the out-state for $w<1/9$\\
Uniform density & $(2+12w)/(1+3w)$ & absent when the resonance has nonzero weight\\
Uniform expansion & $-2/(1+3w)-2$ & equals the out-state\\
Synchronous & $m\le\max(-1,-2/(1+3w))$ & equals the out-state\\
\bottomrule
\end{tabularx}
\caption{Standard gauges for fixed $0<w<1$, Gaussian regular growing scalar modes with bounded compactly supported spectra, and transported tensor initial data.
Failure in the comoving gauge for $w\ge1/9$ and in the uniform-density gauge requires a spectrum continuous across the source resonance surface and $C_{\rm res}(k)>0$.
At $w=1/9$, the comoving residue at finite resolution stays bounded in spectral mean square but does not converge.
If the spectral support stays away from every resonance, the field residue exists instead.}
\label{ap-tab:shell-orders}
\end{table}

Appendix~\ref{ap-sec:uniform-shell-form} makes this counting precise and shows that the growth is sharp.

\subsection{Resonance expansion, threshold and failure}\label{ap-sec:uniform-shell-form}
We fix $0<c_s<1$ and a bounded spectrum supported on $q_{\min}\le q\le q_{\max}$ with $q_{\min}>0$, so that the support meets only bounded pieces of the resonance surfaces, with detunings
\begin{equation}\label{ap-eq:fixed-k-detunings}
 \delta_+=1-\frac{c_s(q+p)}k,\qquad
 \delta_q=1-\frac{c_sq}k,\qquad\delta_p=1-\frac{c_sp}k.
\end{equation}
Near each surface, with $\delta$ its detuning and $\xi$ coordinates along it, we assume that the kernel amplitude of the shift has the uniform expansion
\begin{equation}\label{ap-eq:transverse-normal-form}
 \mathsf A[\mathscr K_{ij}[Q]]\simeq\Lambda_{ij}{}^{lm}q_lq_m\,\mathrm{e}^{\mathrm{i}\delta x}\sum_{(\kappa,n,l)}x^\kappa(\log x)^l\delta^n a_{\kappa nl}(\xi,\delta),
\end{equation}
with smooth coefficients $a_{\kappa nl}$, up to a remainder consisting of lower-order $(\kappa-n,l)$ terms of the same form and an $O(x^{-N})$ term for some $N>1$, both uniform near the surface.
The time integral of such a term is dominated by $x\sim1/\max(|\delta|,\epsilon)$, since at later times either the phase $\mathrm{e}^{\mathrm{i}\delta x}$ or the regulator $\mathrm{e}^{-\epsilon x}$ cuts it off.
A term $x^\kappa(\log x)^l\delta^n$ with $\kappa>-1$ therefore contributes to the residue at resolution $\epsilon$ an amount of order
\begin{equation}\label{ap-eq:abel-scaling}
 \epsilon\,|\delta|^n\max(|\delta|,\epsilon)^{-\kappa-1}\Bigl(\log\frac1{\max(|\delta|,\epsilon)}\Bigr)^l,
\end{equation}
which scales as
\begin{equation}
 \epsilon^{-(\kappa-n)}\bigl(\log\tfrac1\epsilon\bigr)^l\quad\text{in the resonance layer }|\delta|\lesssim\epsilon,\qquad
 \epsilon\,|\delta|^{n-\kappa-1}\quad\text{outside it},
\end{equation}
while terms with $\kappa\le-1$ contribute $O(\epsilon)$ up to a logarithm.
The resonance order $m$ and its logarithmic order $l_*$ are therefore determined by the largest $(\kappa-n,l)$ among the terms with nonzero coefficients.
After squaring and integrating over the detuning with the bounded spectral weight on the surface, the layer and the pairs outside it contribute
\begin{equation}
 \epsilon\cdot\epsilon^{-2m}=\epsilon^{1-2m}\qquad\text{and}\qquad
 \epsilon^2\int_\epsilon\delta^{-2m-2}\,\dd\delta,
\end{equation}
respectively.
The second integral is dominated by $\delta\sim\epsilon$ for $m>-1/2$, while for $m<-1/2$ it is dominated by finite $\delta$ and scales as $\epsilon^2$.
Hence
\begin{equation}\label{ap-eq:fixed-k-upper-bound}
P_{\Rres_\epsilon[Q]}(k)
 \le C_k\,\epsilon^{\min(1-2m,\,2)}\bigl(1+\log\tfrac1\epsilon\bigr)^{2l_*+\iota_m},\qquad
 0<\epsilon\le\tfrac12,
\end{equation}
with $\iota_m=1$ at $m=-1/2$, where the layer and the pairs at finite detuning contribute at the same order, $\iota_m=2$ for $m\le-1$, and $\iota_m=0$ otherwise.
For $m<1/2$, the shift therefore has zero residue in spectral mean square.
This bound uses only the expansion~\eqref{ap-eq:transverse-normal-form}.

The growth above the threshold is sharp.
In the layer, the time integral is exactly
\begin{equation}\label{ap-eq:abel-power}
\begin{gathered}
 \epsilon\int_0^\infty\mathrm{e}^{(\mathrm{i}\delta-\epsilon)x}x^\kappa\,\dd x
 =\epsilon^{-\kappa}g_\kappa(\sigma),\\
 g_\kappa(\sigma)=\frac{\Gamma(\kappa+1)}{(1-\mathrm{i}\sigma)^{\kappa+1}},\qquad
 \sigma=\frac\delta\epsilon,\quad\kappa>-1,
\end{gathered}
\end{equation}
so a term $x^\kappa\delta^n$ gives the profile $\epsilon^{-(\kappa-n)}\sigma^ng_\kappa(\sigma)$.
If the spectrum is continuous across each resonance surface $\Sigma$, the layer yields, for $m>-1/2$,
\begin{equation}\label{ap-eq:fixed-k-sharp-asymptotic}
\begin{aligned}
 P_{\Rres_\epsilon[Q]}(k)
 &=\epsilon^{1-2m}\bigl(\log\tfrac1\epsilon\bigr)^{2l_*}
 [C_{\rm res}(k)+o(1)],\\
 C_{\rm res}(k)
 &=\sum_\Sigma\int\dd\xi\,W_{\Sigma,k}(\xi)
 \int_{\mathbb R}\dd\sigma\,
 \|\mathbf H_\Sigma(k,\xi,\sigma)\|^2.
\end{aligned}
\end{equation}
Here, in the coordinates $(\xi,\delta)$ and evaluated on $\Sigma$, the measure of the spectrum~\eqref{eq:kernel-outstate-assembly} is
\begin{equation}
 W_{\Sigma,k}(\xi)\,\dd\xi\,\dd\delta=2[\dd q]\,P_\zeta(q)P_\zeta(p)\,\bigl\|\Lambda_{ij}{}^{lm}q_lq_m\bigr\|^2,
\end{equation}
and the coefficients $a_{\kappa nl}$ are those of the scalar amplitude, with the factor $\Lambda_{ij}{}^{lm}q_lq_m$ removed.
The profile $\mathbf H_\Sigma$ is the coefficient of $\epsilon^{-m}(\log\frac1\epsilon)^{l_*}$ in the two real components of $\Rres_\epsilon[Q_{ij}]$ in the layer.
It is given by
\begin{equation}
\begin{gathered}
 \mathbf H_\Sigma=(-\operatorname{Im}\mathcal F_\Sigma,\operatorname{Re}\mathcal F_\Sigma),\qquad
 \|\mathbf H_\Sigma\|^2=|\mathcal F_\Sigma|^2,\\
 \mathcal F_\Sigma\equiv\sum_{\kappa-n=m,\,l=l_*}a_{\kappa nl}(\xi,0)\,\sigma^ng_\kappa(\sigma).
\end{gathered}
\end{equation}
All terms on the same surface are summed before squaring, because they can cancel.
Cross terms between the gauge resonance surfaces $c_sq=k$ and $c_sp=k$, which can intersect, are of order $\epsilon^{2-2\max(m,0)}$ and hence subleading.
This suppression arises because the two layers of width $\epsilon$ overlap only in a region of area $O(\epsilon^2)$ around their intersection in the $(u,v)$ plane.
Thus, whenever $C_{\rm res}(k)>0$ (meaning that the leading profile is nonzero on a subset of a resonance surface with positive measure), the spectrum grows without bound for $m>1/2$ and for $m=1/2$ with $l_*>0$.
The field residue then does not exist.
At $m=1/2$ and $l_*=0$, the spectrum stays bounded, but the residue still fails to converge, since the layer contributions at resolutions $\epsilon$ and $2\epsilon$ differ by a nonvanishing $O(1)$ term as $\epsilon\to0$.

\subsection{Identification of an existing field residue}\label{ap-sec:fixed-k-space}
This subsection shows that a field residue, whenever it exists, equals the out-state, as stated in Sec.~\ref{sec:kernel-def}.
We use the Gaussian input of Sec.~\ref{sec:averaging} and quadratic fields built from the same realization, with the symmetric kernels~\eqref{eq:statistical-pair-kernel}.
For noncompact spectra in radiation domination, we use the bounds of Appendix~\ref{ap-app:gaussian-resonance}.
By Eq.~\eqref{ap-eq:fixed-k-isometry}, if the field ${\mathscr U}_\epsilon$ of Eq.~\eqref{eq:fixed-k-spectral-convergence} converges in spectral mean square, its kernels converge with the weight $P_\zeta(q)P_\zeta(p)$ to a unique symmetric limit.
For almost every pair with nonzero spectral weight, that limit equals the kernel residue $\mathscr K_{0,ij}$, the pairwise limit of the kernels (Appendix~\ref{app:residue-existence}).
This holds because a mean-square limit coincides with the pairwise limit along a subsequence $\epsilon_n\to0$.
This identifies an existing field residue with the kernel residue, without proving that it exists.

On a single physical solution, transporting the tensor initial data gives
\begin{equation}
 Y^X_{ij}-Y^{X'}_{ij}=Q^X_{ij}-Q^{X'}_{ij}.
\end{equation}
For intrinsic gauges on a constant-$w$ background with $0<w<1$, Sec.~\ref{sec:generator-frequencies} shows that this difference has zero kernel residue almost everywhere.
Equation~\eqref{ap-eq:fixed-k-isometry} then implies that any two existing field residues are equal.
More generally, if a gauge shift $Q_{ij}$ has zero kernel residue, any existing field residue of $Y^{\rm N}_{ij}+Q_{ij}$ must coincide with the out-state assembled from $\mathscr K^{\rm N}_{0,ij}$.
This does not require the Newtonian field residue itself to exist.
If that residue does exist, convergence for $Y^{\rm N}_{ij}+Q_{ij}$ is equivalent to $\Rres_\epsilon[Q_{ij}]\to0$ in spectral mean square.

Convergence for almost every pair is weaker than convergence of the field, just as asymptotic limits in scattering theory hold only weakly \cite{Haag:1958vt,Haag:1996hvx}.
For $m>1/2$ with $C_{\rm res}(k)>0$, the kernels converge for every pair off the resonance surfaces, while the spectrum at resolution $\epsilon$ grows without bound.

\subsection{Resonance orders of the standard gauges}\label{ap-app:proofs}
This subsection derives the resonance orders $m$ listed in Table~\ref{ap-tab:shell-orders}.
The time generators of the flat, comoving, uniform-density, and uniform-expansion slicings are standard \cite{Malik:2008im,Hwang:2017oxa,Lu:2020diy}, and follow from Table~\ref{tab:normal-rules} with the late-time Newtonian potential
\begin{equation}
 \Phi\sim\eta^{-(\gamma+1)}\cos(c_sq\eta+\theta),
\end{equation}
where $\theta$ is a fixed phase.
Their leading late-time terms are
\begin{equation}
\begin{gathered}
 \alpha_X\simeq A_X(q)\,\eta^{s_X}\cos(c_sq\eta+\theta_X),\\
 s_{\rm flat}=-\gamma,\qquad s_{\rm com}=1-\gamma,\qquad
 s_{\rm ud}=2-\gamma,\qquad s_{\rm ue}=-\gamma-1,
\end{gathered}
\end{equation}
with $A_X(q)\neq0$ for $q>0$.
For these pure time shifts, $\beta=0$, and the shift~\eqref{eq:gauge-shift-ordinary} reduces to
\begin{equation}
 \Delta h_{ij}=-\Lambda_{ij}{}^{lm}\partial_l\alpha\,\partial_m\alpha .
\end{equation}
Since $Q^X_{ij}=a\Delta h_{ij}$ with $a\propto\eta^\gamma$, its envelope grows as $\eta^{2s_X+\gamma}$, giving
\begin{equation}\label{ap-eq:intro-orders}
m_{\rm flat}=-\gamma,\qquad m_{\rm com}=2-\gamma,\qquad
 m_{\rm ud}=4-\gamma,\qquad m_{\rm ue}=-\gamma-2.
\end{equation}
The product $\partial_l\alpha_X\partial_m\alpha_X$ contains the sum-frequency term
\begin{equation}
 \tfrac12A_X(q)A_X(p)\,\eta^{2s_X}\cos[c_s(q+p)\eta+2\theta_X],
\end{equation}
which resonates on the source resonance surface.
Its TT projection carries the factor $\Lambda_{ij}{}^{lm}q_lq_m$, which vanishes only for collinear pairs.
The Wronskian projection multiplies a component of frequency $\omega k$ by $\mathrm{i}(\omega-1)$, which equals $-2\mathrm{i}$ for the resonant component $\omega=-1$ on that surface.
The leading coefficient is therefore nonzero for every noncollinear resonant pair, and the remaining terms have strictly lower powers uniformly near the surface.
For uniform expansion this check is not needed, since $m_{\rm ue}<1/2$ for every $w$.
Table~\ref{ap-tab:shell-orders} follows, with failure requiring a spectrum continuous across the source resonance surface and $C_{\rm res}(k)>0$.
In particular, $m_{\rm com}<1/2$ is equivalent to $w<1/9$.

For the synchronous generators~\eqref{eq:sync-general-generators}, fix $C$ and $d$ bounded on the support of the spectrum.
Since $a\Phi$ has an oscillatory $\eta^{-1}$ envelope, its integral tends to a constant, and the difference from that constant oscillates as $O(\eta^{-1})$.
Thus $\alpha$ contains a zero-frequency $\eta^{-\gamma}$ term and an oscillating $O(\eta^{-\gamma-1})$ term.
The zero-frequency part of $\beta$ approaches a constant for $\gamma>1$, grows as $\log\eta$ for $\gamma=1$, and grows as $\eta^{1-\gamma}$ for $\gamma<1$.
Its oscillating part is $O(\eta^{-\gamma-1})$.
In Eq.~\eqref{eq:gauge-shift-ordinary}, the terms combining a zero-frequency part with an oscillating part have frequencies $\pm c_sq$ or $\pm c_sp$, so they resonate on the gauge resonance surfaces and obey
\begin{equation}\label{eq:sync-resonance-order}
 m_{\rm sync}\le\max(-1,-\gamma),
\end{equation}
with at most one logarithm at $\gamma=1$.
Products of two oscillating parts do not raise the exponent, while pure zero-frequency terms are off resonance.
This bound includes fixed residual modes and lies below the threshold $m=1/2$.
Appendix~\ref{app:histories} treats initial regularity separately.

The estimates in this appendix assume a bounded spectrum supported away from $q=0$ and $q=\infty$.
Extending them to noncompact spectra requires, for each gauge, infrared and ultraviolet bounds on its shift analogous to those of Appendix~\ref{ap-app:gaussian-resonance} for the Newtonian source, which do not carry over automatically.

\FloatBarrier
\section*{Notation index}
{\footnotesize\setlength{\LTcapwidth}{\textwidth}\setlength{\LTleft}{0pt}\setlength{\LTright}{0pt}
\begin{longtable}{@{\extracolsep{\fill}}l >{\raggedright\arraybackslash}p{0.66\textwidth} l@{}}
\caption{Notation index for the symbols used throughout the paper.
The last column gives the equation or section in which each symbol is defined.}
\label{tab:notation}\\
\toprule
Symbol & Meaning & Definition\\
\midrule
\endfirsthead
\toprule
Symbol & Meaning & Definition\\
\midrule
\endhead
\bottomrule
\endfoot
$\eta$, $'$ & Conformal time. A prime denotes a derivative with respect to $\eta$. & \eqref{eq:background}\\
$a$ & Scale factor of the flat FLRW background. & \eqref{eq:background}\\
$\mathcal H$ & Conformal Hubble rate, $\mathcal H\equiv a'/a$. & \eqref{eq:background}\\
$x$ & Dimensionless time at fixed external momentum, $x\equiv k\eta$, with $\partial_x=k^{-1}\partial_\eta$. & Sec.~\ref{sec:setup}\\
$h_{ij}$ & Second-order tensor perturbation, $h_{ij}\equiv\tfrac12a^{-2}g_{ij}^{(2)\TT}$. & \eqref{eq:h-def}\\
$Y_{ij}$ & Rescaled tensor perturbation, $Y_{ij}\equiv ah_{ij}$, the tensor Mukhanov--Sasaki variable up to a constant normalization. & \eqref{eq:general-Y}\\
$\Phi$ & First-order Bardeen potential of the growing scalar mode. & Sec.~\ref{sec:general-background}\\
$\Phi_*$ & Initial scalar amplitude, $\Phi_*=3(1+w)\zeta_*/(5+3w)$, equal to $2\zeta_*/3$ in radiation domination and $3\zeta_*/5$ in matter domination. & \eqref{eq:Phi-star}\\
$\zeta$ & Primordial curvature perturbation. & Sec.~\ref{sec:general-background}\\
$\Lambda_{\rm c}$ & Scalar cutoff of the spectrum support, $\Lambda_{\rm c}^{-1}\le q/q_*\le\Lambda_{\rm c}$. & Sec.~\ref{sec:averaging}\\
$\Lambda_{ij}{}^{lm}$ & Transverse-traceless projector at the external momentum. & \eqref{eq:TT-projector}\\
$u$, $v$ & Internal momenta in units of $k$, $u\equiv p/k$ and $v\equiv q/k$. & Sec.~\ref{sec:setup}\\
$[\dd q]$ & Momentum measure, $[\dd q]\equiv\dd^3q/(2\pi)^3$. & Sec.~\ref{sec:setup}\\
$\mathscr K_{ij}[\cdot]$ & Symmetric kernel of a quadratic field, multiplying $\zeta_{\mathbf q}\zeta_{\mathbf p}$. & \eqref{eq:statistical-pair-kernel}\\
$P_{\mathscr U}$ & Power spectrum of a Fourier field $\mathscr U$. & \eqref{eq:field-power-spectrum}\\
$\alpha$, $\beta$ & Components of the first-order scalar gauge generator $\xi_1=(\alpha,\partial^i\beta)$ from Newtonian gauge to gauge $X$. & Sec.~\ref{sec:redefinitions}\\
$Q^X_{ij}$ & Gauge shift of the rescaled tensor perturbation, $Q^X_{ij}\equiv Y^X_{ij}-Y^{\rm N}_{ij}$. & Sec.~\ref{sec:redefinitions}\\
$\Qcal$ & Linear space of fields quadratic in the scalar input. & \eqref{eq:gauge-problem}\\
$u_c$, $u_s$, $W$ & Real free modes and their Wronskian, $W[f,g]\equiv f\partial_xg-g\partial_xf$. & Secs.~\ref{sec:general-background}, \ref{sec:outstate-def}\\
$\mathsf h[\cdot]$ & Wronskian components $(\mathsf h_c,\mathsf h_s)$ of a tensor perturbation against the free modes. & \eqref{eq:wronskian-pair}\\
$\varphi$ & Free mode paired with $Y_{ij}$ in $\mathsf A$, $\varphi\equiv u_c+\mathrm{i}u_s\simeq\mathrm{e}^{\mathrm{i}x}$ at late times. & \eqref{eq:psi-def}\\
$\mathsf A[\cdot]$ & Complex free amplitude of a tensor perturbation, $\mathsf A[Y_{ij}]\equiv W[\varphi,Y_{ij}]$. & \eqref{eq:complex-amplitude}\\
$\mathsf c_{ij}$ & Complex amplitude of a single mode, $\mathsf c_{ij}=\mathsf a_{s,ij}-\mathrm{i}\mathsf a_{c,ij}$. & Sec.~\ref{sec:outstate-def}\\
$\mathsf a_{ij}$ & Real amplitudes $(\mathsf a_{c,ij},\mathsf a_{s,ij})$ of a free solution. & \eqref{eq:real-out-amplitudes}\\
$\mathsf a^{{\rm N},{\rm ini}}_{ij}$, $\mathsf a^{\rm ind}_{ij}$ & Free amplitude of the transported initial data and out-amplitude of the purely induced Newtonian tensor perturbation. & \eqref{eq:intrinsic-invariance}\\
$\widehat s_{ij}$ & Transform of the source projected on $\varphi$, $\widehat s_{ij}(z)\equiv k^{-2}\int_0^\infty\mathrm{e}^{\mathrm{i}zx}\varphi s_{ij}\,\dd x$. & \eqref{eq:transform-identity}\\
$Y^\out_{ij}$, $Y^\refr_{ij}$ & Out-state and forced response, $Y_{ij}=Y^\out_{ij}+Y^\refr_{ij}$. & \eqref{eq:outstate}, \eqref{eq:forced-remainder}\\
$\mathscr K_{0,ij}$ & Kernel residue, the residue of the pair transform at the light-cone pole. & \eqref{eq:kernel-outstate-limit}\\
$\delta_\nu$ & Detuning of a frequency component from the tensor frequency, $\delta_\nu=1+\omega_\nu$. & Sec.~\ref{sec:kernel-def}\\
$\delta_0$ & Detuning gap: pairs with $\min_\nu|\delta_\nu|<\delta_0$ are discarded. & \eqref{eq:gapped-history}\\
$\epsilon$ & Resolution of the adiabatic approach $z=\mathrm{i}\epsilon\to0$ to the light-cone pole. & \eqref{eq:residue-definition}\\
$\Rres_\epsilon$ & Out-amplitude at resolution $\epsilon$, whose limit $\epsilon\to0^+$ is the residue. & \eqref{eq:residue-at-resolution}\\
$m$ & Resonance order, the growth exponent of a gauge shift on a resonance surface, which sets the $m<1/2$ threshold. & Appendix~\ref{ap-app:fixed-k-proof}\\
$b$ & Background exponent, $b\equiv(1-3w)/(1+3w)$, so that $a\propto\eta^{1+b}$. & Sec.~\ref{sec:observables}\\
$\gamma$ & Background exponent, $\gamma\equiv2/(1+3w)=1+b$, so that $a\propto\eta^\gamma$, which is also the decay exponent of the resonant source. & Sec.~\ref{sec:lsz}\\
\end{longtable}
}

\FloatBarrier
\clearpage
\bibliographystyle{JHEP}
\bibliography{ref_inspire_jhep}
\end{document}